\documentclass[%
 reprint,
superscriptaddress,
 amsmath,amssymb,
 aps,
prb,
]{revtex4-2}

\usepackage{booktabs}
\usepackage{multirow}
\usepackage{color}
\usepackage{graphicx}
\usepackage{dcolumn}
\usepackage{bm}
\usepackage{hyperref}
\usepackage[mathlines]{lineno}
\usepackage{siunitx}
\usepackage{amsmath}

\begin{document}

\title{Universally optimizable strategy for magnetic gaps towards high-temperature quantum anomalous Hall states via magnetic-insulator/topological-insulator building-blocks}

\author{Zhe Li}%
 \email{lizhe21@iphy.ac.cn}
\affiliation{%
	Beijing National Research Center for Condensed Matter Physics, and Institute of Physics, Chinese Academy of Sciences, Beijing 100190, China
}%

\author{Feng Xue}
\affiliation{%
	College of Physics \& Optoelectronic Engineering, Jinan University, Guangzhou 510632, China
}%

\author{Xin-Yi Tang}
\affiliation{%
  State Key Laboratory of Low Dimensional Quantum Physics, Department of Physics, Tsinghua University, Beijing 100084, China
}%

\author{Xiyu Hong}
\affiliation{%
	State Key Laboratory of Low Dimensional Quantum Physics, Department of Physics, Tsinghua University, Beijing 100084, China
}%

\author{Yang Chen}
\affiliation{%
	State Key Laboratory of Low Dimensional Quantum Physics, Department of Physics, Tsinghua University, Beijing 100084, China
}%

\author{Xiao Feng}
\affiliation{%
  State Key Laboratory of Low Dimensional Quantum Physics, Department of Physics, Tsinghua University, Beijing 100084, China
}%

\affiliation{%
	Beijing Academy for Quantum Information Sciences, Beijing 100193, China
}%

\affiliation{%
	Frontier Science Center for Quantum Information, Beijing 100084, China
}%

\affiliation{%
	Hefei National Laboratory, Hefei 230088, China
}%

\author{Ke He}
\affiliation{%
	State Key Laboratory of Low Dimensional Quantum Physics, Department of Physics, Tsinghua University, Beijing 100084, China
}%

\affiliation{%
	Beijing Academy for Quantum Information Sciences, Beijing 100193, China
}%

\affiliation{%
	Frontier Science Center for Quantum Information, Beijing 100084, China
}%

\affiliation{%
	Hefei National Laboratory, Hefei 230088, China
}%

\date{\today}

\begin{abstract}
Optimizing the magnetic Zeeman-splitting term, specifically the magnetic gap of the topological surface states (TSSs), is a crucial issue and central challenge in advancing higher-temperature quantum anomalous Hall (QAH) states. In this work, we demonstrate a counterintuitive, nonmonotonic relationship between the magnetic gap and the hybridization strength in ferromagnetic-insulator (FMI)/topological-insulator (TI) sandwich structures. Concretely, insufficient hybridization strength fails to induce a substantial magnetic gap; while excessive hybridization incandesces the competition between kinetic and Coulomb exchange interactions, thereby reducing the gap. Strong hybridization strength also spatially delocalizes the TSSs, diminishing the effective orbital coupling between TSS-based $p$ and magnetic $d$ orbitals, which further weakens kinetic and Coulomb exchange interaction strength. Moreover, modifying the stacking order offers an experimentally viable approach to optimizing magnetic gaps, enabling the tunability of Chern numbers, chirality and maximizing global gaps. These findings unveil a universal guiding principle for optimizing magnetic gaps in FMI-TI proximity-based QAH systems, offering valuable insights for advancing experimental implementations in this field.
\end{abstract}

\maketitle
\section{introduction}
 \label{i:introduction}
 
Pursuing large band gap, high critical temperature is the basic and pivotal topic in designing quantum anomalous Hall (QAH) systems, which contains dissipationless quantized current generated by the momentum-space-based Berry flux monopole \cite{weng2015quantum,haldane1988model,qi2006topological,yu2010quantized,chang2013experimental,mogi2015magnetic,ou2018enhancing,hasan2010colloquium,qi2011topological}. Behaved as the most prominent quantum effect in topology, the highlighted issue of overcoming its ultralow-temperature requirement \cite{chang2013experimental} attracts enormous attentions. Researchers have made great efforts to increase the critical temperature of the long-ranged magnetism in various QAH systems these years \cite{mogi2015magnetic,ou2018enhancing,hasan2010colloquium,qi2011topological,li2019intrinsic,otrokov2019unique,zhang2019topological,gong2019experimental,deng2020quantum,liu2020robust,ge2020high,bai2024quantized,li2020tunable,zhang2020large,tang2023intrinsic,kobialka2022dynamical,hou2019magnetizing,eremeev2013magnetic,men2013magnetic,grutter2021magnetic,fang2023exchange,zhang2018strong,zou2020intrinsic}. For example, by replacing the diluted magnetic-doped topological insulators (TIs) with the intrinsic magnetic-element-intercalated TIs \cite{li2019intrinsic,otrokov2019unique,zhang2019topological,gong2019experimental,deng2020quantum,liu2020robust,ge2020high,bai2024quantized,li2020tunable,zhang2020large,tang2023intrinsic,kobialka2022dynamical}; by establishing heavily magnetic-doped insulator/TI heterostructures that benefit from the van der Waals (vdW) based nature of these two-dimensional materials \cite{mogi2015magnetic}; by directly designing and constructing non-doped ferromagnetic insulator (FMI)/TI heterostructures \cite{hou2019magnetizing,eremeev2013magnetic,men2013magnetic,grutter2021magnetic,fang2023exchange,zhang2018strong,zou2020intrinsic,cgt-alegria2014large,cgt-mogi2019large,cgt-yao2019record,YFO-jiang2015independent,YFO-jiang2016enhanced,TFO-tang2017above,LCO-zhu2018proximity,Fe3O4-pereira2020topological,Cr2O3-wang2019observation,EuS-katmis2016high,EuS-lee2016direct,EuS-wei2013exchange,CrSb-he2017tailoring,CrSb-he2018topological,MnTe-he2018exchange,GaMnAs-lee2018engineering,ZnCrTe-watanabe2019quantum}, and so on. Furthermore, merging magnetism and topology within the $d$-orbitals, utilizing $d$-$d$ correlated-electron-based topology, breaks away from the traditional approach of coupling between magnetism and topology, unlocking the potential for devices to operate at even higher temperatures. These predicted systems including $YX_3$ ($Y$ = transition metal elements, $X$ = halogen elements) \cite{you2019two,li2023tunable}, Ni(Pd)As(Sb, Bi)O$_3$ \cite{li2022chern,wu2023robust}, LiFeSe \cite{li2020high}, V$_2$W(Mo)S(Se)$_4$ \cite{jiang2024monolayer}, etc. Nevertheless, none of these candidates has been successfully fabricated under the experimental regime, nor has been characterized by high-temperature quantized dissipationless conductance, hindered by the strict demand for fabrication conditions. Hence, optimizing physical characters of the traditional FMI/TI heterostructures or magnetic TI systems still remains the major base towards experimental investigations on QAH effects.

In contrast to tremendous investments in magnetic critical temperatures of numerous candidate QAH systems, a comprehensive understanding of how to modulate, enhance, and optimize the Zeeman-splitting mass term, or referred to as, “the magnetic gap”, is still lacking. Typically, previous works were focus on the two separate mechanisms contributing to the mass terms, including $p$-$d$ orbital hybridizations via kinetic exchange and Coulomb exchange \cite{xu2022controllable,kacman2001spin}, which, in most cases, exhibit opposite chirality. The size of the magnetic gap, however, depends not only on the chirality of different exchange mechanisms but also on the real-space distributions of topological surface states (TSSs). Evolved with the coupling strength between magnetism and topology, the distribution of TSSs cast an even complicated guiding factor towards modulating magnetic gaps.  Previously, researchers built magnetic insulator (MI)/TI heterostructures to investigate its TSS rearrangement along out of the stacking directions \cite{otrokov2017highly,tss-li2024realization,tss-shoman2015topological,tss-wu2013tuning,tss-zhang2012exotic,tss-zhang2017electronic}, among which X. Li \textit{et al} \cite{tss-li2024realization} systematically computed and discussed the pinning, floating and diving behaviors of TSS under the MI/TI proximity basis, and provided the candidate materials that engender the above three phenomenon respectively, giving a vital conclusion that moderate hybridization strength between MI and TI arrives the floating of TSS into the MI layer which is inclined to create large gapped QAH states.  Here, via utilizing the representative FMIs such as Cr$_2$$X_2$Te$_6$ ($X$ = Si, Ge, Sn) and CrI$_3$ \cite{huang2017layer,gong2017discovery,liu2019anisotropic,xue2023thickness}, and TIs composed of Bi(Sb)$_2$Te$_3$ and Mn(Ni)Bi$_2$Te$_4$ \cite{li2019intrinsic,otrokov2019unique,zhang2019topological,gong2019experimental,deng2020quantum,liu2020robust,ge2020high,bai2024quantized,li2020tunable,tang2023intrinsic,kobialka2022dynamical,zhang2009topological,chen2009experimental,liu2010model}, and by implementing multiform continuous tuning methods, such as artificially varying the vdW-spacing distances between FMI and TI, or manipulating stacking-order-shifts, we capture	 the non-monotonic evolution properties of TSSs, and then, the size of the magnetic gap. The competition between kinetic exchange and Coulomb exchange also limits the enlargement of the magnetic gap. Consequently, there exists a maximum value of the magnetic gap along with the optimization of the hybridization strength between FMI and TI. Noteworthily, with its multifunctional containments encompassing magnetism, topology and vdW-layered nature, MnBi$_2$Te$_4$ serves as a widely studied basis among intrinsic magnetic topology, magnetic substrate and Floquet engineering\cite{li2019intrinsic,deng2020quantum,liu2020robust,ge2020high,bai2024quantized,li2020tunable,zhang2019topological,gong2019experimental,kobialka2022dynamical,otrokov2019unique,zhu2023floquet,fan2024circularly}, etc.  When Mn(Ni)Bi$_2$Te$_4$ is selected as the TI layer, the opposite chirality between Cr and Mn(Ni), combined with an experimentally operable stacking-order shift method, enables continuous modulation of both the size and chirality of the magnetic gap, resulting in Chern-number and chirality tunable characteristics. For the first time, our revealments provide a comprehensive guiding principle for how to optimize the magnetic gaps of most of the traditional QAH systems that experimentally available up to now, which finds an extensive use in assisting researchers to design and modulate both large-gapped and high-critical-temperature QAH materials. 

The rest of this paper is arranged as follows. In Sec. \ref{ii:methods}, we list all the computational and data-processing methods related in this work. In Sec. \ref{iii:vdW-spacing}, via artificially adopting vdW-spacing-distance variable method, we systematically reveal and analyze the non-monotonic character of the evolution behaviors between magnetic gaps and the FMI-TI hybridization strength. In Sec. \ref{iv:stacking}, based on the experimentally executable technique of stacking-order-shift, we map the distributions of magnetic gaps, global gaps and Chern numbers in typical FMI/TI heterostructures, verifying that both the gaps and Chern numbers are continuously tunable and optimizable within an experimentally achievable regime. Section \ref{v:Tc} illustrates the magnetic critical temperatures of typical FMI/TI systems discussed in this work. We measure the robustness of them under multifold conditions including stacking-orders and biaxial strains. Section \ref{vi:conclusions} summarizes the main findings and design strategies towards optimizing the magnetic gaps in most traditional FMI/TI-proximate QAH systems.

In Appendix \ref{A:structural}, in order to confirm the structural stability, phonon spectrums of FMIs and TIs that mentioned in this work under various in-plane lattice constants are depicted by density perturbation functional theory (DFPT) assisted with frozen-phonon method. Then, we unfold the evolution performances of the hybridization gaps and the $\Gamma$-point gaps versus (vs.) the vdW-spacing distances based upon the building-blocks discussed in the main text in Appendix \ref{B:hybridization}.   In order to further verify the robustness of the non-monotonic evolution characters within the magnetic gaps, we make trials by varying the vdW corrections, and by adopting the bi-heterostructure basis as FMI/TI building blocks in Appendix \ref{C:Robustness}.   In Appendix \ref{D:M-Gap_others}, we discuss the behavior of the magnetic gap (i. e. the distribution of TSSs) vs. the vdW-spacing distances of FMI/TI heterostructures under various conditions, including antiferromagnetic (AFM) couplings and other forms of heterostructures. Besides, the signs and strengths of interlayer magnetic couplings are listed in Tables \ref{tab4:FMI-FMIs}-\ref{tab6:MBT-MBTs}. Stacking-order-dependent parameters, involving the $\Gamma$-point gaps, the hybridization gaps and the TSS distribution for some discussed systems, are enumerated as mapping distributions in Appendix \ref{E:stacking-order}. Appendix \ref{F:Tc} discusses further information and analysis of exchange coefficients and critical temperatures in Cr$_2$$X_2$Te$_6$/TI-based heterostructures.

\section{methods}
 \label{ii:methods}

All the first-principle computations implemented on the systems discussed in this work are based on density functional theory (DFT), all executed with Vienna ab initio simulation package (VASP) \cite{kresse1996efficient}. The generalized gradient approximation (GGA) functional is utilized in all the computations with Perdew-Burke-Ernzerhof (PBE) functional \cite{perdew1996generalized}. Considering the larger cutoff-energy demands of 3$d$-transitional elements (including Cr, Mn and Ni), we uniformly use 400eV as the plane-wave cutoff energy for all the systems. For these 3$d$-transitional elements, in order to treat the Hubbard $U$ terms created by the highly correlated 3$d$ electrons, we choose $U$ as 4 for Cr, Mn and Ni referring to the previous works \cite{li2019intrinsic,li2020tunable,tang2023intrinsic,li2024multimechanism}, and test the robustness of magnetic-gap evolution behaviors by varying $U$ as 1.0, 1.5, 2.0, 3.0, 4.0 and 5.0 in the 3$d$ orbitals of Cr. For the sake of imitating the two-dimensional (2D) characters, 15Å of vacuum layer is set perpendicular to the vdW layers. For relaxation and static self-consistent steps, we choose $\Gamma$-point based 7$\times$7$\times$1 k-mesh grid for most general systems, and 5$\times$5$\times$1 k-mesh grid for the fast-positioning of the Chern number and the extremum of magnetic gaps. For self-consistent steps within that of magneto crystalline anisotropy energies (MAEs), we carry on 15$\times$15$\times$1 k-mesh grid. The convergent criterion is chosen as 0.01eV/Å for structure optimization and 1$\times$$10^{-6}$eV for electronic-step iteration.  DFT-3D method with Becke-Johnson damping function (by setting “IVDW = 11”) is adopted to treat the vdW corrections \cite{grimme2010consistent} as the conventional method in the most investigations and discussions. For the sake of testing the influences stemming from the selection of vdW corrections, we also employ other vdW corrections containing DFT-D2 \cite{ivdw10-grimme2006semiempirical} (by setting “IVDW = 10”), DFT-D3 method with Becke-Johnson damping function \cite{ivdw12-grimme2011effect} (by setting “IVDW = 12”), Tkatchenko-Scheffler method \cite{ivdw20-tkatchenko2009accurate} (by setting “IVDW = 20”), Tkatchenko-Scheffler method with iterative Hirshfeld partitioning \cite{ivdw21-bucko2013improved,ivdw21-buvcko2014extending} (by setting “IVDW = 21”),  Many-body dispersion (MBD) energy method \cite{ivdw202-tkatchenko2012accurate} (by setting “IVDW = 202”) and DFT-ulg method \cite{ivdw3-kim2012universal} (by setting “IVDW = 3”). For the band-structure data-processing, we use VASPKIT code \cite{wang2021vaspkit} to accomplish it; meanwhile we implement the package of PHONOPY \cite{togo2015first} to obtain the space group (SG) of the system, to structurally correct the lattice to its optimized symmetry, and to do data-processing along phonon-spectrum computations operated with DFPT. For several lattice-stretched monolayer FMIs, including CrI$_3$, Cr$_2$Si$_2$Te$_6$ and Cr$_2$Sn$_2$Te$_6$, we further adopt frozen-phonon method to confirm the structural stability.

In order to obtain critical temperatures ($T_{\rm C}$) of long-range magnetism, Monte Carlo (MC) simulations are employed with Heisenberg spin model via adopting our group-made package, assisted by the \textit{mcsolver} package \cite{liu2019magnetic} that is applied to quickly get the critical temperatures of systems under various stacking-orders. We have checked that the outcoming $T_{\rm C}$ simulated from the above two methods under the same system coincides exactly with the deviation smaller than 2K. In the MC simulations, we choose 16$\times$16$\times$1 superlattices with larger than $10^6$ steps of Wolff algorithm to reach the thermal equilibrium. The resolution of the temperature is chosen as 1K to seek the value of $T_{\rm C}$.

For the tight-binding model fitting process, we choose Wannier90 package \cite{mostofi2014updated} to get the tight-binding Hamiltonian (TBH) based on the maximally localized Wannier functions \cite{marzari1997maximally,souza2001maximally}, and utilizing WannierTools package \cite{wu2018wanniertools} to further check the topological edge states and the concrete Chern numbers.

\section{Non-monotonic Evolution Behaviors}
 \label{iii:vdW-spacing}

In this section, for the sake of retaining inversion symmetry which is most suitable for FMI/TI heterostructure investigations, we build FMI/TI/FMI sandwich structures to facilitate magnetic-gap analysis protected by inversion symmetry. In these systems, FMI contributes magnetic Zeeman-splitting interactionsC, meanwhile the medial TI layer contributes topological nontrivial characters with band inversions at $\Gamma$-point. We choose this sandwich model in consideration of simplifying the evolutions of band gaps, and the magnetic-gap model into the classical model of R. Yu \emph{et al}. \cite{yu2010quantized}, facilitated to analyze. Besides, the revealment and guiding principles gained in this work can also be applied to most $\Gamma$-point-based magnetic topological materials. Artificially, we select the method of vdW-spacing-distance variation between FMI and TI to continuously modulate the FMI-TI hybridization strength. 

In subsection \ref{iii-A:non-monotonic}, firstly we adopt Cr$_2$$X_2$Te$_6$/six-layer-Bi$_2$Te$_3$ (BT)/Cr$_2$$X_2$Te$_6$ and CrI$_3$/six-layer-Bi$_2$Te$_3$ (BT)/CrI$_3$ as paradigmatic systems to depict their non-monotonic evolution outcomes of magnetic gaps. In subsection \ref{iii-B:distribution}, via charge-density-distribution computations, we cast the distributions of TSS (related to the four bands at the $\Gamma$ point in the vicinity of the Fermi level) along the out-of-vdW-plane direction and carefully observe its real-space redistributions depending on the hybridization strength between FMI and TI. Focus on the robustness of the non-monotonic behaviors, in subsection \ref{iii-C:lattice_constants} we verify that imposing moderate biaxial strains, or adjusting in-plane lattice constants into the system causes no damage to this feature. Considering magnetic topological insulator that intercalated as the medial layer, and taking Cr$_2$Ge$_2$Te$_6$/trilayer-MnBi$_2$Te$_4$/Cr$_2$Ge$_2$Te$_6$ (abbreviated as CGT/TL-MBT/CGT) as the landmark representation, we carefully explore the magnetic-gap evolvement containing competitive contributions rooting from the Cr-based and Mn-based magnetic Zeeman splitting mass terms in subsection \ref{iii-D:competitions}. We give a final summary of the Sec. \ref{iii:vdW-spacing} compared to previous related works in subsection \ref{iii-E:discussions}.

\begin{center}
	\begin{figure*}
		\centering
		\includegraphics[width=0.92\linewidth]{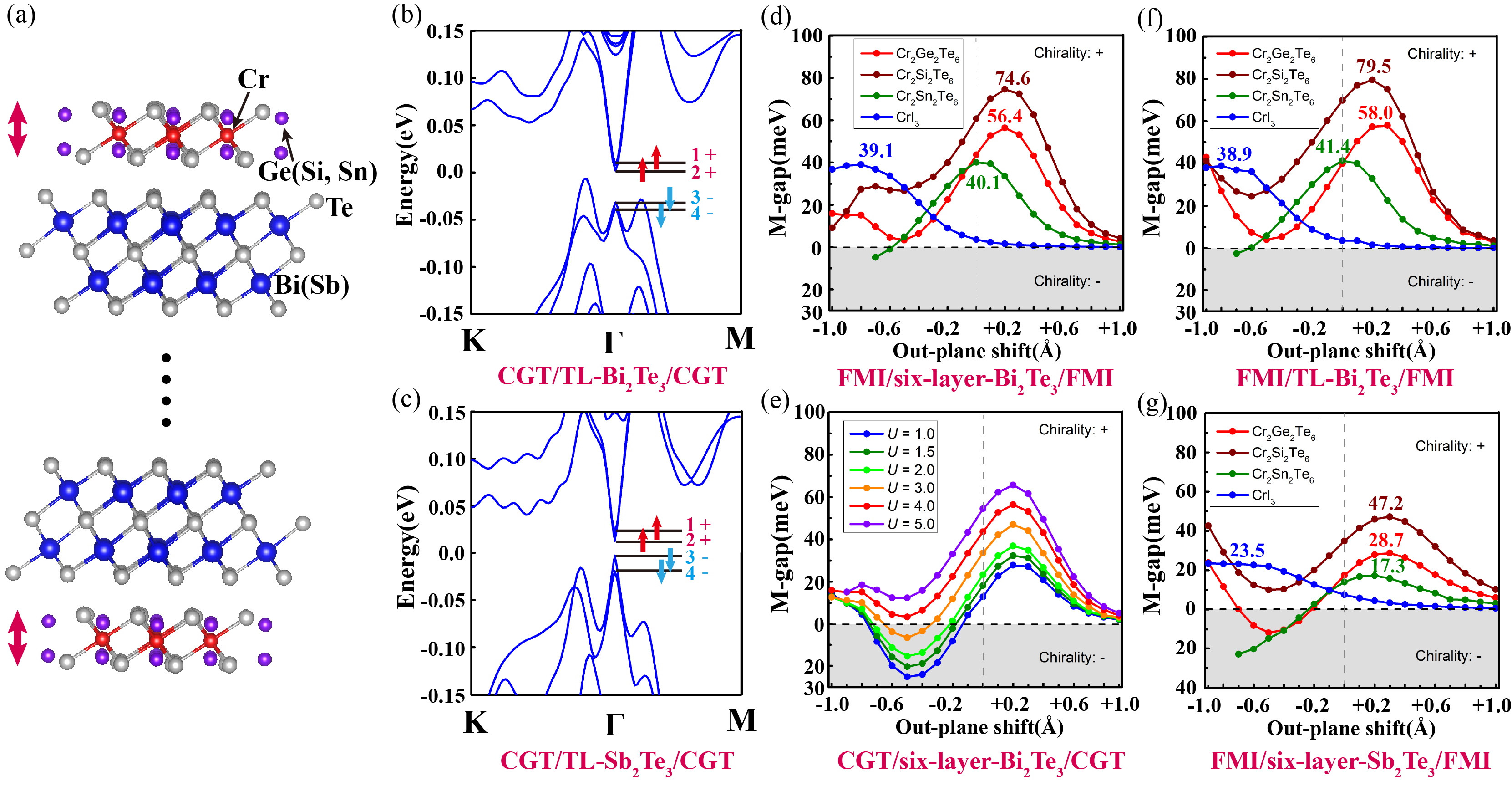}
		\caption{Structures, four-band illustrations and vdW-spacing-distance induced magnetic-gap evolutions in FMI/several-layer-Bi(Sb)$_2$Te$_3$/FMI. (a) Side-view bond-connected-ball structures of FMI/several-layer-Bi(Sb)$_2$Te$_3$/FMI. Red, purple, blue and gray balls are related to Cr, Ge(Si, Sn), Bi and Te atoms. (b) and (c) reveals the band-structure of CGT/TL-BT/CGT and CGT/TL-ST/CGT respectively. Blue and red short-arrows denote the down and up spins, corresponding to the odd and even parities respectively. (d) Magnetic-gap evolutions of FMI/six-layer-BT/FMI along with vdW-spacing distances in which FMI = Cr$_2$$X_2$Te$_6$ ($X$ = Si, Ge, Sn) and CrI$_3$. The “out-plane shift” means the out-plane-shift values of FMI that comparing to the balanced position, the latter drawn with light-gray dotted line. White and gray zones stand for the positive and negative chirality of the magnetic Zeeman splitting respectively. (e) Magnetic-gap evolutions of CGT/six-layer-BT/CGT along with vdW-spacing distances under various values of $U$ that implemented on 3$d$ orbitals of Cr. (f) and (g) depict the magnetic-gap evolutions under building-blocks of FMI/TL-BT/FMI and FMI/six-layer-ST/FMI.}
		\label{fig1:m_gap}
	\end{figure*}
\end{center}

\subsection{Non-monotonic evolving behaviors of magnetic gap}
 \label{iii-A:non-monotonic}

The sandwich structures adopted in this subsection is illustrated in Fig. \ref{fig1:m_gap}(a), in which red, purple, blue and gray atoms stand for Cr, $X$ ($X$ = Si, Ge, Sn), Bi(Sb) and Te atoms respectively. The red double-arrows note the out-of-plane shift of Cr$_2$$X_2$Te$_6$ in order to engineer the FMI-TI hybridizations. For convenience, firstly we opt the in-plane lattice constant as \textbf{\textit{a}} = 4.38Å in the following FMI/TI/FMI heterostructures and execute relaxations via fixing lattice constants, aligning to the condition of thick layers of BT that serve as the medial layer. For the sake of testing whether the lattice constants impact the non-monotonic property, we carry out relative trials in sub-section \ref{iii-C:lattice_constants}. Absolutely no-virtual frequency exists in this lattice constants of all the selected FMIs, phonon spectrums displayed in Appendix \ref{A:structural} with DFPT and frozen-phonon methods (see Figs. \ref{fig13:phonon-CGT}-\ref{fig15:phonon-TI}).

Employing R. Yu \textit{et al}.’s model \cite{yu2010quantized}, we show the four-band of near-Fermi-energy $\Gamma$-point within CGT/TL-BT/CGT in Fig. \ref{fig1:m_gap}(b) and CGT/TL-Sb$_2$Te$_3$/CGT (CGT/TL-ST/CGT) in Fig. \ref{fig1:m_gap}(c), and name the serial number as 1, 2, 3, 4 along the energy-decreasing sequence. “+” and “-” are corresponding to the even and odd parities of the $p_z$ orbitals respectively. Being analogous to the magnetic gap, the hybridization gap is created by the hybridization between top and bottom TSS of a 2D TI. According to the above notation, we can define the gap, the hybridization gap and the magnetic gap as follows according to parity-sequence shown in CGT/TL-BT/CGT and CGT/TL-ST/CGT:

\begin{equation}
\Delta E_{\rm hybrid} = \frac{1}{2}(E_{\rm 1+}-E_{\rm 2+}+E_{\rm 3-}-E_{\rm 4-})
 \label{eq1}
\end{equation}
\begin{equation}
\Delta E_{\rm mag} = \frac{1}{2}(E_{\rm 1+}+E_{\rm 2+}-E_{\rm 3-}-E_{\rm 4-})
 \label{eq2}
\end{equation}
\begin{equation}
\Delta E_{\rm gap} = \Delta E_{\rm mag}-\Delta E_{\rm hybrid}
 \label{eq3}
\end{equation}

$\Delta E_{\rm gap}$ is the gap at the $\Gamma$-point. Eqs.\ref{eq1}-\ref{eq3} enumerate the relations among the gap, the magnetic gap and the hybridization gap under “++- -” parities which is noted as $C$ = +1 that relying on the criterion put forward in Ref. \cite{xu2022controllable}. Eqs.\ref{eq1}-\ref{eq3} valid also in the case $C$ = -1 with parities “- -++”. Nevertheless, for the case of $C$ = 0 that with “+-+-” or “-+-+”, Eqs. \ref{eq1}-\ref{eq3} should be revised into (take "+-+-" as example):

\begin{equation}
\Delta E_{\rm hybrid} = \frac{1}{2}(E_{\rm 1+}+E_{\rm 2-}-E_{\rm 3+}-E_{\rm 4-})
 \label{eq4}
\end{equation}
\begin{equation}
\Delta E_{\rm mag} = \frac{1}{2}(E_{\rm 1+}-E_{\rm 2-}+E_{\rm 3+}-E_{\rm 4-})
 \label{eq5}
\end{equation}
\begin{equation}
\Delta E_{\rm gap} = \Delta E_{\rm hybrid}-\Delta E_{\rm mag}
 \label{eq6}
\end{equation}

Hence, conveniently we’re in a position to obtain the magnetic gaps and the hybridization gaps according to the Eqs. \ref{eq1}-\ref{eq6} derived from R. Yu \textit{et al}.’s model \cite{yu2010quantized}, benefiting from the simple $\Gamma$-point-based, four-band-based topology in the FMI/several-layer-Bi(Sb)$_2$Te$_3$/FMI.

In view of suppressing the hybridization gap \cite{liu2010oscillatory} and maintaining higher lattice-symmetry (SG: 162) in the sandwich structure, we construct FMI/six-layer-BT/FMI, exhibiting the evolvement between the magnetic gaps and vdW-spacing distances in Fig. \ref{fig1:m_gap}(d). Red, brown, olive and blue curves are linked to the FMI as CGT, Cr$_2$Si$_2$Te$_6$, Cr$_2$Sn$_2$Te$_6$ and CrI$_3$ respectively. In this setup, we shift the two edge FMI layers symmetrically with the inversion-symmetry protected. The vdW-spacing distance is defined as the width of vdW-spacing that subtracts the width at the balanced position, by setting the latter as 0.0Å. Specifically, the positive (negative) value means stretching (compressing) the vdW-spacing distance. Intuitively, the hybridization strength between FMI and TI increases when the vdW-spacing distance decreases. But counterintuitively, the maximum value of the magnetic gap emerges at certain positions in each curve of heterostructures: 56.4meV, 74.6meV, 40.1meV and 39.1meV when FMI = CGT, Cr$_2$Si$_2$Te$_6$, Cr$_2$Sn$_2$Te$_6$ and CrI$_3$ respectively. Noticeably, for the case FMI = CGT and Cr$_2$Si$_2$Te$_6$, the maximum of value is situated at +0.2Å, slightly further than the balanced position; while for the case of Cr$_2$Sn$_2$Te$_6$, coincidently it appears at the balanced position. Further increasing the hybridization strength acutely cuts down the magnetic gaps. For the case FMI = CrI$_3$, although the magnetic gap reaches to the peak at -0.8Å, the non-monotonic behavior is not absent after going below this shifting position. This behavior is strongly immune to the value of Hubbard $U$, the numbers of the layers in medial TI, and the different TI materials chosen in the medial part, exhibited in Figs. \ref{fig1:m_gap}(e)-\ref{fig1:m_gap}(g), indicating that the nonmonotonic character of the magnetic gap manifests itself a universal physical phenomenon independent of certain materials. This phenomenon may sit behind all the FMI/TI-based QAH systems, hereinafter discussed in the following sections.

Replacing six-layer BT into TL BT will moderately increasing the magnetic gaps [Fig. \ref{fig1:m_gap}(f)] compared to those shown in Fig. \ref{fig1:m_gap}(d), due to the stronger influence from the FMI layer located in the other edge. However, thinner layers of TI leads to substantially enhanced value of the hybridization gap, hindering to obtain the large-gapped Chern-insulating phases, displayed in Fig. \ref{fig16:H-BT}(c) of Appendix \ref{B:hybridization} and Fig. \ref{fig24:CGT-3ST-stacking}(c), Fig. \ref{fig24:CGT-3ST-stacking}(f) of Appendix \ref{E:stacking-order}. It’s noteworthy to notice that the negative chirality arises in some conditions under the larger hybridization regime [Figs. \ref{fig1:m_gap}(d)-\ref{fig1:m_gap}(g)], suggesting that kinetic exchange and Coulomb exchange contributing to the opposite chirality of the magnetic gap \cite{xu2022controllable} may coexist and compete with each other. Furthermore, comparing the case TI = six-layer-ST [Fig. \ref{fig1:m_gap}(g)] to that of the case TI = six-layer-BT [Fig. \ref{fig1:m_gap}(d)], totally the magnetic gaps shrink, stemming from the even higher extension of TSS in ST [see Figs. \ref{fig2:tss}(a) and \ref{fig2:tss}(b) in the next sub-section.]

\subsection{Non-monotonic evolving behaviors of TSS distributions}
 \label{iii-B:distribution}

Understanding the real-space TSS distributions is crucial for the magnetic-gap evolution explorations. Majorly two mechanisms of $p$-$d$ hybridization, including kinetic exchange and Coulomb exchange co-contribute to the magnetic gap. For kinetic exchange interactions, the strength becomes as: $J_{\rm kinetic}\propto  \frac{t^2}{U}$ \cite{pavarini2012correlated}, within which $t$ is the electron hopping strength, while $U$ is the Hubbard $U$ term. Distinctly, for Coulomb exchange interactions, the strength can be written as $J_{\rm Coulomb} \propto J_{pd} $\cite{xu2022controllable,pavarini2012correlated}, in which $J_{pd}$ is the $p$-$d$ electron wave-function integral. Kinetic exchange is more long-ranged than Coulomb exchange according to the previous analysis \cite{xu2022controllable}. Under the conditions of FMI/TI hybridization systems, the electron hopping term $t$ decreases exponentially, while the electron wave-function integral $J_{pd}$ declines sharply to zero when the $p$-$d$ hybridization length increases. Hence, the real-space distribution of TSS directly and majorly decides the strength of kinetic and Coulomb exchange strength, and then, the magnetic gap.

\begin{center}
	\begin{figure*}
		\centering
		\includegraphics[width=0.88\linewidth]{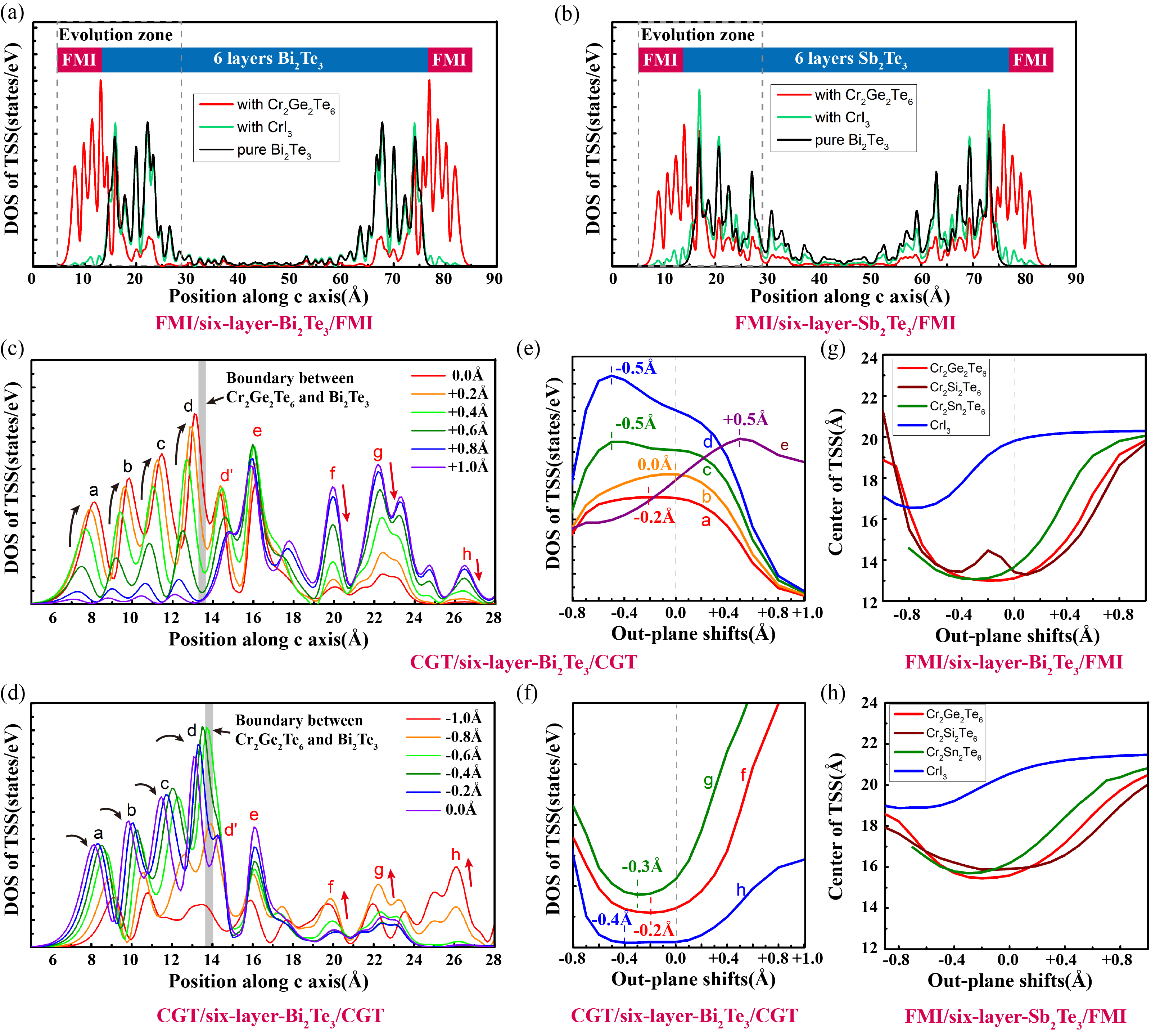}
		\caption{Evolutions of the real-space distribution of TSS in FMI/six-layer-Bi(Sb)$_2$Te$_3$/FMI. Real-space distributions along out-of-vdW-plane direction of TSS in pure TI, CGT/TI/CGT and CrI$_3$/TI/CrI$_3$ in which (a) TI = six-layer-BT and (b) TI = six-layer-ST. Black, red and green curves stand for pure-TI, CGT/TI/CGT and CrI$_3$/TI/CrI$_3$ cases respectively. The red and blue bars above the curves illustrate the related position of FMI and TI along c axis. (c) and (d) present the multicurve evolutions of TSS distributions of CGT/six-layer-BT/CGT within the real-space range that marked in (a) and (b) with gray dashed rectangles. (c) is related to large-vdW-spacing regime while (d) is related to small-vdW-spacing regime respectively. Gray narrow zone is the boundary zone between CGT and BT. Short-curve frames denote the evolution directions when the vdW-distance shrinks. Black-marked “\textbf{a}”, “\textbf{b}”, “\textbf{c}”, “\textbf{d}” are corresponding to the four peaks of TSS within CGT, simultaneously red-marked “\textbf{d'}”, “\textbf{e}”, “\textbf{f}”, “\textbf{g}” and “\textbf{h}” are the six peaks hidden within BT. (e) the peak-value evolutions of “\textbf{a}”, “\textbf{b}”, “\textbf{c}”, “\textbf{d}” and “\textbf{e}” depending on vdW-spacing distances. (f) is similar to (e), but for the three peaks of “\textbf{f}”, “\textbf{g}” and “\textbf{h}”. (g) the weighted mean position evolutions of one half of TSS depending on vdW-spacing distances under the cases of FMI/six-layer-BT/FMI. Red, brown, olive and blue curves severally are homologous to the FMI layer as CGT, Cr$_2$Si$_2$Te$_6$, Cr$_2$Sn$_2$Te$_6$ and CrI$_3$. (h) is similar to (g), but is relevant to the cases of FMI/six-layer-ST/FMI.}
		\label{fig2:tss}
	\end{figure*}
\end{center}

If a topological-trivial insulator is placed near to a TI, the TSS can be extended into the trivial insulator with finite distance \cite{liu2010oscillatory,bernevig2006quantum,otrokov2017highly}. Apparently, comparing TSS distributions of pure six-layer-BT [black curve in Fig. \ref{fig2:tss}(a)] with that of CGT/six-layer-BT/CGT [red curve in Fig. \ref{fig2:tss}(a)], the most part of TSS is extracted from the layer of six-layer-BT into CGT-layer. CrI$_3$ acts analogously with CGT [green curve in Fig. \ref{fig2:tss}(a)], but the extraction of TSS is much weaker compared to that of CGT, the most part of TSS still located in the TI-layer. Changing the TI-layer into six-layer-ST causes no damage to this behavior, but more extended distribution of TSS, seen from Fig. \ref{fig2:tss}(b). We gain an elementary conclusion that the higher hybridization strength between FMI and TI, the stronger extraction of TSS from TI into FMI, and the more extended distribution of TSS emerges. Concretely, the element of I has larger electronegativity (2.66) than that of Te (2.10), therefore the Te-I hybridization strength is much weaker than that of Te-Te, leaving much weaker extraction of TSS from the TI-layer.

Next, let’s focus on the TSS-distribution evolvement based on the vdW-spacing distances. Firstly, take one edge (by choosing the left edge) of CGT/six-layer-BT/CGT as a typical object. At the large vdW-distance regime shown in Fig. \ref{fig2:tss}(c), the TSS fleetly extracted into CGT and grows, we denote “\textbf{a}”, “\textbf{b}”, “\textbf{c}”, “\textbf{d}” as the peaks in CGT layer while “\textbf{d'}”, “\textbf{e}”, “\textbf{f}”, “\textbf{g}” and “\textbf{h}” as the peaks in TI-layer for further detailed study. Obviously, the peaks in CGT and that in BT near to the CGT-BT boundary [marked with gray zone in Figs. \ref{fig2:tss}(c) and \ref{fig2:tss}(d)] increases firstly (peaks “\textbf{a}” to “\textbf{e}”), then reaches the maximum at different vdW-spacing distance, finally falls after rise [see Fig. \ref{fig2:tss}(e)]. Conversely, the peaks deeply amongst BT-layer (peaks), i.e. peaks “\textbf{f}” to “\textbf{h}”, go through declining, reaching minimum, then re-rising process, displayed in Fig. \ref{fig2:tss}(f). More distinctly, we count the weighted mean position evolutions of one half of the TSS in FMI/six-layer-BT/FMI in Fig. \ref{fig2:tss}(g) and FMI/six-layer-ST/FMI in Fig. \ref{fig2:tss}(h) respectively, firmly authenticate that TSS distributions along the out-of-vdW-plane direction goes through an extraction-backflow process that is independent with certain materials of FMI and TI along with the growing of FMI-TI hybridization strength. It’s facile to be comprehended when the vdW-spacing distance decreases, i.e., when the FMI-TI hybridization strength grows, the extraction of TSS increases. Nonetheless, when the FMI-TI hybridization strength turns into too strong, a further gaining in the hybridization strength brings about more delocalized electron behaviors and more extended distributions of TSS, which hamper the increasing distribution in the CGT layer, resulting in backflow characters. Naturally to be apprehended, the non-monotonic distributing behaviors of TSS constitute the basic factor of non-monotonic evolution property of the magnetic gaps.

The positions of the vdW-spacing distance that the minimum value of weighted-positions of TSS reaches is corresponding to the largest extraction of TSS into FMI layer, which depends on the hybridization strength between the two adjacent atomic layers around the vdW interval, or we call it inter-vdW hybridization strength. For instance, in Fig. \ref{fig2:tss}(g), for the case FMI = CGT and Cr$_2$Sn$_2$Te$_6$, the largest extraction of TSS appears at -0.2Å and -0.3Å respectively. However, when FMI = CrI$_3$, it appears at -0.7Å. Weaker inter-vdW hybridization strength demands much smaller vdW-spacing distance to reach the largest extraction of TSS. A weak fluctuation appears in the case FMI = Cr$_2$Si$_2$Te$_6$ due to the small electronic convergence error, leaving two existed minimum, but the whole process still obeys the extraction-backflow rule.

It's deserved to be mentioned that neither the maximum of peaks “\textbf{a}” to “\textbf{e}” in Fig. \ref{fig2:tss}(e), nor the minimum weight-position of TSS [red curve in Fig. \ref{fig2:tss}(g), -0.2Å] appears in the position of the vdW-spacing distance that is same to those of the largest magnetic gaps [red curve shown in Fig. \ref{fig1:m_gap}(d), +0.2Å], taking CGT as the exemplified FMI. This discrepancy indicates that the extraction-backflow of TSS distribution doesn’t undertake the only factor motivating the non-monotonic character of magnetic gaps.

\begin{center}
	\begin{figure}
		\centering
		\includegraphics[width=1\linewidth]{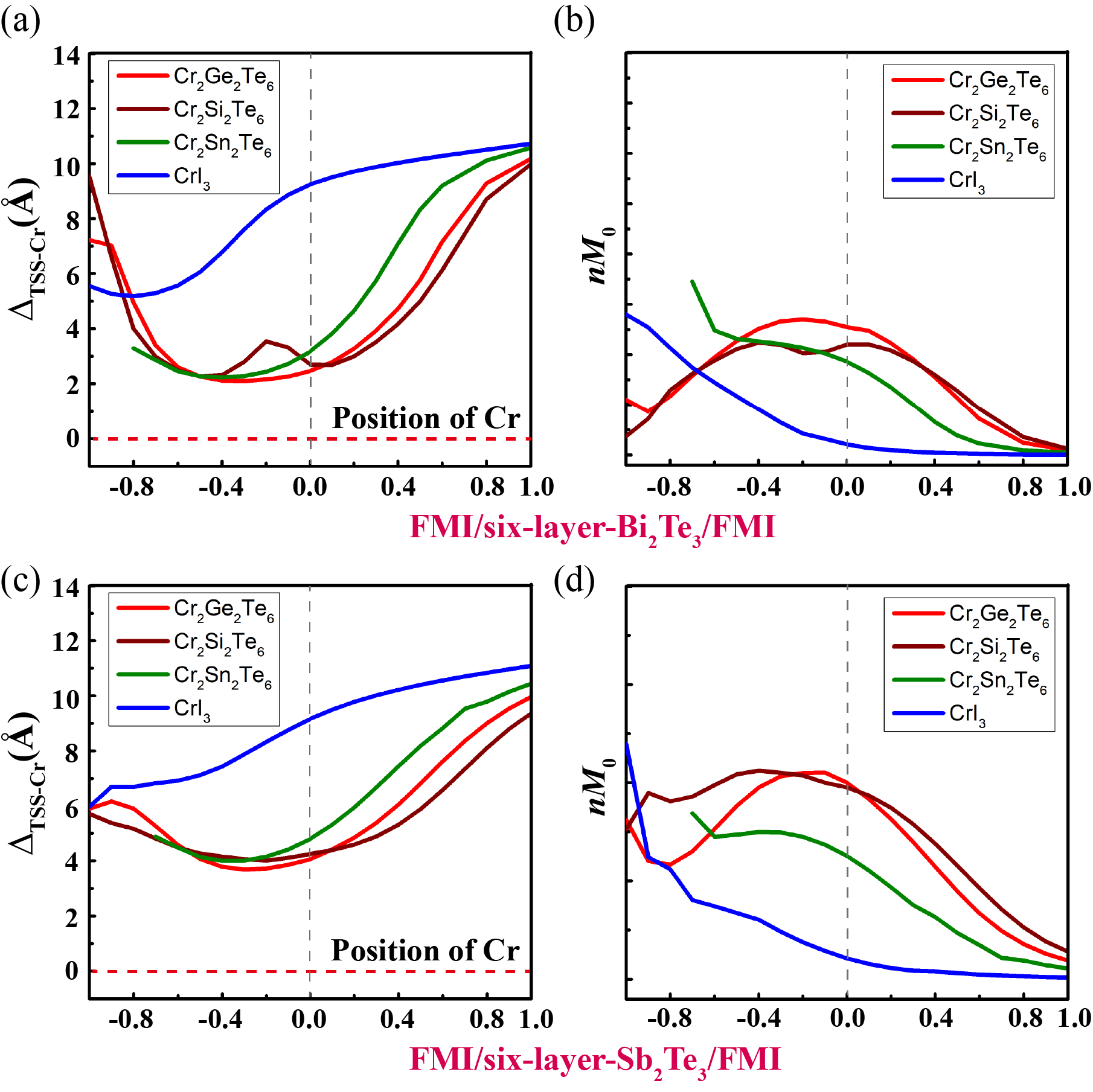}
		\caption{Evolutions of $\Delta_{\rm TSS}$ and $\textbf{\textit{n}}\textbf{\textit{M}}_\textbf{0}$ in FMI/six-layer-Bi(Sb)$_2$Te$_3$/FMI. The evolvement of  $\Delta_{\rm TSS}$ under the systems of (a) FMI/six-layer-BT/FMI and (b) FMI/six-layer-ST/FMI. Red, brown, olive and blue curves are related to the FMI as CGT, Cr$_2$Si$_2$Te$_6$, Cr$_2$Sn$_2$Te$_6$ and CrI$_3$ respectively. Light-gray vertically dashed line denotes the balanced vdW-distance, meanwhile the red transverse dashed line denotes the position of Cr. (c) and (d) are similar to that of (a) and (b) respectively, but display the evolvement of $\textbf{\textit{n}}\textbf{\textit{M}}_\textbf{0}$.}
		\label{fig3:nM0}
	\end{figure}
\end{center}

Hereinafter, we’ll qualitatively and separately discuss the two major mechanisms. For kinetic exchange, multiple and complicated contributions from $d$-electrons on various elements hamper the extraction of Cr-based ingredient in four bands at the $\Gamma$-point around the Fermi level. Therefore, we only qualitatively measure the distances between the weighted-mean-position of TSS and the position of Cr along the out-of-vdW-plane direction, displayed in Figs. \ref{fig3:nM0}(a) and \ref{fig3:nM0}(c). We define this value as a variable: $\Delta_{\rm TSS}$, in which the positive (negative) value is related to the weighted-mean-position of TSS located at the right (left) side of Cr atom. For Coulomb exchange, we select the index: $\textbf{\textit{n}}\textbf{\textit{M}}_\textbf{0}$, defined by Z. Xu \textit{et al}. \cite{xu2022controllable}, in which $\textbf{\textit{n}}$ is the overlapping integral between 3$d$-orbital of Cr and TSS in the real-space, and $\textbf{\textit{M}}_\textbf{0}$ stands for the magnetic moment of the chosen 3$d$-element. The evolutions of $\textbf{\textit{n}}\textbf{\textit{M}}_\textbf{0}$ vs. vdW-spacing distance are displayed in Figs. \ref{fig3:nM0}(b) and \ref{fig3:nM0}(d).

Similarly, the value of $\Delta_{\rm TSS}$ overall undergoes a decreasing-increasing process in all four candidates of FMI/six-layer-BT/FMI [Fig. \ref{fig3:nM0}(a)]. Corresponding positions of the minimum value of $\Delta_{\rm TSS}$ are particularized in Table \ref{tab1:vdW_spacing}. For the case of Cr$_2$Sn$_2$Te$_6$/six-layer-BT/Cr$_2$Sn$_2$Te$_6$, values below the vdW-spacing distance of -0.8Å are absent due to the highly metallic property of the band structure, which is hard to extract the four-band model to obtain the TSS-distributions. The candidates of FMI/six-layer-ST/FMI [Fig. \ref{fig3:nM0}(c)] generate the similar behavior except for the condition that FMI = CrI$_3$, in which the minimum is hidden below the vdW-spacing distance of -0.9Å. By contraries, the evolution behavior of $\textbf{\textit{n}}\textbf{\textit{M}}_\textbf{0}$ undergoes an increasing-decreasing type, with the maximum value located close to the minimum of $\Delta_{\rm TSS}$, the comparison shown in Table \ref{tab1:vdW_spacing}. Summarized in the work of Z. Xu \textit{et al}. \cite{xu2022controllable}, 3$d$ electrons in Cr with +3 valence state dedicate positive chirality of mass term from kinetic exchange and negative chirality from Coulomb exchange, forming competitive nature of the magnetic-gap evolution peculiarity. 

Recall that Coulomb exchange is a short-range interaction comparing with kinetic exchange interaction that relatively long-ranged, when decreasing the vdW-spacing distance under far range (+0.4Å to +1.0Å, taking CGT as the example), kinetic exchange increases quickly while Coulomb exchange rises relatively slowly, manifesting the former mechanism the dominant factor. Hence, in far regime the magnetic gap increases fleetly with positive chirality, see Figs. \ref{fig1:m_gap}(d)-\ref{fig1:m_gap}(g). Next, under the medial-distance regime (-0.5Å to +0.3Å), also taking CGT as the example, distinctly seen from Figs. \ref{fig3:nM0}(a) and \ref{fig3:nM0}(b) that kinetic exchange rises slowly and approaches into the maximum value, while the short-ranged Coulomb exchange grows quickly. The magnetic gap firstly reaches the maximum (+0.2Å), then declines apace to nearly zero under the case of FMI/six-layer-BT/FMI [Fig. \ref{fig1:m_gap}(d)], and even down to the negative chirality in FMI/six-layer-ST/FMI [Fig. \ref{fig1:m_gap}(g)], within the latter of which Coulomb exchange conquers kinetic exchange. Now, capable of explaining the discrepancy of the vdW-spacing positions between the magnetic-gap maximum and the minimum value of $\Delta_{\rm TSS}$, we approach the conclusion that both the extraction-backflow character of TSS-distributions and the competition between kinetic and Coulomb exchange interactions co-contribute the non-monotonic evolution behaviors of the magnetic gaps in the above systems.

\begin{table}
    \caption{\label{tab1:vdW_spacing} vdW-spacing positions of minimum values of $\Delta_{\rm TSS}$ and the maximum values of $\textbf{\textit{n}}\textbf{\textit{M}}_\textbf{0}$ in different systems. All the vdW-spacing positions are shown relatively to the balanced positions in each case.}
    \begin{ruledtabular}
    \begin{tabular}{cccc}
        TI & FMI       & Minimum of TSS Center      & Peak of $\textbf{\textit{n}}\textbf{\textit{M}}_\textbf{0}$    \\  
        \colrule
          \multirow{4}{*}{BT}& CGT & -0.3Å & -0.2Å \\
          & Cr$_2$Si$_2$Te$_6$ & -0.5Å, +0.1Å & -0.4Å, 0.0Å \\
            & Cr$_2$Sn$_2$Te$_6$ & -0.4Å & \textless-0.7Å \\
            & CrI$_3$ & -0.8Å & \textless-1.0Å \\
        \hline
         \multirow{4}{*}{ST} & CGT & -0.3Å & -0.2Å \\
        & Cr$_2$Si$_2$Te$_6$ & -0.2Å & -0.4Å \\
        & Cr$_2$Sn$_2$Te$_6$ & -0.4Å & \textless-0.7Å \\
        & CrI$_3$ & \textless-1.0Å & \textless-1.0Å \\
       
        \end{tabular}
    \end{ruledtabular}
    \end{table}

At the very near regime, i.e., vdW-spacing distances below -0.5Å, both kinetic and Coulomb exchange interactions decline, originating from the backflow of TSS, making the magnetic gap emerge a more intricate evolving behavior. Considering the severe requirement to equivalently reach this regime under experimental conditions, we ignore to analyze it here. 

When kinetic exchange and Coulomb exchange interactions engender the same-sign chirality of Zeeman splitting, for instance, we build a sandwich model of NiBi$_2$Te$_4$/six-layer-BT/NiBi$_2$Te$_4$ (NiBT/six-layer-BT/NiBT), the maximum value of the magnetic gap meets well with the maximum value of $\textbf{\textit{n}}\textbf{\textit{M}}_\textbf{0}$ and arrives even later than that of the minimum value of $\Delta_{\rm TSS}$, seen from Figs. \ref{fig23:NiBT-6BT}(b) and \ref{fig23:NiBT-6BT}(d) in Appendix \ref{D:M-Gap_others}. This outcome plays an auxiliary to the conclusion that whether the chirality of the two exchange mechanisms meets the same also dominates the vdW-spacing position at which the extremum of the magnetic gap is situated.

\subsection{Robustness of the non-monotonic character}
 \label{iii-C:lattice_constants}

\begin{center}
	\begin{figure}
		\centering
		\includegraphics[width=1\linewidth]{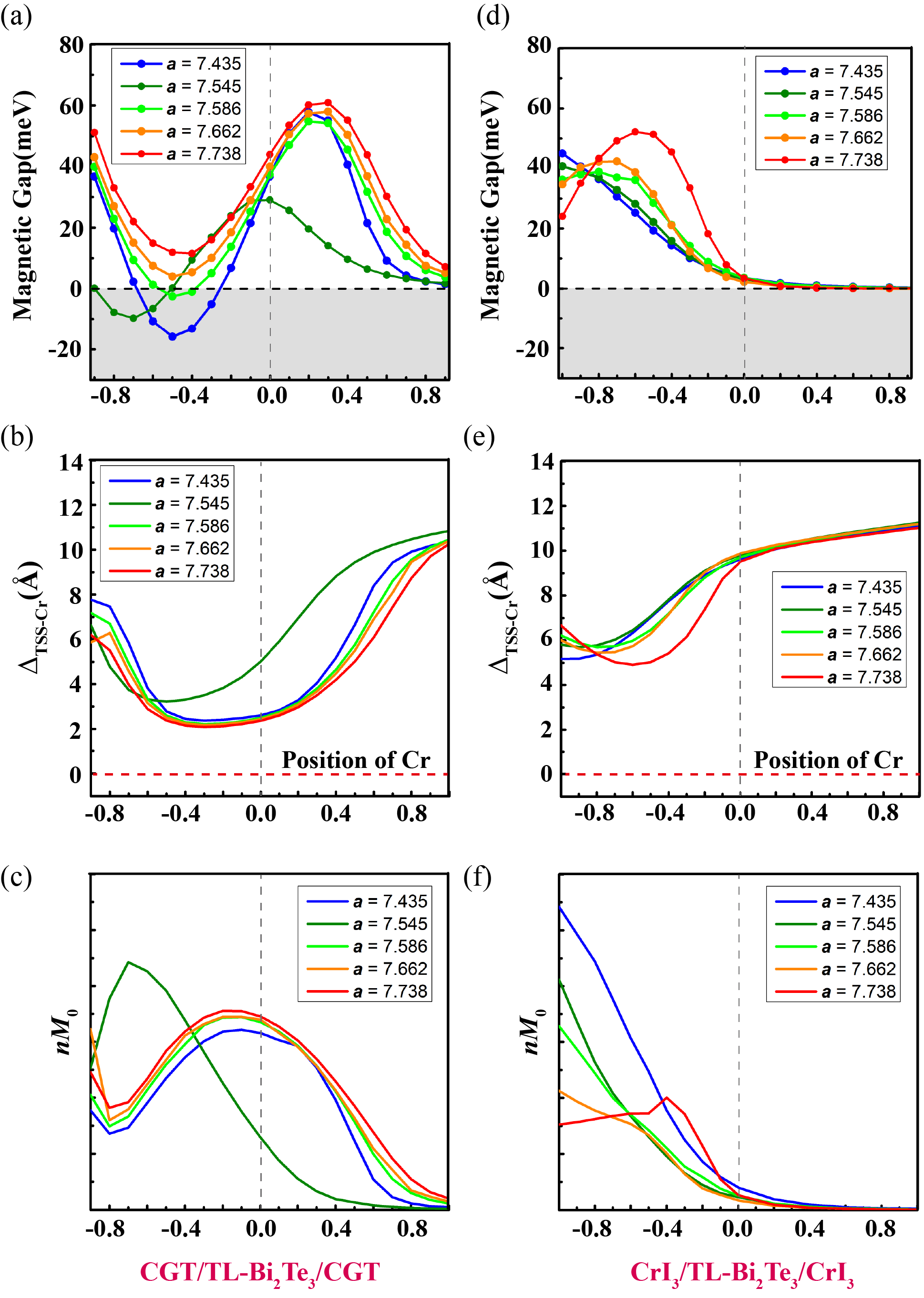}
		\caption{Evolutions of magnetic gaps, $\Delta_{\rm TSS}$ and $\textbf{\textit{n}}\textbf{\textit{M}}_\textbf{0}$ in CGT/TL-BT/CGT and CrI$_3$/TL-BT/CrI$_3$ under various in-plane lattice constants. Evolvements of magnetic gaps (a), the values of $\Delta_{\rm TSS}$ (b) and the values of $\textbf{\textit{n}}\textbf{\textit{M}}_\textbf{0}$ are depicted in the basis of CGT/TL-BT/CGT under in-plane lattice constants of 7.435Å, 7.545Å, 7.586Å, 7.662Å and 7.738Å respectively, delineated by blue, olive, green, orange and red curves respectively. White and gray zones in (a) stand for the positive and negative chirality of magnetic gaps. Light-gray dashed vertical line in (a)-(c) denotes the balanced vdW-distance position, in the meantime the red dashed transverse line in (b) is the position of Cr atoms. (d)-(f) are similar to those of (a)-(c), but under the basis of CrI$_3$/TL-BT/CrI$_3$.}
		\label{fig4:lattice_constants}
	\end{figure}
\end{center}

Lattice constants play a vital role in magnetic topological properties in broad variety of systems. Selecting two options of FMI as CGT and CrI$_3$, and the only option of TI as TL BT, after freely relaxation with cell-volume restriction, we’ve acquired the freely-relaxed in-plane lattice constants \textit{\textbf{a}} = 7.455Å for CGT/TL-BT/CGT and \textit{\textbf{a}} = 7.545Å for CrI$_3$/TL-BT/CrI$_3$, derived from that of the $\sqrt{3}\times\sqrt{3}$ in-plane superlattice of bulk-BT as -1.7\% and -0.5\% respectively. In Fig. \ref{fig4:lattice_constants}, we delineate the evolvement of magnetic gaps, the values of $\Delta_{\rm TSS}$ and $\textbf{\textit{n}}\textbf{\textit{M}}_\textbf{0}$ under in-plane lattice constants of 7.435Å, 7.545Å, 7.586Å, 7.662Å and 7.738Å respectively, relating to -2.0\%, -0.5\%, 0.0\%, +0.5\% and +2.0\% of lattice derivations compared to that of the $\sqrt{3}\times\sqrt{3}$ in-plane superlattice of bulk-BT. Evidently, in vdW-spacing-distance evolution regime, the non-monotonic behavior of magnetic gaps, the extraction-backflow nature of TSS and the increasing-decreasing property of $\textbf{\textit{n}}\textbf{\textit{M}}_\textbf{0}$ persist under all the in-plane lattice constants and under both the cases of CGT/TL-BT/CGT and CrI$_3$/TL-BT/CrI$_3$. For the case of CrI$_3$/TL-BT/CrI$_3$ under \textit{\textbf{a}} = 7.435Å and 7.545Å, the maximum magnetic gap is concealed below the vdW-spacing distance of -0.9Å, consequently we can’t see their non-monotonic performances. 

Despite the robustness of non-monotonic behavior immune to moderate biaxial strains, the maximum value of magnetic gap and the relevant vdW-spacing position substantially suffer from the in-plane lattice constants, for instance, Coulomb exchange can conquer kinetic exchange under the smaller vdW-spacing distance regime when \textit{\textbf{a}} = 7.435Å, 7.545Å or 7.586Å in CGT/TL-BT/CGT, seen from Fig. \ref{fig4:lattice_constants}(a). Larger in-plane lattice constant induces the maximum magnetic gap to even larger values, depicted in Figs. \ref{fig4:lattice_constants}(a) and \ref{fig4:lattice_constants}(b). In-plane-lattice constant variations modulate the TSS distributions in the heterostructure, adjusting both the vdW-spacing positions and the maximum value of the magnetic gaps without breaking the non-monotonic evolving law.

Moreover, we’ve also carried out trials of this robust non-monotonic characters by employing various vdW correction methods. Mentioned in the Sec. \ref{ii:methods}, the seven selected vdW correction methods coming along with no vdW correction are utilized based on CGT/TL-BT/CGT system, with the result shown within Fig. \ref{fig18:vdW_correction} in Appendix \ref{C:Robustness}. Distinctly, implementing and varying the vdW corrections cast almost no influence on the non-monotonic evolving behaviors of the magnetic gaps, and even their maximum values, manifests the forceful robustness of this character against various vdW corrections.

Considering the experimentally more suitable bi-heterostructure model: CGT/TL-BT, mainly discussed in Figs. \ref{fig19:bi-hetero}(a)-\ref{fig19:bi-hetero}(c) of Appendix \ref{C:Robustness}, we’ve confirmed that the non-monotonic evolving behavior of the magnetic gaps is still valid, with the final magnetic totally descend compared to sandwich structures, hampered by the non-CGT edge. Detailed discussions are developed in Appendix \ref{C:Robustness}.

\subsection{Competitions of multiple mass terms}
 \label{iii-D:competitions}

\begin{center}
	\begin{figure*}
		\centering
		\includegraphics[width=0.89\linewidth]{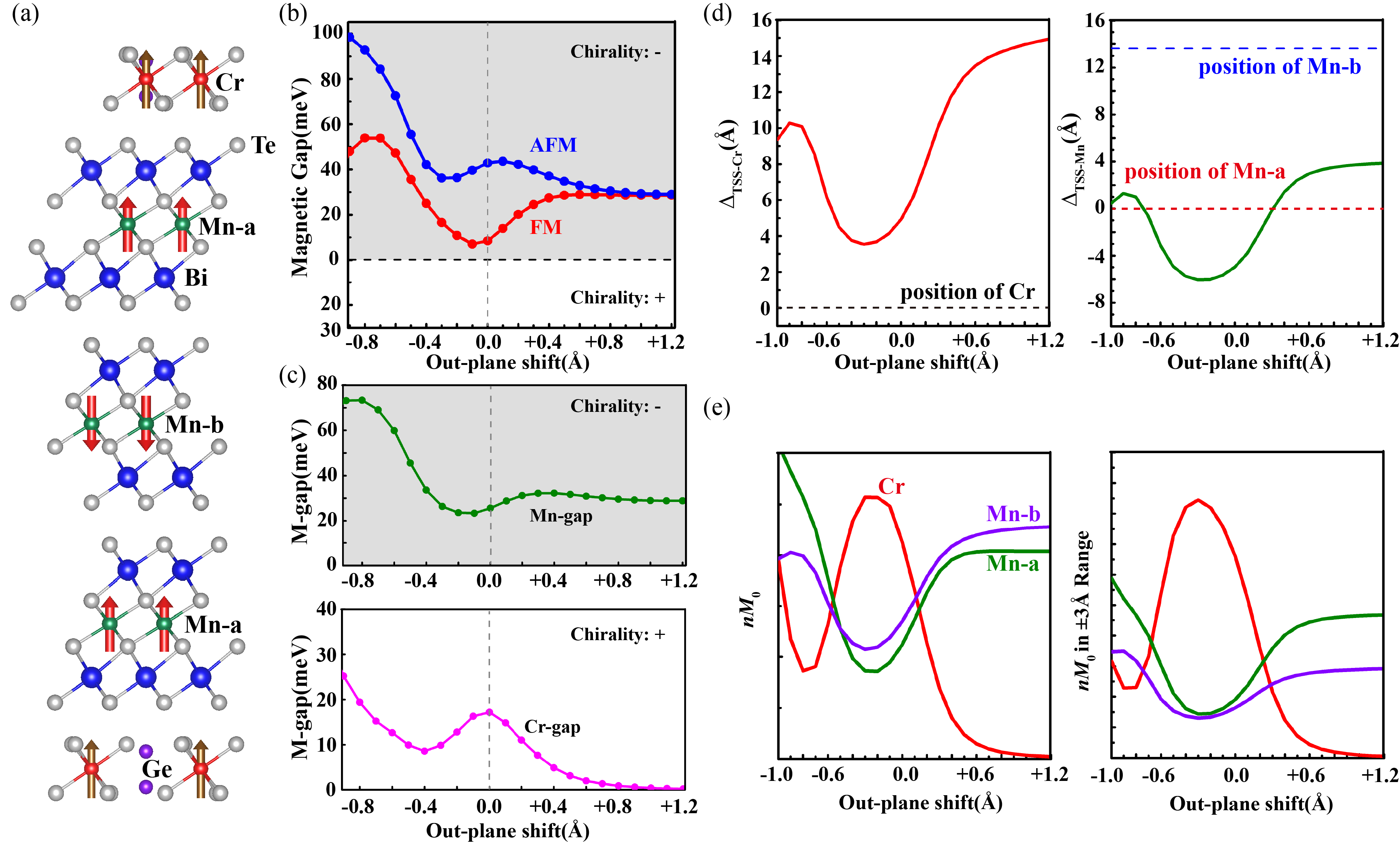}
		\caption{Basic structures, evolutions of magnetic gaps, $\Delta_{\rm {TSS-Cr}}$, $\Delta_{\rm {TSS-Mn}}$, $\textbf{\textit{n}}\textbf{\textit{M}}_\textbf{0}$ and $\langle\textbf{\textit{n}}\textbf{\textit{M}}_\textbf{0}\rangle_{\rm ave}$ in CGT/TL-MBT/CGT via continuously tuning vdW-spacing distances. (a) Side view of CGT/TL-MBT/CGT in which green, blue, gray, red and purple balls are corresponding to Mn, Bi, Te, Cr, Ge atoms. Red and brown arrows are related to magnetic moments on Mn and Cr atoms respectively. (b) the magnetic gaps under the FM (red curve) and the AFM (blue curve) coupling configurations. Gray and white zones denote the negative and positive chirality. (c) The magnetic gaps contributed by Mn (green curve in the upper panel) and Cr (pink curve in the lower panel). Gray dashed vertical lines stand for the balanced vdW-spacing distance with 0.0Å. (d) Evolutions of $\Delta_{\rm {TSS-Cr}}$ (red curves in the left panel) and $\Delta_{\rm {TSS-Mn}}$ (green curves in the right panel) via various vdW-spacing distances. Black dashed transverse line in the left panel denotes the position of Cr atomic layer, while red and blue dashed transverse lines in the right panel denote the positions of marginal Mn layer and medial Mn layer respectively. Positive (negative) values are corresponding to the positions of $\Delta_{\rm {TSS-Cr}}$ or $\Delta_{\rm {TSS-Mn}}$ located at the right (left) side of Cr or Mn atoms. (e) $\textbf{\textit{n}}\textbf{\textit{M}}_\textbf{0}$ (curves in the left panel) and $\langle\textbf{\textit{n}}\textbf{\textit{M}}_\textbf{0}\rangle_{\rm ave}$ (curves in the right panel) evolve along with the vdW-spacing distances. Red, green and purple curves stand for Cr, marginal Mn (noted as “Mn-\textbf{a}”) and medial Mn (noted as “Mn-\textbf{b}”) based $\textbf{\textit{n}}\textbf{\textit{M}}_\textbf{0}$ and $\langle\textbf{\textit{n}}\textbf{\textit{M}}_\textbf{0}\rangle_{\rm ave}$ values respectively.}
		\label{fig5:CGT-MBT}
	\end{figure*}
\end{center}

In this sub-section, the TI layer replaced by MBT, multiple origins of magnetic atomic layers provide a more abundant competing mechanisms for investigation and discussion. In detail, chosen as typical 3$d$-orbital magnetic elements, Cr and Mn contributes the opposite sign of mass terms forming the competitive nature of mechanisms existing within the magnetic gaps, makes it more complicated to be analyzed and modulated. However, there are several motivations manifest it more meaningful for manipulating and optimizing the magnetic gaps despite of its intricate nature of multiple mass terms. Firstly, although CGT itself has the possibility to create the negative chirality in certain artificial conditions discussed in Sec. \ref{iii-A:non-monotonic}, it’s hard to be realized experimentally. Establishing the two magnetic element sources with the opposite chirality makes both the chirality and the Chern-number theoretically and experimentally tunable. For example, in Sec. \ref{iv-B:CGT-MBT} that will be discussed later, via fabricating Cr$_2$Si$_2$Te$_6$/TL-MBT/Cr$_2$Si$_2$Te$_6$ building blocks, it’s facilitated to see the Chern number varying as -1, 0, +1 under stacking-order shifts. Secondly, these systems have multiple critical temperatures devoted by different magnetic exchange terms. For example, CGT/TL-MBT/CGT has three critical temperatures. Increasing the external temperatures make the system experiences three kinds of magnetic gaps until the largest critical temperature reaches, supplying another tunable degree of freedom. Thirdly, for experimentally more suitable bi-heterostructures, i. e., CGT/MBT, whose gap of TSS is determined by the smaller one comparing those at the two edges, the MBT layer maintains the large enough magnetic gap in non-CGT edge, while the CGT-neared edge still contributes the final tunable magnetic gap. Hence, choosing Chern insulator as the TI layer is the optimum selection in the experimentally suitable FMI/TI bi-heterostructure building blocks.

Constructing CGT/TL-MBT/CGT sandwich structure due to its high symmetry (SG: 162), the side view of the structure is shown in Fig. \ref{fig5:CGT-MBT}(a) with the ground-state magnetic configurations of FM interlayer coupling between CGT and MBT \cite{li2020tunable}, and the AFM interlayer coupling within MBT layers \cite{li2019intrinsic,li2020tunable}. Figure \ref{fig5:CGT-MBT}(b) displays the magnetic gap curves under FM (red curve) and AFM (blue curve) couplings between CGT and MBT respectively. Noticeably, in all the conditions shown in Fig. \ref{fig5:CGT-MBT}(b) the AFM-coupling creates larger magnetic gaps than the FM-coupling, hinting that Cr and Mn dedicate the opposite chirality of magnetic gaps, consistent with the conclusions shown in Z. Xu \textit{et al}. \cite{xu2022controllable}. It’s facilitated to obtain the separated magnetic gaps generated by Cr and Mn elements individually, exhibited severally in Fig. \ref{fig5:CGT-MBT}(c), that the Cr-based magnetic gap performs a very similar role with that of CGT/TL-BT/CGT, meanwhile Mn-based magnetic gap acts contrarily, i.e. with decreasing-increasing nature. With the opposite-sign-induced compensation, Mn and Cr comprise the total magnetic gap evolving as the slightly-increasing, decreasing, increasing process, with the minimum value near to 0 at -0.1Å.  

We define $\Delta_{\rm {TSS-Cr}}$ ($\Delta_{\rm{TSS-Mn}}$) as the distance between the weighted-mean-position of TSS and Cr-atoms (Mn-atoms in the marginal MBT-layer), shown in Fig. \ref{fig5:CGT-MBT}(d) in two panels individually. Distinctly, $\Delta_{\rm {TSS-Cr}}$ and $\Delta_{\rm{TSS-Mn}}$ also behave analogously as a decreasing-increasing phenomenon, the weighted-mean-position of TSS crossing through the marginal Mn layer (denoted as “Mn-\textbf{a}”) and stay far away from the medial Mn layer (denoted as “Mn-\textbf{b}”). Therefore, for kinetic exchange interaction, the strength undergoes increasing-decreasing process for Cr; slightly increasing, sharply decreasing and sharply increasing process for marginal Mn; slightly fluctuating for medial Mn respectively.

The left panel of Fig. \ref{fig5:CGT-MBT}(e) reveals the behaviors of $\textbf{\textit{n}}\textbf{\textit{M}}_\textbf{0}$ on positions of Cr (red curve), marginal Mn (olive curve) and medial Mn (purple curve) separately. Due to the extraction-backflow nature of TSS, $\textbf{\textit{n}}\textbf{\textit{M}}_\textbf{0}$ experiences increasing-decreasing process around the Cr layer and decreasing-increasing process around both the marginal and the medial Mn layers. For Mn, Coulomb exchange interaction dominates the negative chirality of its magnetic gaps, resulting in its decreasing-increasing nature. But for this system, AFM-coupling of TL MBT makes the medial Mn contribute the opposite sign of magnetic gaps, performing a compensated role to the marginal Mn atoms. In order to including all the $p$-$d$ orbital overlapping integrals to roundly measure Coulomb exchange strength, in the right panel of Fig. \ref{fig5:CGT-MBT}(e) we define a mean-averaged value, $\langle\textbf{\textit{n}}\textbf{\textit{M}}_\textbf{0}\rangle_{\rm ave}$, integrating TSS around certain range along the out-of-vdW-plane direction. Here, we choose $\pm$3.0Å range to roundly meet the atomic-layer distances forward and behind. Evidently $\langle\textbf{\textit{n}}\textbf{\textit{M}}_\textbf{0}\rangle_{\rm ave}$ is totally smaller around the medial Mn than that of the marginal Mn. Hence, the residue contributions from the marginal Mn-atoms, compensated by both Cr-atoms and medial Mn-atoms, dominates the final magnetic gaps possessing the negative chirality. 

Establishing FMI/MBT/FMI heterostructure supplies a more abundant modulated freedom of magnetic gaps especially originating from the multiple magnetic elements with the same or opposite signs of chirality. It’s lightly to find that when the interlayer magnetic coupling with MBT aligned to FM, larger negative chirality of magnetic gaps forms. We compute and discuss FMI/TL-MBT/FMI within which FMI = Cr$_2$Si$_2$Te$_6$, Cr$_2$Sn$_2$Te$_6$ and CrI$_3$ in Appendix \ref{D:M-Gap_others} (see Figs. \ref{fig20:FMI-MBT-magneticgaps}-\ref{fig21:FMI-MBT-2BT}), which share the analogous features with CGT/TL-MBT/CGT. We also assume a more artificial model of FMI/M(Ni)BT/bilayer(BL)-BT/M(Ni)BT/FMI in Appendix \ref{D:M-Gap_others} (see Fig. \ref{fig22:FMI-NiBT-2BT}) to exclude the influence from the medial magnetic layers, enhancing the magnetic gaps up to larger values.

Besides, considering bi-heterostructure model: CGT/TL-MBT, mainly discussed in Figs. \ref{fig19:bi-hetero}(d)-\ref{fig19:bi-hetero}(f) of Appendix \ref{C:Robustness}, similar to that of CGT/TL-BT, the non-monotonic evolving behavior of the magnetic gaps also takes effect robustly. Benefited from the Chern insulating layer mentioned in the third motivation, its final magnetic gap doesn’t descend from that of CGT/TL-MBT/CGT. Consequently, choosing Chern insulator as the TI layer is more reasonable in FMI/TI bi-heterostructures.

\subsection{Discussions of the non-monotonic evolving behavior}
\label{iii-E:discussions}

The robustness of the non-monotonic evolving behavior on the magnetic gaps along with the increase of hybridization strength indicates a universal physical character that is capable of applications in all the proximity building blocks. M. M. Otrokov \textit{et al} \cite{otrokov2017highly} has discovered the extraction of the TSS from TI layer to MI layer via constructing MBT/BT heterostructures. This property is verified in several related works theoretically and experimentally \cite{tss-li2024realization,tss-shoman2015topological,tss-wu2013tuning,tss-zhang2012exotic,tss-zhang2017electronic}. Moreover, X. Li \textit{et al} \cite{tss-li2024realization} gives a comprehensive research and discussion on the real-space distributing behaviors of TSS on MI/TI heterostructures. They build Mn(V, Ni)Bi(Sb)$_2$Te$_4$/Bi(Sb)$_2$Te(Se)$_3$ heterostructures and reveal the floating, pinning and diving characters of TSS, corresponding to the moderate, weak and strong interactions between MI and TI. These findings also indicate the non-monotonic behaviors of the magnetic gaps behaving along with the MI-TI hybridization strength. In the strong hybridization regime, our revealments still persist part of the TSS distributing within the MI layer, yet the most part of TSS dives deeply into the TI layer.

Furthermore, via continuously tunable methods, we’re in a position to more clearly see the whole process on the extraction-backflow of TSS compared to previous works, and then, furnish an unambiguous guiding principle for optimizing the final magnetic gaps. Mentioned in X. Li \textit{et al} \cite{tss-li2024realization} in like manner, optimizing the final magnetic gaps and then, the gaps, is crucial for solving the problems existing in most magnetic topological systems that the realize temperature of QAH state is much lower than that of their magnetic critical temperature. Hence, our outcomes in this work becomes more impressively to solving this problem, which is further analyzed and discussed in Sec. \ref{v:Tc}.

\section{Modulating the Magnetic Gaps via Stacking-order Shift}
 \label{iv:stacking}

The artificial model of continuously varying the vdW-spacing distances between FMI and TI in Sec. \ref{iii:vdW-spacing} has cozily excavated the hidden physical nature of non-monotonic evolved character within the magnetic gap, but it’s unrealistic for experimental implementations apparently. Profited from its vdW-layered nature, here in this section we employ the frequently-used stacking-order-shift method \cite{li2020tunable,li2023tunable,li2024multimechanism,sivadas2018stacking} to behave likely with tuning vdW-spacing distances but is available under experimental conditions. Stacking-order-shift can make dislocations between the electron orbitals located within the vacuum zone of vdW-interval, performing interlayer-hybridization-strength modulations therein.

In order to maintain the inversion symmetry in the system, we shift the top and bottom FMI layers with opposite directions illustrated in Fig. \ref{fig6:CGT-BT}(a) accordingly, in which red solid arrows drive the top FMI layer while blue dashed arrows drive the bottom one. Retaining inversion symmetry brushes aside disturbing factors when observing the evolutions between the magnetic gaps and the stacking orders. We also adopt FMI/TI bi-heterostructure systems (instead of sandwich), shifting FMI layer to see the magnetic-gap performances in this section. This is the most available building blocks that experimentally operational and tunable. We cast $6\times6$ grids in one unit in-plane shifting cell, computing with structural relaxations by fixing lattice vectors. In the high symmetrical $[100]$ and $[1\overline{1}0]$ directions we cast 9 grids. For experimentally mostly available FMI/TI bi-heterostructure systems, we cast $9\times9$ grids in the whole shifting cell owning even higher resolutions. 

In sub-section \ref{iv-A:CGT-BT}, we construct FMI/TI/FMI sandwich structure and FMI/TI bi-heterostructure in which TI = TL BT or TL ST. Stacking-order-shifts can continuously modulate the magnetic gaps, the $\Gamma$-point gaps and the global gaps in these building-blocks, nevertheless none of candidates contains desired capability including large global gaps, large-ranged nontrivial Chern insulating property and tunable Chern numbers. By importing magnetism in TI layer, achieved with FMI/MBT/FMI or FMI/MBT building blocks, not only the large-value, large-ranged and tunable global-gap nature is engendered, but also the character of step-likely tunable Chern-number and switchable chirality is motivated, discussed in sub-section \ref{iv-B:CGT-MBT}.

\subsection{Optimizing the magnetic and global gaps} 
 \label{iv-A:CGT-BT}

\begin{center}
	\begin{figure*}
		\centering
		\includegraphics[width=0.8\linewidth]{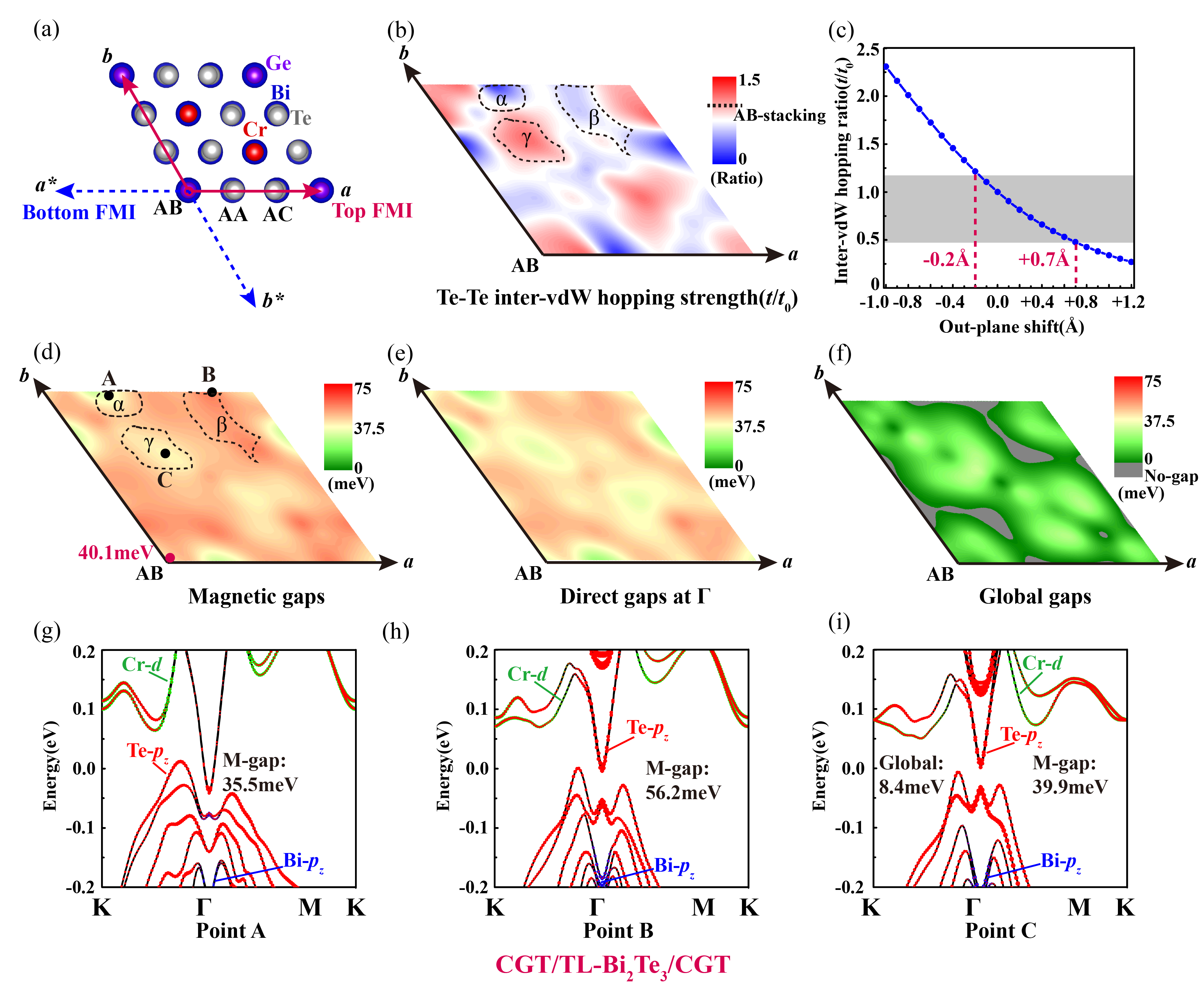}
		\caption{Magnetic-gap manipulations in CGT/TL-BT/CGT induced by stacking-order-shifts. (a) illustrations of how to utilize stacking-order-shift in CGT/TL-BT/CGT by maintaining the inversion symmetry, depicted by top-view. Blue, gray, red and purple balls are related to Bi, Te, Cr and Ge atoms respectively. Red solid and blue dashed arrows stand for the shifting directions of the top and bottom CGT separately. (b) Stacking-order-shift modulated inter-vdW Te-Te interactions shown in mapping distribution. The dashed line in the amplitude bar is corresponding to the interactions under normal-stacking condition. $\alpha$, $\beta$ and $\gamma$ are the zones of weak, moderate and strong couplings of inter-vdW Te-Te interactions. (c) Inter-vdW Te-Te interaction hopping strengths vs. vdW-spacing distances in CGT/TL-BT/CGT. The gray zone covers the interaction strength that can be reached in the stacking-order-shift. Distributions of (d) the magnetic gaps, (e) the $\Gamma$-point gaps and (f) the global gaps under stacking-order-shifts by employing the green-yellow-red color bar. In (d) the normal-stacking magnetic gap: 40.1meV, is noted with red digit. Orbital-projected band structures under (g) Point A, (h) Point B and (i) Point C with stacking positions denoted in (d). Olive, blue and red bubbles act as Cr-3$d$, Bi-6$p$ and Te-5$p$ orbital projections.}
		\label{fig6:CGT-BT}
	\end{figure*}
\end{center} 
 
\begin{center}
	\begin{figure*}
		\centering
		\includegraphics[width=0.8\linewidth]{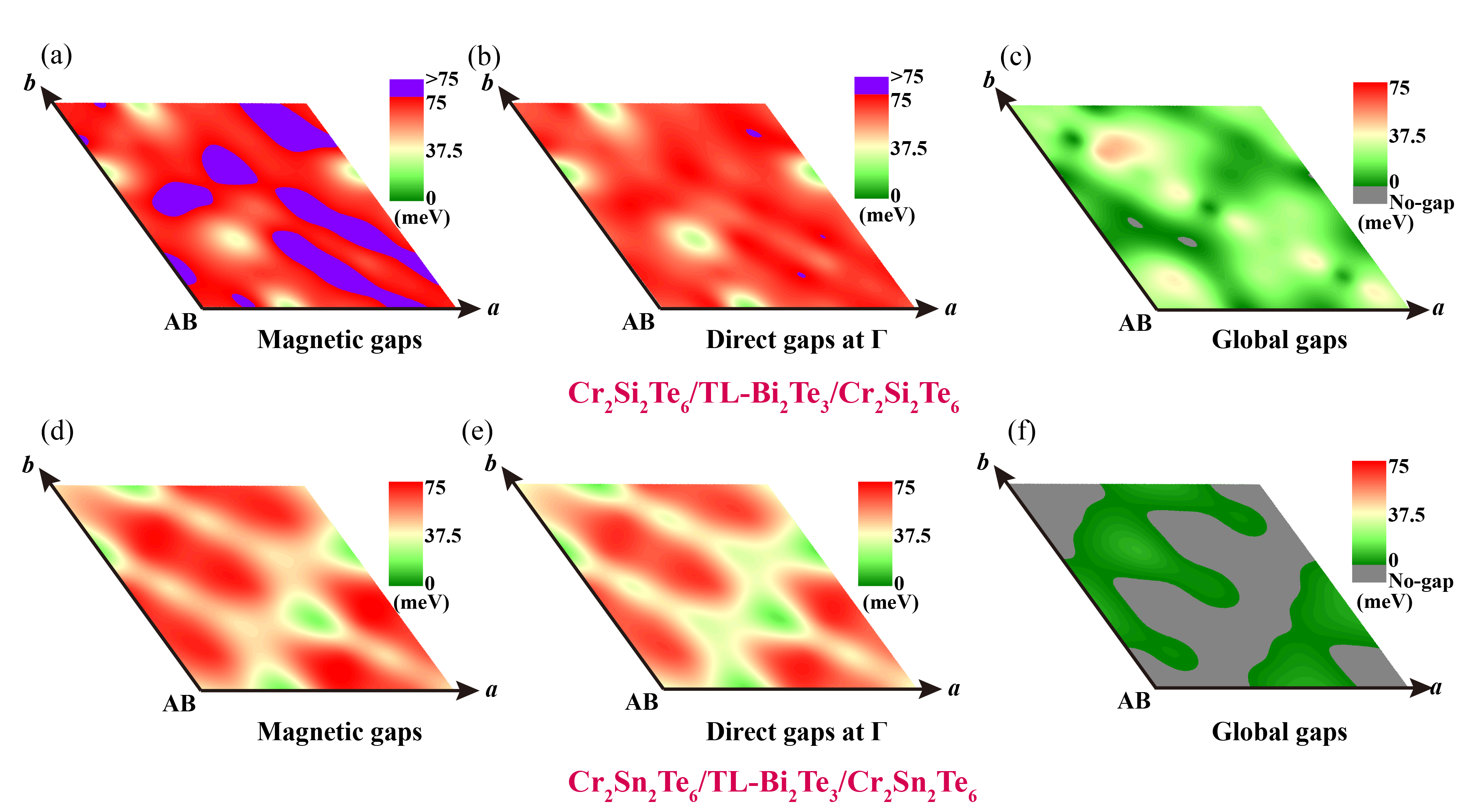}
		\caption{Magnetic-gap manipulations in Cr$_2$Si(Sn)$_2$Te$_6$/TL-BT/Cr$_2$Si(Sn)$_2$Te$_6$ induced by stacking-order-shifts. Distributions of (a) the magnetic gaps, (b) the $\Gamma$-point gaps and (c) the global gaps via employing stacking-order-shifts of FMI layers under the building-block of Cr$_2$Si$_2$Te$_6$/TL-BT/Cr$_2$Si$_2$Te$_6$. Green-yellow-red color bar is adopted, with gray and purple colors stand for “no-gap” zones and “$\textgreater$75meV” zones respectively. (d)-(f) share the similarity to (a)-(c), but under the building-block of Cr$_2$Sn$_2$Te$_6$/TL-BT/Cr$_2$Sn$_2$Te$_6$.}
		\label{fig7:CrSiTe3-BT}
	\end{figure*}
\end{center}

Constructing FMI/TL-BT/FMI firstly, we explore the potentially manipulated capacity of stacking-order-shifts, with the distributions of corresponding parameters exhibited in Figs. \ref{fig6:CGT-BT} and \ref{fig7:CrSiTe3-BT}. Immediate result of inter-vdW stacking-shift possesses the tunable electronic hybridization strength across the vdW-interval, shown with inter-vdW Te-Te hopping-term distributions in Fig. \ref{fig6:CGT-BT}(b). Seen from the color bar, from blue to white then to red, the hopping strength $t$ increases, with the dashed black transverse line across the color bar denoting the hopping strength $t_0$ under the normal stacking. Noticeably, $t_0$ located in the light-red zone of color bar, varying the stacking-orders declines the hopping strength in most of zones due to the breaking of inter-vdW covalence bonds. Fig. \ref{fig6:CGT-BT}(b) confirms the capacity of tunable hybridization strength between FMI and TI. Dashed-line closed regions noted as “$\alpha$”, “$\beta$” and “$\gamma$” are corresponding to weak, moderate and strong Te-Te hopping strength regions respectively (fetched from the TBH method), leaving for future investigations of magnetic gaps.

Comparisons between the methods of stacking-order-shifts and varying vdW-spacing-distance provide a referred tuning range, shown in the gray zone in Fig. \ref{fig6:CGT-BT}(c) with hopping strengths obtaining from the latter method, offering an equivalent range from -0.2Å to +0.7Å. Notably, most regions in stacking-order shifts generate weaker hybridization strength, coinciding with the maximum of the magnetic gap located as +0.2Å that further from the balanced position. Hence, we successfully gain the larger magnetic gaps in most regions shown in Fig. \ref{fig6:CGT-BT}(d), especially in the “$\beta$” zone with the magnetic gaps higher than 50meV accordingly. Both the weaker hybridization “$\alpha$” zone and the stronger hybridization “$\gamma$” zone contain smaller magnetic gaps than those of “$\beta$” zone, coinciding with the non-monotonic behavior between magnetic gaps and the FMI-TI hybridization strengths. The $\Gamma$-point gap distributions [Fig. \ref{fig6:CGT-BT}(e)] shares similar mapping features with that of the magnetic gap, but slightly small values, subtracted by the hybridization gaps. Encountered by the “M”-shaped valence band nature of BT, the global gaps are totally crippled to below 30meV in the whole region, and even metallic phase in some zones colored by gray.
 
Depicted in Figs. \ref{fig6:CGT-BT}(g)-\ref{fig6:CGT-BT}(i), orbital resolved and projected band structures under the stacking orders of “Point A”, “Point B” and “Point C”, signed in the distribution mapping of Fig. \ref{fig6:CGT-BT}(d) in “$\alpha$”, “$\beta$” and “$\gamma$” zones respectively, are displayed obviously with noted magnetic gaps. Distinctly, “Point B” contains the largest magnetic gap (56.2meV) without global gap. “Point C” with the strongest interaction strength generates small global gap (8.4meV) with moderate size of magnetic gap (39.9meV).

Replacing BT by ST in TI layer successfully conquers the “M”-shaped valence band, recovering moderate size of global gaps in whole regions, but the significant enlargement of hybridization gaps almost totally ruins the nontrivial-Chern insulating phases [see Figs. \ref{fig24:CGT-3ST-stacking}(a)-\ref{fig24:CGT-3ST-stacking}(c) in Appendix \ref{E:stacking-order}]. Thicker layer of ST is required to repress the hybridization gaps and recover its Chern insulating features.

Noteworthily, selecting Cr$_2$Si$_2$Te$_6$ as a substitute for CGT magnifies the magnetic gaps, the $\Gamma$-point gaps and the global gaps in the whole shifting range [Figs. \ref{fig7:CrSiTe3-BT}(a)-\ref{fig7:CrSiTe3-BT}(c)]. The metallic region almost vanishes, within which the large global-gap region opens even up to 45meV, the orange ellipse region cherished in the mapping pattern of Fig. \ref{fig7:CrSiTe3-BT}(c). In contrast, selecting Cr$_2$Sn$_2$Te$_6$ as FMI obtains similar behaviors with that of CGT [Figs. \ref{fig7:CrSiTe3-BT}(d)-\ref{fig7:CrSiTe3-BT}(f)], with large gray-colored metallic phase and small regions with little value of global gaps around 5meV, depicted in Fig. \ref{fig7:CrSiTe3-BT}(f). Consequently, only opting Cr$_2$Si$_2$Te$_6$ as the magnetic insulator in this sandwich has potential applications for high temperature Chern insulators under stacking-order-shift regime.

“M”-shaped valence band of BT totally destroys global gaps in bi-heterostructure of CGT/TL-BT, owing to the very weak $p$-$d$ orbital hybridizations between CGT and the FMI-absent edge of TL BT, leaving very small magnetic gaps. The stacking-order-shift results of CGT/TL-BT are displayed in Figs. \ref{fig24:CGT-3ST-stacking}(d)-\ref{fig24:CGT-3ST-stacking}(f) of Appendix \ref{E:stacking-order}, inspiring us the necessity to introduce magnetism within the TI layer itself in bi-heterostructure regime.

\subsection{Characters of tunable Chern number and switchable chirality} 
 \label{iv-B:CGT-MBT}

\begin{center}
	\begin{figure*}
		\centering
		\includegraphics[width=0.96\linewidth]{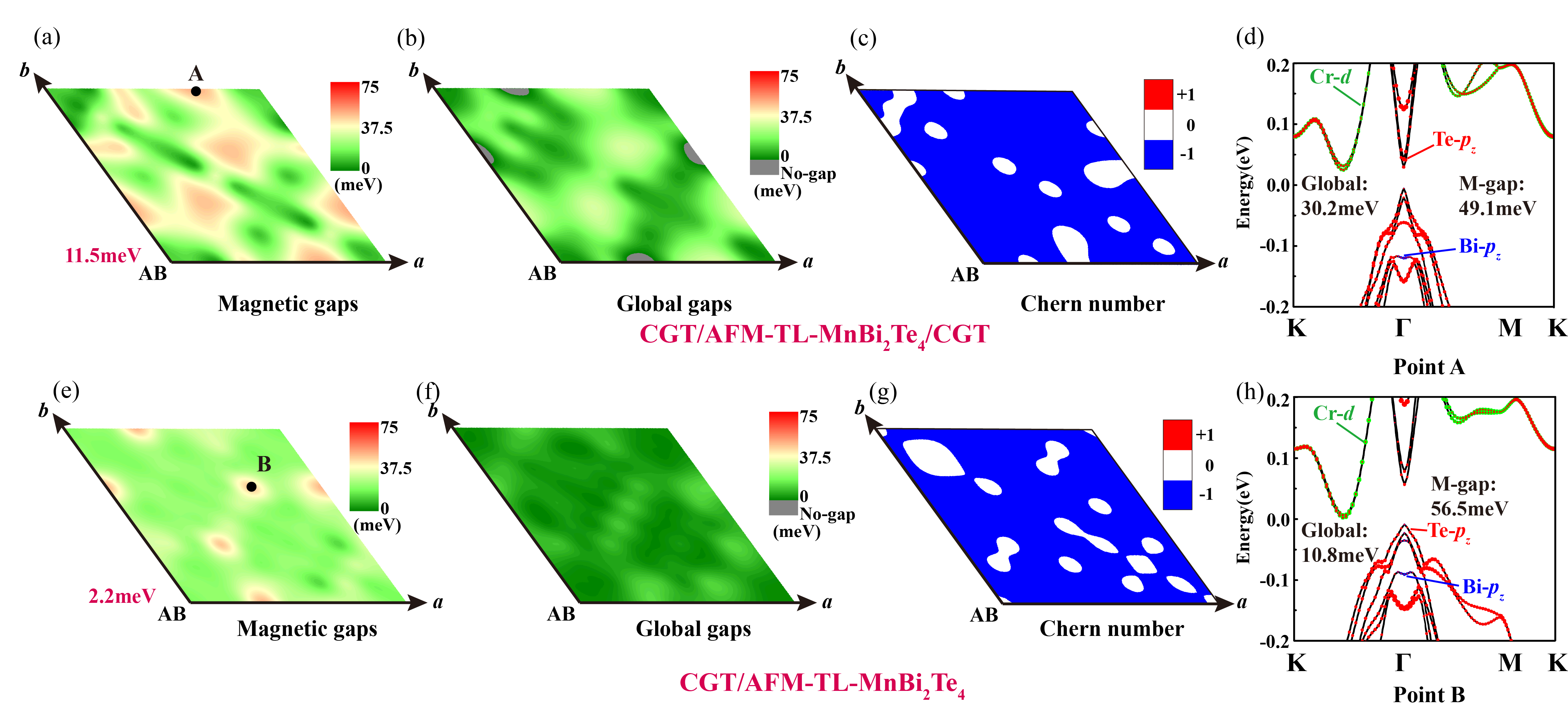}
		\caption{Gaps and Chern-number manipulations in CGT/TL-MBT/CGT and CGT/TL-MBT under the magnetic ground state induced by stacking-order-shifts. Distributions of (a) the magnetic gaps, (b) the global gaps and (c) the Chern numbers via employing stacking-order-shifts of CGT layers adopting the building-block of CGT/TL-MBT/CGT under the magnetic ground state. In (a) and (b) green-yellow-red color bar is adopted, with the gray color stands for “no-gap” zones. In (c), blue and white are relevance to that $C$ = -1, 0 respectively. (d) Orbital-projected band structures of ground-state CGT/TL-MBT/CGT at the stacking-order position of “Point A” denoted in (a). Olive, blue and red bubbles perform as Cr-3$d$, Bi-6$p$ and Te-5$p$ orbital components respectively. (e)-(g) are similar to (a)-(c), but under the building-block of ground-state CGT/TL-BT. (h) depicts the orbital projected band structures similar to that in (d) but at the stacking-order position of “Point B” denoted in (e).}
		\label{fig8:CGT-MBT-stacking}
	\end{figure*}
\end{center}

Introducing magnetism into TI layer manifests itself the ability of broadening the global gap, step-likely tunable Chern number character with switchable chirality in the same time. We play close attention to sandwich-structural CGT/TL-MBT/CGT and bi-heterostructure of CGT/TL-MBT opting AFM interlayer coupling within TL MBT and FM coupling between CGT and MBT, corresponding to the magnetic ground states, referring to Tables \ref{tab4:FMI-FMIs} and \ref{tab5:FMI-MBTs} in Appendix \ref{D:M-Gap_others}. Outcomes of stacking-order-shift manipulated gaps, Chern numbers and orbital projected band structures revealed in Fig. \ref{fig8:CGT-MBT-stacking} contain much more abundant novel behaviors compared to that of Bi(Sb)$_2$Te$_3$-based TI layers.

\begin{center}
	\begin{figure*}
		\centering
		\includegraphics[width=0.96\linewidth]{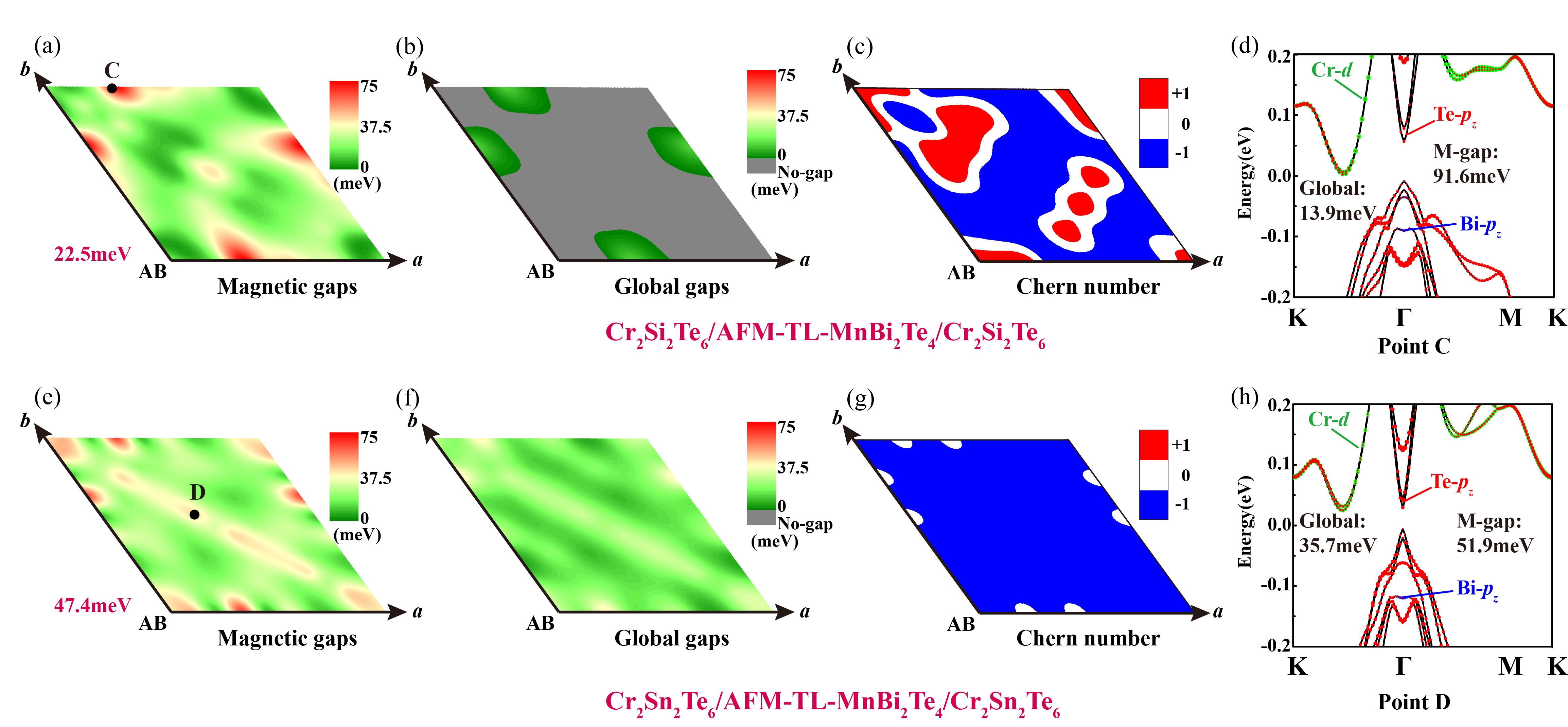}
		\caption{Stacking-order-shift induced gaps and Chern-number manipulations in Cr$_2$Si(Sn)$_2$Te$_6$/TL-MBT/Cr$_2$Si(Sn)$_2$Te$_6$ under the magnetic ground state. Distributions of (a) the magnetic gaps, (b) the global gaps and (c) the Chern numbers via employing stacking-order-shifts of CGT layers adopting the building-block of Cr$_2$Si$_2$Te$_6$/TL-MBT/Cr$_2$Si$_2$Te$_6$ under the magnetic ground state. In (a) and (b) green-yellow-red color bar is adopted, with the gray color stands for “no-gap” zones. In (c), blue, white and red zones are associated with that $C$ = -1, 0, +1, respectively. (d) Orbital-projected band structures of ground-state Cr$_2$Si$_2$Te$_6$/TL-MBT/Cr$_2$Si$_2$Te$_6$ at the stacking-order position of “Point C” denoted in (a). Olive, blue and red bubbles represent Cr-3$d$, Bi-6$p$ and Te-5$p$ orbital components respectively. (e)-(g) are similar to (a)-(c), but under the building-block of ground-state Cr$_2$Sn$_2$Te$_6$/TL-MBT/Cr$_2$Sn$_2$Te$_6$. (h) delineates the orbital projected band structures analogous to that in (d) but at the stacking-order position of “Point D” denoted in (e).}
		\label{fig9:CrSiTe3-MBT-stacking}
	\end{figure*}
\end{center}

The competition between Cr-based and Mn-based magnetic gaps delineates the Chern-number- and chirality-tunable functions in this building block. For the case CGT/TL-MBT/CGT, even if ground-state-nature of MBT that contributes the compensated magnetic gaps attenuated by its AFM-coupling configuration, Mn-based magnetic gap conquers Cr-based one in the whole range of FMI-TI hybridization strengths. Thus, Chern number is modulated only as $C$ = 0 and -1 exhibited in Fig. \ref{fig8:CGT-MBT-stacking}(c). The majority of stacking-order regions weakens the inter-vdW electronic hybridizations, inducing the decreasing of Cr-based magnetic gaps and attenuating the compensating effects, engendering the final magnetic gaps to larger extent. Summarized in Fig. \ref{fig8:CGT-MBT-stacking}(a), most regions leave much larger magnetic gaps than the normal stacking [11.5meV noted in the original point in Fig. \ref{fig8:CGT-MBT-stacking}(a)]. The absence of “M”-shaped valence band creates sizable global gaps above 20meV in most regions, in relevance with the thermal excitation temperature around 232K. Selected as “Point A” in Fig. \ref{fig8:CGT-MBT-stacking}(a) along the [100] direction, the orbital-projected band structure under this stacking-order depicted in Fig. \ref{fig8:CGT-MBT-stacking}(d) opens up the magnetic-gap as 49.1meV and the global gap as 30.2meV severally, much larger than that under normal stacking (11.5meV for the former and 3.7meV for the latter). Thus, in CGT/MBT/CGT-based building block, shifting stacking orders turns into an impactful method on optimizing the magnetic and the global gaps accordingly. Regrettably, the maximum Cr-based magnetic gap fails to totally counteract the Mn-based magnetic gap, leaving only $C$ = -1 in most areas and little regions with trivial nature, shown in Fig. \ref{fig8:CGT-MBT-stacking}(c). 

Bi-heterostructure type CGT/TL-MBT possesses very similar magnetic-gap optimization functions summarized in Figs. \ref{fig8:CGT-MBT-stacking}(e)-\ref{fig8:CGT-MBT-stacking}(g), in which the global gap retains by both the absence of “M”-shaped valence band the Mn-based magnetic gap in the non-CGT edge. “Point B” shown in Fig. \ref{fig8:CGT-MBT-stacking}(e) contains band structures with the magnetic gap up to 56.5meV and the global gap as 10.8meV, the latter of which correspond to about 125K. Without building sandwich structure, via only experimentally tuning stacking-orders in bi-heterostructure we’ve also obtained the exercisable functions of optimizing the magnetic and the global gap stemming from the knowledge of inner factors determining the magnetic gap we’ve concluded in Sec \ref{iii:vdW-spacing}. Behaving alike, no positive Chern number emerges totally in the whole stacking-order-shift range [Fig. \ref{fig8:CGT-MBT-stacking}(c)]. But the only dual-Chern-number (-1 and 0) nature can also be served to the switches as “on” and “off” in electronic applications.

It's worth to be put forward that for the above two cases, the interlayer magnetic coupling between adjacent MBT layers vanishes above $T_{\rm C1}$, i.e., the first critical temperature within TL MBT itself (about 11K as computed in Sec. \ref{v:Tc}), the magnetic-gap contribution of the medial MBT vanishing, lifting the final magnetic gaps and the global gaps totally in the whole stacking-order zone (see Fig. \ref{fig27:High-T CGT-3MBT} in Appendix \ref{E:stacking-order} via the foundation of CGT/TL-MBT/CGT). Therefore, it’s large enough to seek for the high-temperature Chern insulating phases around this magnetic-gap size up to $T_{\rm C2}$ (critical temperature of marginal layers of MBT).

Producing larger positive-chirality magnetic gaps in Cr$_2$Si$_2$Te$_6$, in some stacking-order areas of magnetic-ground-state Cr$_2$Si$_2$Te$_6$/TL-MBT/Cr$_2$Si$_2$Te$_6$, the Cr-based magnetic gap successfully conquers the Mn-based magnetic gaps, creating Chern number as +1 distributed in red regions of Fig. \ref{fig9:CrSiTe3-MBT-stacking}(c), including the normal stacking-order with residue Cr-based magnetic gap as 22.5meV [Fig. \ref{fig9:CrSiTe3-MBT-stacking}(a)]. Stacking-order-induced and step-likely Chern-number tunable and chirality-switchable functions emerge in this setup, going shares with alike functions consisting in that of germanene/MBT \cite{li2024multimechanism,xue2024valley}. However, the global gaps exist only around the “Point C” position of Fig. \ref{fig9:CrSiTe3-MBT-stacking}(a) with the value lower than 14meV. We only display the orbital-projected band structure of “Point C” with the global gap as large as 13.9meV, corresponding to the thermal excitation temperature of about 161K with Chern number as -1. Under this condition, it’s convenient to manipulate the chirality of Chern number, but discommodious to observe the transportation outcomes under most of stacking-order regions.

Conversely, Cr$_2$Sn$_2$Te$_6$ has lower ability conquering Mn-based magnetic gap but higher ability to guarantee the global gap in the whole stacking-order zone, with most regions possessing global gaps above 30meV, seen from Figs. \ref{fig9:CrSiTe3-MBT-stacking}(e)-\ref{fig9:CrSiTe3-MBT-stacking}(f). Only Chern numbers as -1 and 0 can be reached with the latter exists at very small areas along the [100] direction, emerged in Fig. \ref{fig9:CrSiTe3-MBT-stacking}(g). Located along the [1$\overline{1}$0] direction, the band structure of “Point D” denoted in Fig. \ref{fig9:CrSiTe3-MBT-stacking}(e) embodies large global-gapped (35.7meV) superiority of this system. The maintained large-value global-gap property filling in the whole stacking-order zone manifests it an excellent candidate for the transportation measurements, the topological switchable properties (trivial or non-trivial), high temperature Chern insulating candidates, and so on.

\section{Magnetic critical temperatures}
 \label{v:Tc}

Employing MI/TI based proximity system is confirmed theoretically and experimentally to achieve high critical temperatures among previous investigations \cite{hou2019magnetizing,eremeev2013magnetic,men2013magnetic,grutter2021magnetic,fang2023exchange,zhang2018strong,zou2020intrinsic,cgt-alegria2014large,cgt-mogi2019large,cgt-yao2019record,YFO-jiang2015independent,YFO-jiang2016enhanced,TFO-tang2017above,LCO-zhu2018proximity,Fe3O4-pereira2020topological,Cr2O3-wang2019observation,EuS-katmis2016high,EuS-lee2016direct,EuS-wei2013exchange,CrSb-he2017tailoring,CrSb-he2018topological,MnTe-he2018exchange,GaMnAs-lee2018engineering,ZnCrTe-watanabe2019quantum}. In most works, the TI layer is chosen as Bi$_2$Te$_3$, Bi$_2$Se$_3$, Sb$_2$Te$_3$, or Cr-doped into the former candidates. For the MI layer, not only CGT, but also non-vdW-layered oxides including Cr$_2$O$_3$, Fe$_3$O$_4$, Tm$_3$Fe$_5$O$_{12}$, LaCoO$_3$, BaFe$_{12}$O$_9$, and other ordinary transition metal oxides find wide use in proximity systems \cite{grutter2021magnetic,cgt-alegria2014large,cgt-mogi2019large,cgt-yao2019record,YFO-jiang2015independent,YFO-jiang2016enhanced,TFO-tang2017above,LCO-zhu2018proximity,Fe3O4-pereira2020topological,Cr2O3-wang2019observation,EuS-katmis2016high,EuS-lee2016direct,EuS-wei2013exchange,CrSb-he2017tailoring,CrSb-he2018topological,MnTe-he2018exchange,GaMnAs-lee2018engineering,ZnCrTe-watanabe2019quantum}.

Specially, M. Mogi \textit{et al} \cite{cgt-mogi2019large} fabricates the sandwich structures of CGT/(Bi,Sb)$_2$Te$_3$/CGT which persists the long-ranged magnetism up to about 100K. But regrettably, the quantized anomalous Hall resistance is unable to be achieved ($\sim$25.812 k$\Omega$), with only about a little bit more than 1k$\Omega$ at 2.5K can be reached. Among the above chosen MIs, transition metal oxides manifest high critical temperatures, with most of them above 100K \cite{grutter2021magnetic,YFO-jiang2015independent,YFO-jiang2016enhanced,TFO-tang2017above,LCO-zhu2018proximity,Fe3O4-pereira2020topological,Cr2O3-wang2019observation}. But considering the large electronegativity difference between O and Te, Se, Bi, Sb, etc, the hybridization strength is very weak \cite{li2020tunable}, hindering the gaps of TSS and then, the observation of quantized anomalous Hall resistance. Employing CrSb as the MI layer proximate to (Bi,Sb)$_2$Te$_3$ successfully increases the anomalous Hall resistance to about 10k$\Omega$ at 1.9K \cite{CrSb-he2017tailoring,CrSb-he2018topological}, possibly benefitting from the small electronegativity between Sb and Te, and then the larger gap of TSS.

Except for the experimental fabricating process that may induces impurities and defects, the failure to reach the high temperature QAH states near to the magnetic critical temperature is mainly originated from the small gap (similarly, the magnetic gap) of TSS, the misalignment of the gaps between the two edges of TSSs, and the invasion of the bulk state into the TSS-gaps crossing through the Fermi level \cite{grutter2021magnetic,tss-li2024realization}. Therefore, establishing the guidance for optimizing the magnetic gaps, and then, the gaps of TSS, plays the vital rule in recovering the quantized anomalous Hall resistance near to their magnetic critical temperatures. In the following, we focus on the magnetic critical temperatures based on the systems discussed in this work, which is expected to be fully exploited with the maximally optimized gaps.

In this section, via MC simulation methods with Heisenberg model and by drawing heat capacity ($C_{\rm V}$)-temperature ($T$) curves, we’ve synthetically estimated the magnetic critical temperatures of the building blocks that discussed in Secs. \ref{iii:vdW-spacing} and \ref{iv:stacking}. Results affirm that the critical temperatures of long-range magnetism are evidently lifted to a great extent in virtue of stretched in-plane lattice of CGT and the strong FM coupling between CGT and MBT, adequately utilizing the high-temperature potential created under the methods of optimizing magnetic gaps put forward in Sec. \ref{iii:vdW-spacing} and \ref{iv:stacking}.

\begin{center}
	\begin{figure*}
		\centering
		\includegraphics[width=0.9\linewidth]{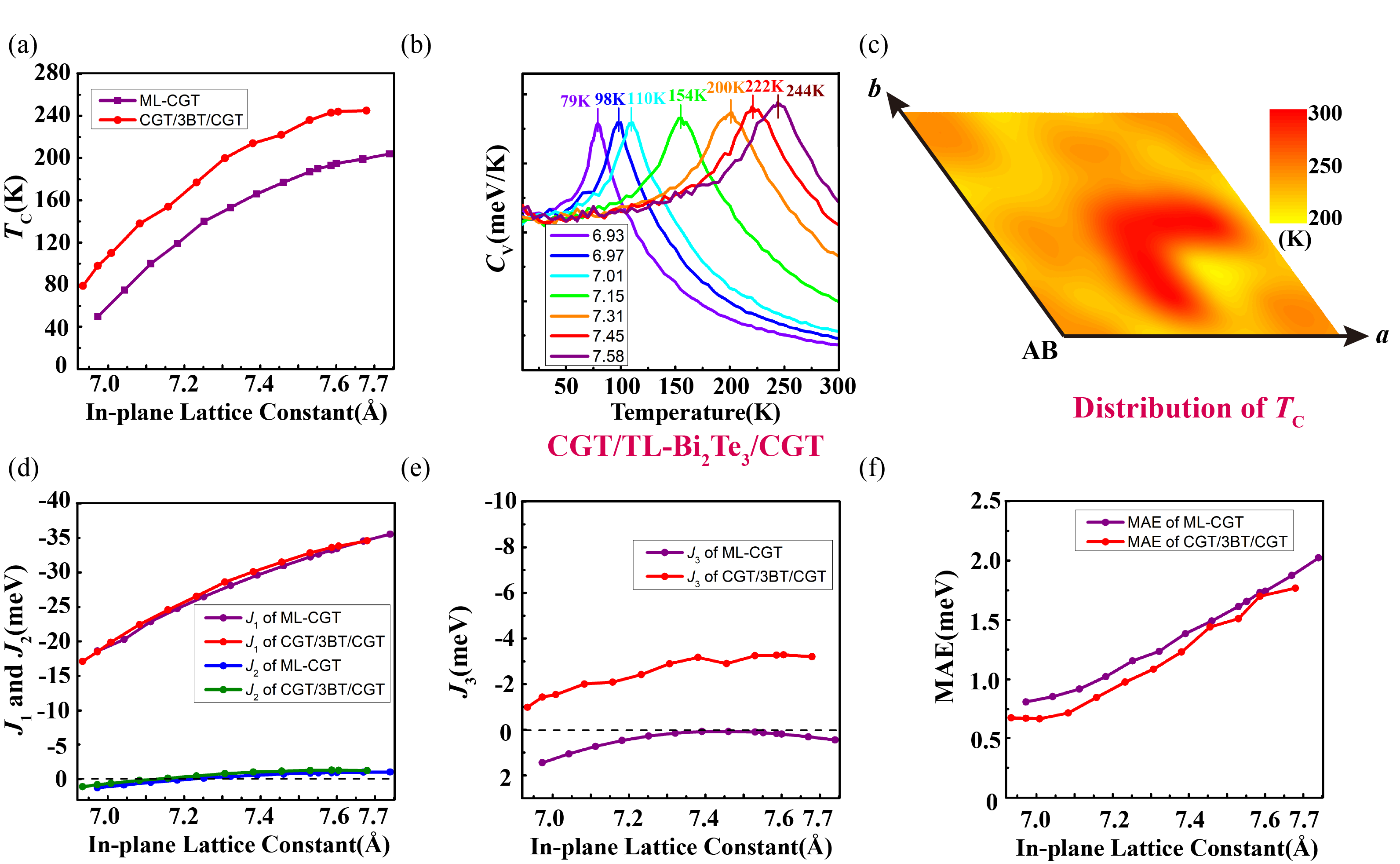}
		\caption{Magnetic exchange coefficients, MAEs and the critical temperatures in various in-plane lattice constants of monolayer CGT and CGT/TL-BT/CGT. (a) Critical temperatures of monolayer CGT (purple curve) and CGT/TL-BT/CGT (red curve) vs. in-plane lattice constants. (b) $C_{\rm V}$-$T$ curves of CGT/TL-BT/CGT under in-plane lattice constants of 6.93Å, 6.97Å, 7.01Å, 7.15Å, 7.31Å, 7.45Å and 7.58Å, depicted with violet, blue, cyan, green, orange, red and purple curves respectively, with the $T_{\rm C}$ noted at the peak of each curve. (c) Distributions of $T_{\rm C}$ by varying stacking-orders of CGT/TL-BT/CGT. From yellow to red colors the value of $T_{\rm C}$ increases. (d) First and second magnetic exchange coefficients of monolayer CGT and CGT/TL-BT/CGT vs. in-plane lattice constants. Purple and red curves denote the $J_1$ of monolayer CGT and CGT/TL-BT/CGT, while blue and olive curves stand for the $J_2$ of monolayer CGT and CGT/TL-BT/CGT. (e) Third magnetic exchange coefficients of monolayer CGT and CGT/TL-BT/CGT vs. in-plane lattice constants, in which purple and red curves denote the $J_3$ of monolayer CGT and CGT/TL-BT/CGT respectively. (f) MAEs’ evolutions of monolayer CGT (purple curve) and CGT/TL-BT/CGT (red curve) along in-plane lattice constants.}
		\label{fig10:CGT-BT-Tc}
	\end{figure*}
\end{center} 

First of all, we’ve revealed that, for FMI/TI heterostructures the high critical temperature stems from the enlargement of magnetic exchange coefficients ($J_1$,$J_2$,$J_3$,$\cdots$) especially for the nearest and the third nearest exchange coefficients ($J_1$ and $J_3$).  For the Heisenberg model adopted in this work, the model can be written as in the cases of monolayer-FMI and FMI/BT heterostructure:

\begin{equation}
H_{\rm Heisenberg} = \sum_{i}A_{i}^{\rm FMI}\left(S_{i}^{z}\right)^{2}+\sum_{i,j}J_{ij}^{\rm FMI}\textbf{\textit{S}}_{\textbf{\textit{i}}}\cdot \textbf{\textit{S}}_{\textbf{\textit{j}}}
 \label{eq7}
\end{equation}

The first term in the right of Eq.\ref{eq7} is the MAE of Cr in FMI layer, and the second term denotes the magnetic exchange interactions including infinite orders. $J_{ij}^{\rm FMI}$ is positive with AFM couplings while negative with FM couplings. For convenience we choose to compute three orders of magnetic interactions via various supercell-based magnetic configurations, omitting higher-order interactions that almost have no contribution to $T_{\rm C}$, in the following critical temperature estimations. 

For the Heisenberg model in the case of FMI/TL-MBT/FMI, we write:

\begin{equation}
\begin{aligned}
H_{\rm Heisenberg} & = \sum_{i}A_{i}^{\rm FMI}\left(S_{i}^{z}\right)^{2}+\sum_{i}A_{i}^{\rm MBT}\left(S_{i}^{z}\right)^{2} \\ & +\sum_{i,j\rm \in FMI}J_{ij}^{\rm FMI}\textbf{\textit{S}}_{\textbf{\textit{i}}}\cdot \textbf{\textit{S}}_{\textbf{\textit{j}}} + \sum_{i,j\rm \in MBT}J_{ij}^{\rm MBT}\textbf{\textit{S}}_{\textbf{\textit{i}}}\cdot \textbf{\textit{S}}_{\textbf{\textit{j}}} \\
& + \sum_{i\rm \in FMI,j\rm \in MBT}J_{ij}^{\rm FMI-MBT}\textbf{\textit{S}}_{\textbf{\textit{i}}}\cdot \textbf{\textit{S}}_{\textbf{\textit{j}}} \\ & + \sum_{ij \in   \rm MBT}J_{ij}^{\rm interlayer-MBT}\textbf{\textit{S}}_{\textbf{\textit{i}}}\cdot \textbf{\textit{S}}_{\textbf{\textit{j}}}
 \label{eq8}
\end{aligned}
\end{equation}

Analogically, the first two terms provide the MAEs of Cr in FMI layer and Mn in MBT layer respectively; the third and the fourth terms are corresponding to the intralayer magnetic exchange coefficients within FMI and MBT layers. The fifth and the sixth terms represent the interlayer magnetic couplings between FMI and marginal MBT layers in the former, and between marginal MBT and medial MBT layers in the latter respectively. For the intralayer coupling, we select third-order approximate coefficients, while for the interlayer coupling, we choose two-order approximate coefficients to estimate the critical temperatures in FMI/MBT heterostructures.

\begin{center}
	\begin{figure*}
		\centering
		\includegraphics[width=0.54\linewidth]{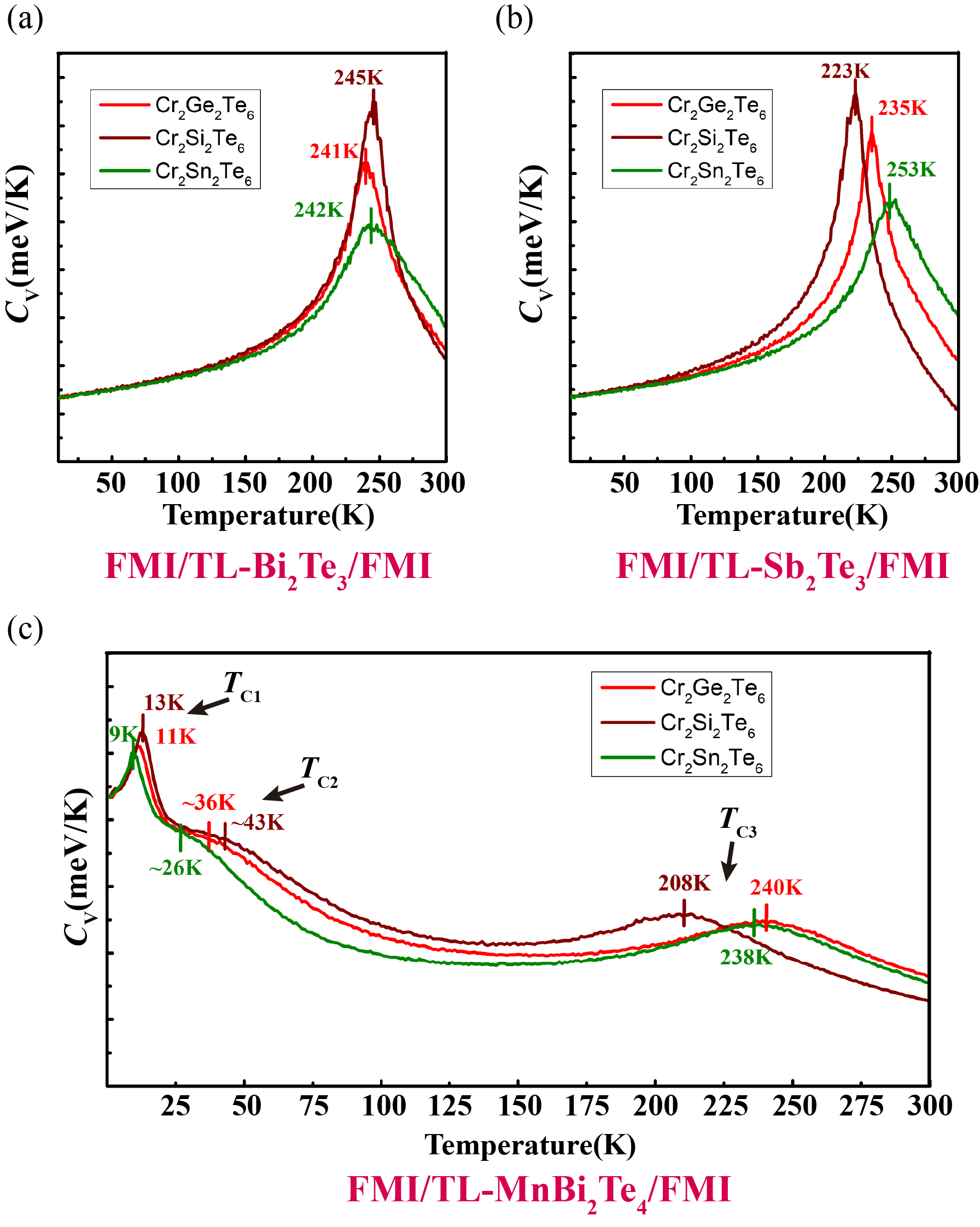}
		\caption{$C_{\rm V}$-$T$ curves of (a) FMI/TL-BT/FMI, (b) FMI/TL-ST/FMI and (c) FMI/TL-MBT/FMI. Red, brown and olive curves stand for the cases when FMI = CGT, Cr$_2$Si$_2$Te$_6$, Cr$_2$Sn$_2$Te$_6$ respectively. The critical temperatures of each curve are noted above the peak.}
		\label{fig11:FMI-3BT-FMT-Cv}
	\end{figure*}
\end{center}

Evidently, the Cuire temperature of monolayer CGT varies acutely to the in-plane lattice constants. Seen from Fig. \ref{fig10:CGT-BT-Tc}(a), the stretching strain from 0\% to +10\% lifts the Curie temperature from 50K to about 204K in monolayer CGT (purple curve); in the meantime, for CGT/TL-BT/CGT, $T_{\rm C}$ increases from 98K to 245K with the lattice constants driving from -7\% to +3\% [red curve in Fig. \ref{fig10:CGT-BT-Tc}(a)], with the in-plane lattice constant range same to that of the monolayer CGT. Considering the lattice mismatch between CGT and the $\sqrt{3}\times\sqrt{3}$ superlattice of BT and MBT that will drive lattice stretch in CGT to about +7\% $\sim$ +10\% depending on the thickness of TI layer, constructing CGT/BT or CGT/MBT heterostructure is a handy way to amplify the Curie temperatures in CGT by about two or three times. $C_{\rm V}-T$ curves under various in-plane lattice constants are displayed in Fig. \ref{fig10:CGT-BT-Tc}(b) with multiple colors. The values of $T_{\rm C}$ denoted at the peak of the curve with the same color, we’ve verified the $T_{\rm C}$-lifting effects created via the lattice stretch. Moreover, stacking-order-shifts cast fluctuations on $T_{\rm C}$ of CGT, shown in Fig. \ref{fig10:CGT-BT-Tc}(c), in the range of 200K$\sim$300K, indicating this lattice-stretch-induced $T_{\rm C}$-lifting effect immune to stacking-orders but influenced by different small lattice distortions of CGT. The extremely high $T_{\rm C}$ region ($\sim$290K) is related to the high $J_3$ region, revealed in Fig. \ref{fig30:J-mapping}(c) of Appendix \ref{F:Tc}.

\begin{table}
	\caption{\label{tab2:Tc}Critical temperatures of monolayer CGT and CGT grown on BT (same to that of CGT/TL-BT/CGT) varies with the in-plane lattice constants. Among these circumstances, the red digits mark the in-plane lattice constants of bulk BT (7.59Å, $\sqrt{3}\times \sqrt{3}$ supercell) and bulk MBT (7.55Å, $\sqrt{3}\times \sqrt{3}$ supercell) respectively.}
	\begin{ruledtabular}
		\begin{tabular}{ccc}
			\multicolumn{1}{c}{CGT} &
			\multicolumn{1}{c}{\textbf{\textit{a}}(Å)} &
			\multicolumn{1}{c}{$T_{\rm C}$(K)} 
			\\ \hline
			\multirow{6}{*}{Monolayer} & 6.97  & 50   \\
			& 7.25 & 140 \\
			& 7.46        & 177    \\   & \textcolor{red}{7.55}  &  \textcolor{red}{190} \\  &  \textcolor{red}{7.59}  & \textcolor{red}{193} \\  &  7.74  &  204 \\
			\hline
			\multirow{6}{*}{Grown on BT} & 6.97  & 98  \\
			& 7.23 & 177 \\  &  7.46  & 222 \\ & 7.53 & 236 \\ & \textcolor{red}{7.59} & \textcolor{red}{244} \\ & 7.67 & 245 \\
		\end{tabular}
	\end{ruledtabular}
\end{table}

\begin{center}
	\begin{figure*}
		\centering
		\includegraphics[width=0.68\linewidth]{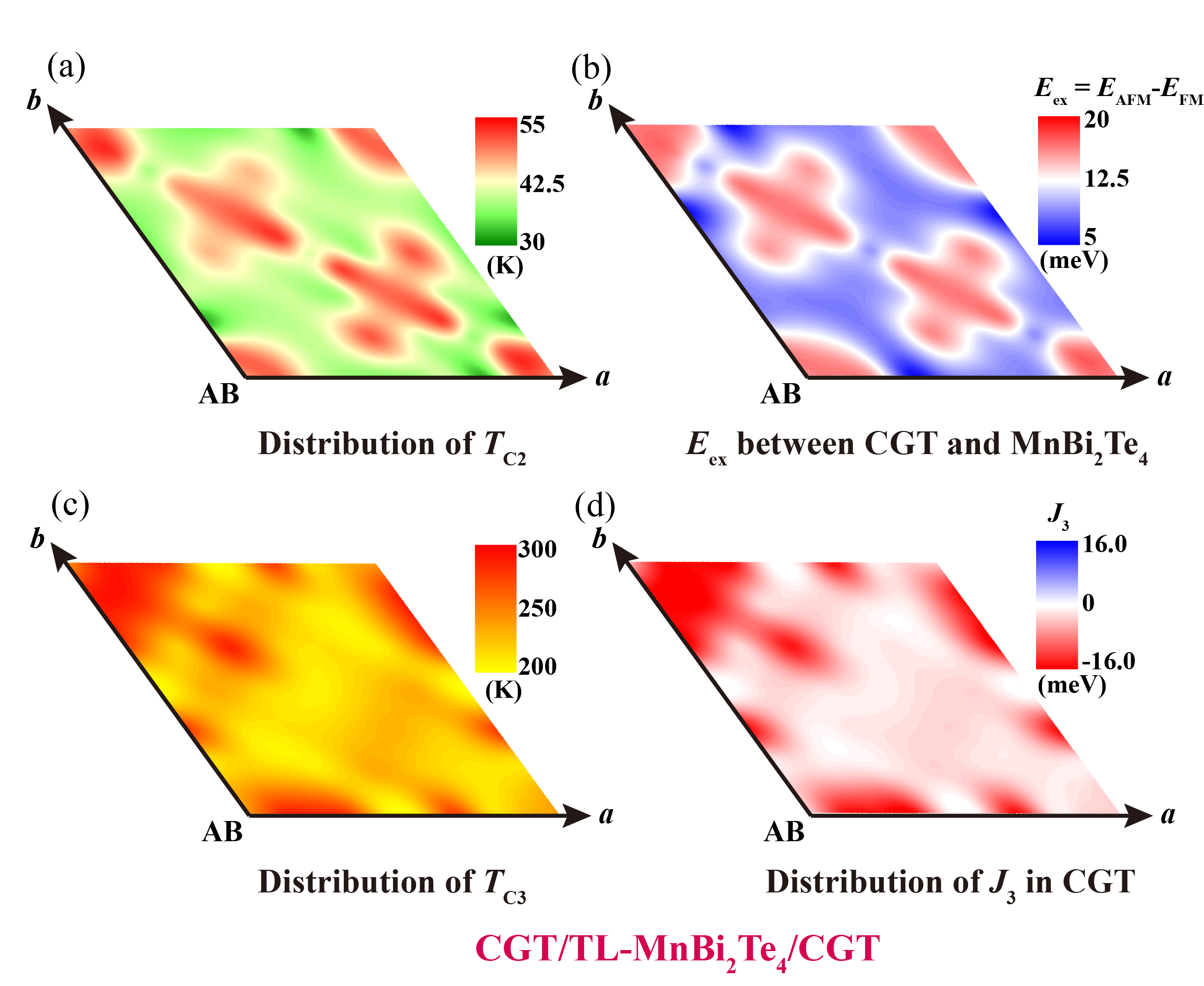}
		\caption{Distributions of $T_{\rm C}$ and magnetic exchange coefficients in CGT/TL-MBT/CGT manipulated by stacking-order-shifts. (a) $T_{\rm C2}$ of CGT/TL-MBT/CGT distributions manipulated by stacking-order-shifts. From green to yellow then to red colors, the value of temperature increases. (b) The value of interlayer magnetic couplings ($E_{\rm ex}$) between CGT and MBT distributes along different stacking-order-shifts. From blue to white then to red colors the strength of exchange increases. (c) $T_{\rm C3}$ of CGT/TL-MBT/CGT distributions modulated by stacking-orders. From yellow to red, the value of temperature increases. (d) The value of $J_3$ in the CGT layer distributes along different stacking-order-shifts. Red color is related to negative (FM-coupling) values while blue color is related to positive (AFM-couplings) values.}
		\label{fig12:FMI-MBT-Tc2}
	\end{figure*}
\end{center}

Noteworthy, in the same lattice constant condition, monolayer CGT that grown on BT possesses larger $T_{\rm C}$ than that of free-standing one, compared in Fig. \ref{fig10:CGT-BT-Tc}(a). This discrepancy is confirmed with the lifting of negative-valued, third-nearest magnetic exchange coefficient as $J_3$, shown in Fig. \ref{fig10:CGT-BT-Tc}(e), adding more far-linked FM-coupling contributions into the further enlargement of $T_{\rm C}$. $J_1$, $J_2$ and MAE suffer little by the emergence of TI-substrate [Figs. \ref{fig10:CGT-BT-Tc}(d) and \ref{fig10:CGT-BT-Tc}(f), Fig. \ref{fig30:J-mapping} in Appendix \ref{F:Tc}]. Hence, growing CGT on typical TI substrate is a suitable strategy to significantly enhance the critical temperatures of CGT by lattice-stretch induced entirely magnetic-exchange-coefficients magnification and the substrate induced third-nearest-interaction promotion. We’ve listed monolayer-CGT under circumstances of free-standing and grown on BT separately in Table \ref{tab2:Tc} under various lattice constants. Marked with red digits, 7.59Å and 7.55Å are corresponding to the $\sqrt{3}\times\sqrt{3}$ superlattice, in-plane lattice constants of BT and MBT accordingly.

Figure \ref{fig11:FMI-3BT-FMT-Cv} delineates the $C_{\rm V}$-$T$ curves under various FMIs with TI layer as TL BT [Fig. \ref{fig11:FMI-3BT-FMT-Cv}(a)], TL ST [Fig. \ref{fig11:FMI-3BT-FMT-Cv}(b)] and TL MBT [Fig. \ref{fig11:FMI-3BT-FMT-Cv}(c)]. Replacing CGT by Cr$_2$Si$_2$Te$_6$ or Cr$_2$Sn$_2$Te$_6$ casts little changes on critical temperatures. For the cases of TI = TL BT or TL ST, only one critical temperature exists corresponding to the Curie temperature of Cr in FMI. However, for the condition TI = TL MBT, things become more complicated, for three critical temperatures emerge due to the great disparity between different kinds of magnetic exchange strengths in Eq.\ref{eq8}. Marked in Fig. \ref{fig11:FMI-3BT-FMT-Cv}(c) with $T_{\rm C1}$, $T_{\rm C2}$ and $T_{\rm C3}$ separately, the first critical temperature ($T_{\rm C1}$) with the peaks at around 11K is linked to the Néel temperature within TL MBT itself, i.e. the critical temperature of medial MBT that is restrained only by the intralayer magnetic couplings of Mn-layer and the weak interlayer magnetic coupling between MBT layers. The temperature raised above $T_{\rm C1}$, the long-range-magnetism is perished in medial MBT, while the marginal MBT still harbors ferromagnetism protected by much stronger interlayer magnetic couplings between MBT and Cr$_2$$X_2$Te$_6$ layer \cite{li2020tunable}. Easy to be comprehend, the second critical temperature ($T_{\rm C2}$) is linked to the Curie temperature of marginal MBT conquering the MBT-FMI interlayer couplings, marked in the points of inflexion in Fig. \ref{fig11:FMI-3BT-FMT-Cv}(c), mainly located from 26K to 43K. Above $T_{\rm C2}$, the whole MBT layer develops into paramagnetic phase contributing no magnetic gap, leaving the edged FMI layer still in the FM state. The third critical temperature ($T_{\rm C3}$) is relevant to the Curie temperature of FMI itself, sharing very closed values with the $T_{\rm C}$ of FMI/BT/FMI and FMI/ST/FMI. Notably, the interlayer magnetic couplings between Cr$_2$$X_2$Te$_6$ and MBT fails to lift the critical temperature of marginal MBT entirely to the extent of Cr$_2$$X_2$Te$_6$ itself, stemming from the much weaker intralayer magnetic exchange (Mn-Mn) strength within MBT.

Setting external temperatures between $T_{\rm C1}$ and $T_{\rm C2}$ in the FMI/MBT/FMI building block obtains larger magnetic gaps in virtue of the absence of the reversed magnetic-gap contributed from the medial MBT layer, depicted in Fig. \ref{fig27:High-T CGT-3MBT} of Appendix \ref{F:Tc} that employing the representative of CGT/TL-MBT/CGT. Based on the same building-block, Fig. \ref{fig12:FMI-MBT-Tc2}(a) displays the distributions of $T_{\rm C2}$ vs. stacking-order-shifts between CGT and MBT. Obviously, $T_{\rm C2}$ varies sensitively to the exchange strength of CGT-MBT interlayer magnetic couplings, coinciding with the distributions of the latter shown in Fig. \ref{fig12:FMI-MBT-Tc2}(b). Remarkably, the value of $T_{\rm C2}$ ranges from 30K to 55K, higher than that of pure MBT but much lower than lattice-stretched CGT itself. The stacking-order-shift induced variation of $T_{\rm C3}$ is drawn in Fig. \ref{fig12:FMI-MBT-Tc2}(c), performing very similar with that of CGT/TL-BT/CGT [Fig. \ref{fig10:CGT-BT-Tc}(c)], the higher value regions corresponding to the larger value regions of $J_3$ of intralayer magnetic couplings in CGT [Fig. \ref{fig30:J-mapping}]. In temperatures between $T_{\rm C2}$ and $T_{\rm C3}$, only CGT produces the positive chirality of magnetic gaps with small values shown in the lower panel of Fig. \ref{fig5:CGT-MBT}(c) in Sec. \ref{iii:vdW-spacing}.

In summary, the Curie temperatures of Cr$_2$$X_2$Te$_6$ arrive even above 200K in most conditions, commendably exploit the high-temperature potential of the global gaps optimized by the guidance of managing the magnetic gaps revealed in this work. The extremely enhanced high Curie temperature of Cr$_2$$X_2$Te$_6$ is supported from the major two factors: the stretch of in-plane lattice constants of CGT acutely causes entire enlargement of intralayer magnetic exchange coefficients, which can be facilitated by larger lattice constants of typical TI substrate; choosing BT as TI-layer even promotes the value of $J_3$ in Cr$_2$$X_2$Te$_6$, lifting Curie temperatures to even larger degree. Three critical temperatures exist when selecting TL MBT as TI layer, in which the second temperature ($T_{\rm C2}$) determines the upper limit of the magnetic-gap contributions from MBT, located at around 26K$\sim$50K under various shifted stacking-orders.

\section{Conclusion and guidance}
 \label{vi:conclusions}

In this work, by constructing FMI/TI/FMI sandwich structures that preserve inversion symmetry and systematically modulating the van der Waals (vdW) distances between FMI and TI layers, we have counterintuitively revealed a nonmonotonic dependence of the magnetic gap on the FMI-TI hybridization strength across various candidate structures. By analyzing the spatial distribution of TSSs and the competition between kinetic and Coulomb exchange interactions, we uncover the underlying mechanisms for optimizing magnetic and global gaps in FMI-TI proximity-based QAH systems. Our findings offer the following guidelines for researchers addressing both theoretical and experimental challenges in this field:

Extraction-backflow Process of TSS: Increasing FMI-TI hybridization strength induces an "extraction-backflow" process in the real-space distribution of TSS along the out-of-vdW-plane direction. Excessive hybridization, however, causes an overextended TSS distribution, pushing a portion back into the TI layer and subsequently suppressing magnetic gap enhancement.

Competing Exchange Mechanisms in Specific FMIs: In FMIs like Cr$_2$$X_2$Te$_6$ and CrI$_3$, kinetic exchange and Coulomb exchange engender the opposite sign of chirality in magnetic gaps, forming the competition relationship between each other. In the weak hybridization regime, the longer-range kinetic exchange dominates, yielding a positive-sign magnetic gap. As the hybridization strength increases, Coulomb exchange strength increases faster than kinetic exchange strength, and drives the value of the residue positive-signed magnetic gap down. Thus, the maximum magnetic gap is reached prior to the maximum TSS extraction along increasing hybridization.

Additional Tunability Brought by Magnetic-TI: Introducing magnetic elements into the TI layer provides additional degrees of freedom into manipulating the magnetic gap. For instance, Mn in MBT generates a larger negative-chirality magnetic gap than Cr. In FMI/MBT/FMI systems with FMI = Cr$_2$$X_2$Te$_6$ or CrI$_3$, increased hybridization strength conversely results in a decreasing-then-increasing process of the magnetic gap, consistent with the extraction-backflow behavior of TSS.

Stacking-Order Modulation: Adjusting the FMI-TI stacking order experimentally allows for continuous tuning of hybridization strength, optimizing magnetic gaps in systems such as FMI/BT, FMI/ST, and FMI/MBT. A slightly weaker hybridization from adjusted stacking yields substantially larger magnetic and global gaps.

Manipulating Chern Number and Chirality: Given the opposite chirality between FMI and MBT, tuning stacking order in FMI/MBT structures can yield Chern-number-tunable ($\pm 1$, 0) and chirality-switchable functions, notably with Cr$_2$Si$_2$Te$_6$. For CGT and Cr$_2$Sn$_2$Te$_6$, the magnetic gap opened by Cr can’t conquer that opened by Mn under all the circumstances, supporting Chern numbers only as -1 and 0.

Maintaining the Global Gap: The "M"-shaped valence band in typical BT-based TIs obstructs sizable global gaps across most stacking configurations, yielding few regions with global gaps exceeding 10 meV for high-temperature applications. Nevertheless, MBT gets rid of this restriction, achieving the sizable global gaps ($\textgreater$20meV) in most stacking regions by selecting FMI as CGT and Cr$_2$Sn$_2$Te$_6$, well-suited for seeking high temperature Chern insulators.

Curie Temperature Enhancement: Stretching the lattice or using BT-based substrates increases Cr$_2$$X_2$Te$_6$’s Curie temperature to above 200 K, compatible with the global gaps achieved through the optimization methods in this work. For FMI/MBT-based structures, MBT maintains negative-chirality magnetic gaps below the second critical temperature ($\sim$26–50 K), constrained by weak intralayer Mn-Mn magnetic exchange. The guidance of optimizing the magnetic gaps is expected to perfectly exploit the high magnetic critical temperatures in these systems.

These seven guiding principles provide a comprehensive roadmap for optimizing magnetic and global gaps, thus increasing the stability range of QAH states in FMI-TI proximity structures in experimentally accessible and controllable ways. We anticipate that these insights will significantly advance the design of high-temperature QAH materials and support future investigations in this promising area.

\begin{acknowledgments}
    We thank for Prof. Y. Xu, Prof. H. Weng, Prof. Q. Wu, Dr. J. Li and Dr. Z. Xu for helpful discussions. We also thank for Dr. Z. Xu for technical supporting. This work was supported by the National Natural Science Foundation of China (92065206). Part of the numerical calculations has been done on the supercomputing system in the Huairou Materials Genome Platform.

    Z. Li and F. Xue contributed equally to this work.
\end{acknowledgments}
\newpage

\appendix

\section{Structural stability \label{A:structural}}

\begin{center}
	\begin{figure}
		\centering
		\includegraphics[width=1\linewidth]{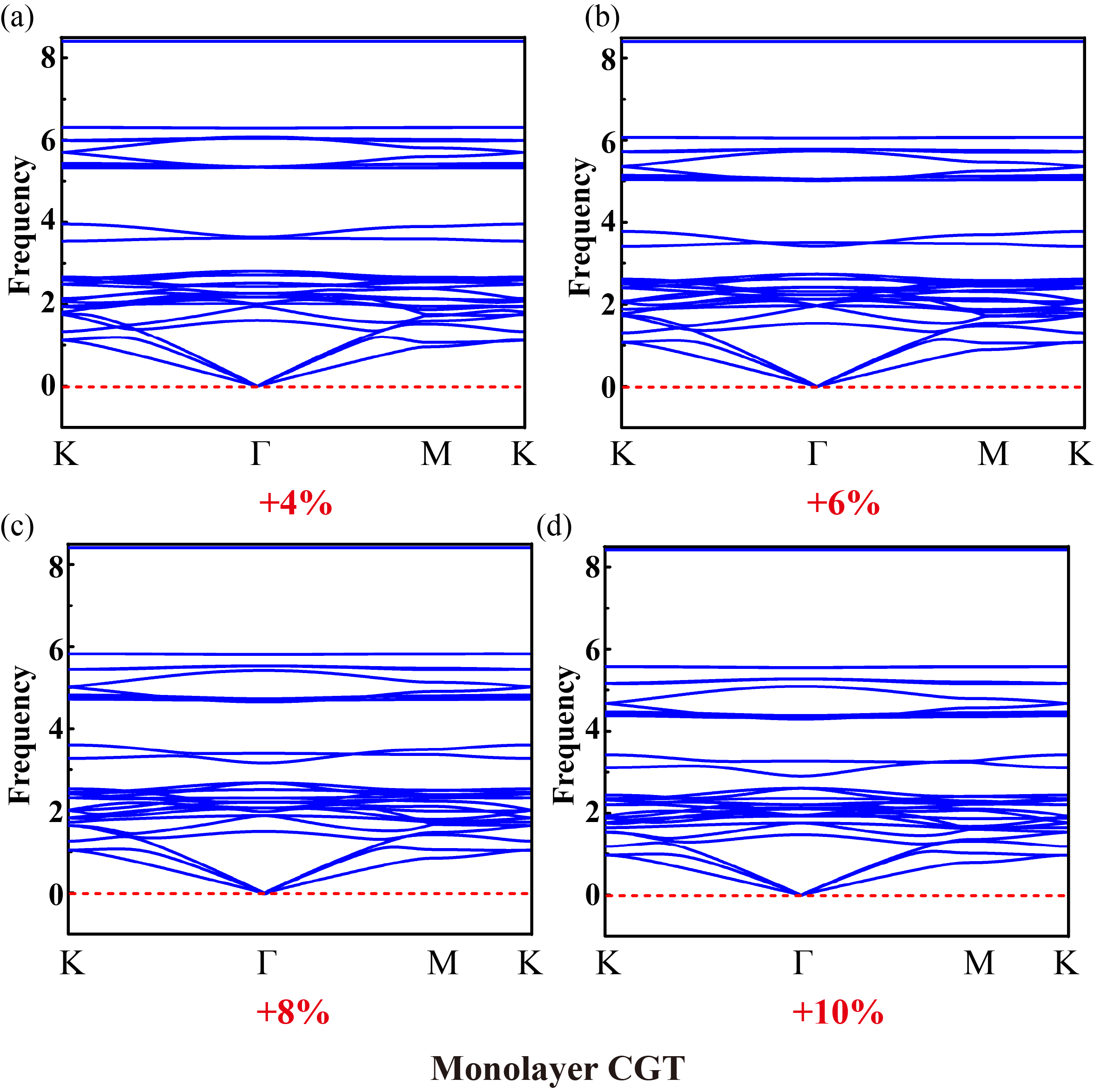}
		\caption{Phonon spectrums of monolayer CGT under in-plane lattice stretches of (a) +4\%; (b) +6\%; (c) +8\%; (d) +10\% respectively.}
		\label{fig13:phonon-CGT}
	\end{figure}
\end{center}

\begin{center}
	\begin{figure*}
		\centering
		\includegraphics[width=0.9\linewidth]{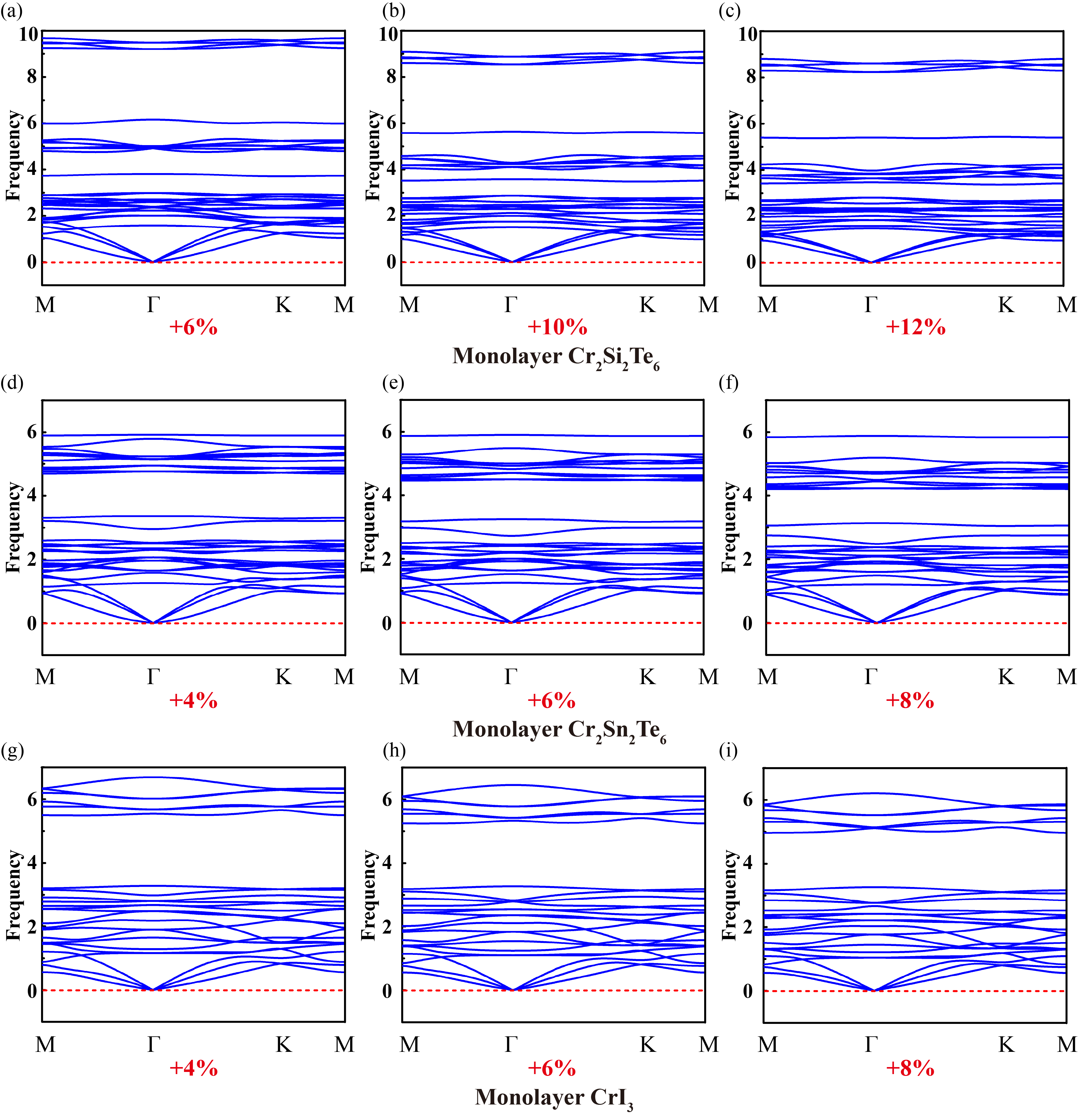}
		\caption{Phonon spectrums of other monolayer FMIs participating in the building blocks discussed in the main text. (a)-(c) are related to monolayer Cr$_2$Si$_2$Te$_6$ with biaxial strains of +6\%, +10\%, +12\% respectively. (d)-(f) are related to those of monolayer Cr$_2$Sn$_2$Te$_6$, but under lattice stretches of +4\%, +6\% and 8\%. (g)-(i) are related to those of monolayer CrI$_3$, under lattice stretches of +4\%, +6\% and 8\%.}
		\label{fig14:phonon-FMI}
	\end{figure*}
\end{center}

\begin{center}
	\begin{figure*}
		\centering
		\includegraphics[width=1\linewidth]{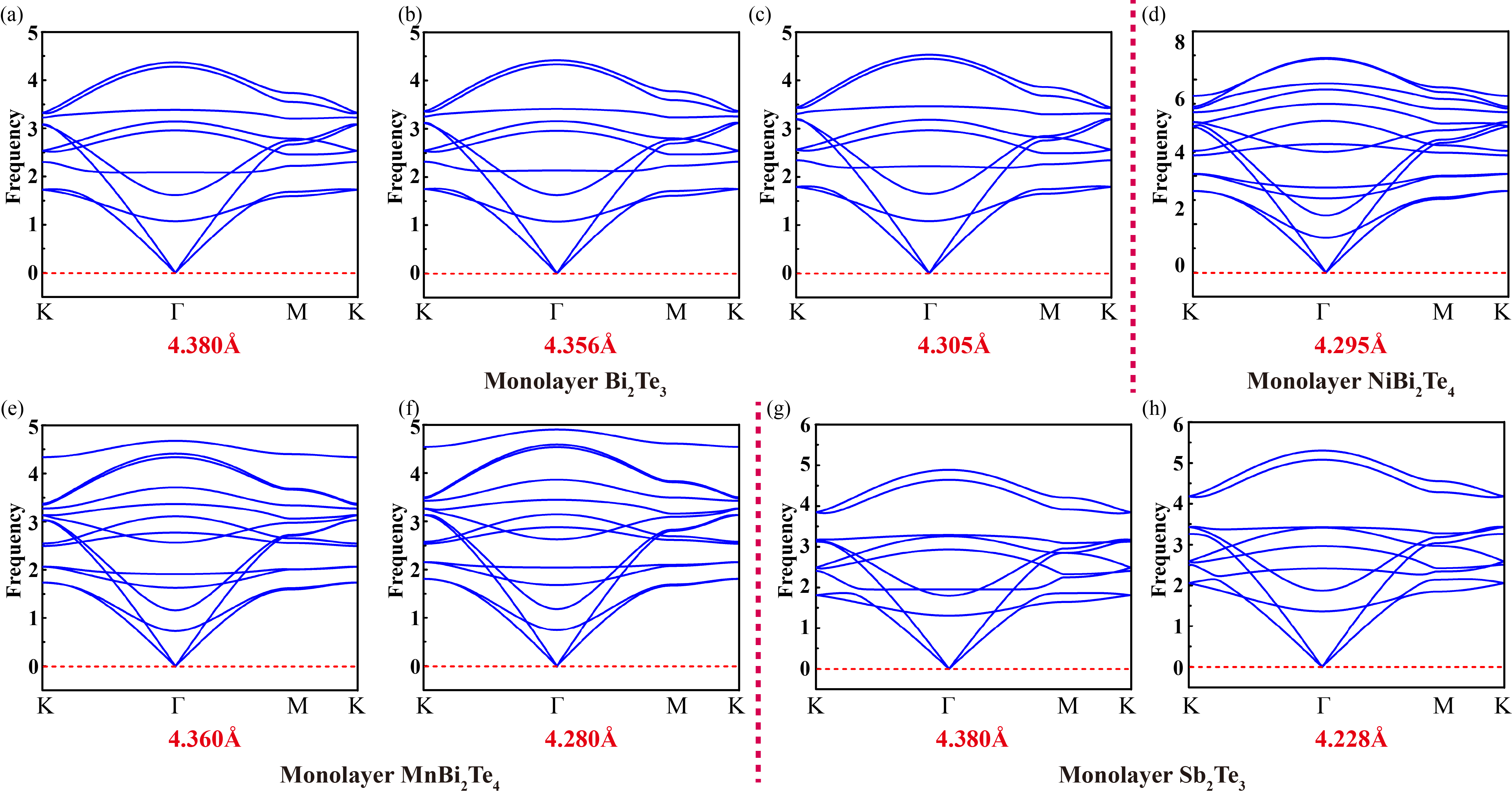}
		\caption{Phonon spectrums of other monolayer TIs participating in the building blocks discussed in the main text. (a)-(c) are related to monolayer BT with in-plane lattice constants of 4.380Å, 4.356Å and 4.305Å respectively. (d) is related to that of monolayer NiBT under the case \textbf{\textit{a}} = 4.295Å. (e) and (f) are linked to those of monolayer MBT under the cases \textbf{\textit{a}} = 4.360Å and 4.280Å. (g) and (h) are corresponding to those of monolayer ST in which \textbf{\textit{a}} = 4.380Å and 4.228Å respectively.}
		\label{fig15:phonon-TI}
	\end{figure*}
\end{center}

We compute the magnetic-gap evolutions mainly by setting the in-plane lattice constants of its bulk form, except for the case TI = ST in which we also choose \textbf{\textit{a}} = 4.380Å that same to the case TI = BT. In the building blocks of CGT/TL-BT/CGT, CrI$_3$/TL-BT/CrI$_3$, CGT/TL-ST/CGT, and CGT/TL-MBT/CGT, we also carry out free relaxation process by fixing the cell volume, with the optimized in-plane lattice constants listed in Table \ref{tab3:lattice}. Weak in-plane lattice compress is induced by CGT into the TI layer because the latter owns the thicker construction feature to resist the biaxial strains than that of monolayer CGT.

The structural stability of the materials under various biaxial strains derived from the substrate is a crucial factor towards the building blocks discussed in the main text, applied in analysis of the magnetic-gap performance under multiform of factors and experimental optimizations via stacking-order-shifts. First of all, the in-plane lattice constants of CGT mainly fall in the values of 7.586Å, corresponding to the $\sqrt{3}$ times that of BT; 7.456Å, corresponding to that of free-relaxed CGT/TL-BT/CGT; 7.323 Å, corresponding to that of free-relaxed CGT/TL-ST/CGT; and 7.413Å, corresponding to that of free-relaxed CGT/TL-MBT/CGT, all of which are distributed within lattice stretches below +8.8\%. Hence, we provide phonon spectrums of monolayer CGT under lattice stretch as +4\%, +6\%, +8\% and +10\%, displayed in Figs. \ref{fig13:phonon-CGT}(a)-\ref{fig13:phonon-CGT}(d) respectively. Obviously, no virtual frequency exists totally under all these circumstances, at least confirming the structural stability of CGT with in-plane lattice stretches below +10\%. Similarly, the structural stability is also affirmed in FMIs of Cr$_2$Si$_2$Te$_6$ [Figs. \ref{fig14:phonon-FMI}(a)-\ref{fig14:phonon-FMI}(c)], Cr$_2$Sn$_2$Te$_6$ [Figs. \ref{fig14:phonon-FMI}(d)-\ref{fig14:phonon-FMI}(f)] and CrI$_3$ [Figs. \ref{fig14:phonon-FMI}(g)-\ref{fig14:phonon-FMI}(i)], all of them measured by phonon spectrums that cover the maximum lattice constant involved in the aforementioned building blocks.

\begin{table}
	\caption{\label{tab3:lattice}In-plane lattice constants of free-relaxed CGT/TL-BT/CGT, CrI$_3$/TL-BT/CrI$_3$, CGT/TL-ST/CGT and CGT/TL-MBT/CGT. }
	\begin{ruledtabular}
		\begin{tabular}{ccc}
			\multicolumn{1}{c}{Systems} &
			\multicolumn{1}{c}{\textbf{\textit{a}}(Å)} &
			\multicolumn{1}{c}{\textbf{\textit{a}}/$\sqrt{3}$(Å)} 
			\\ \hline
			CGT/TL-BT/CGT & 7.456  & 4.305   \\
			CrI$_3$/TL-BT/CrI$_3$ & 7.545 & 4.356 \\
			CGT/TL-ST/CGT & 7.323 & 4.228 \\
			CGT/TL-MBT/CGT & 7.413 & 4.280 \\
		\end{tabular}
	\end{ruledtabular}
\end{table}

Remarkably, heterostructure-like building blocks also implement biaxial stresses in TI layers. We select certain lattice constants of BT, ST, MBT and NiBT that appear in this work, and also verifying the character of structural stability under all these conditions with absolutely no virtual frequency, displayed in Fig. \ref{fig15:phonon-TI}. For the case of NiBT, we only choose the lattice constant of the case CGT/NiBT/BL-BT/NiBT/CGT that mentioned and discussed in Appendix \ref{D:M-Gap_others}.

\section{Evolutions of the hybridization gaps and $\Gamma$-point gaps along various vdW-spacing distances \label{B:hybridization}}

\begin{center}
	\begin{figure*}
		\centering
		\includegraphics[width=0.8\linewidth]{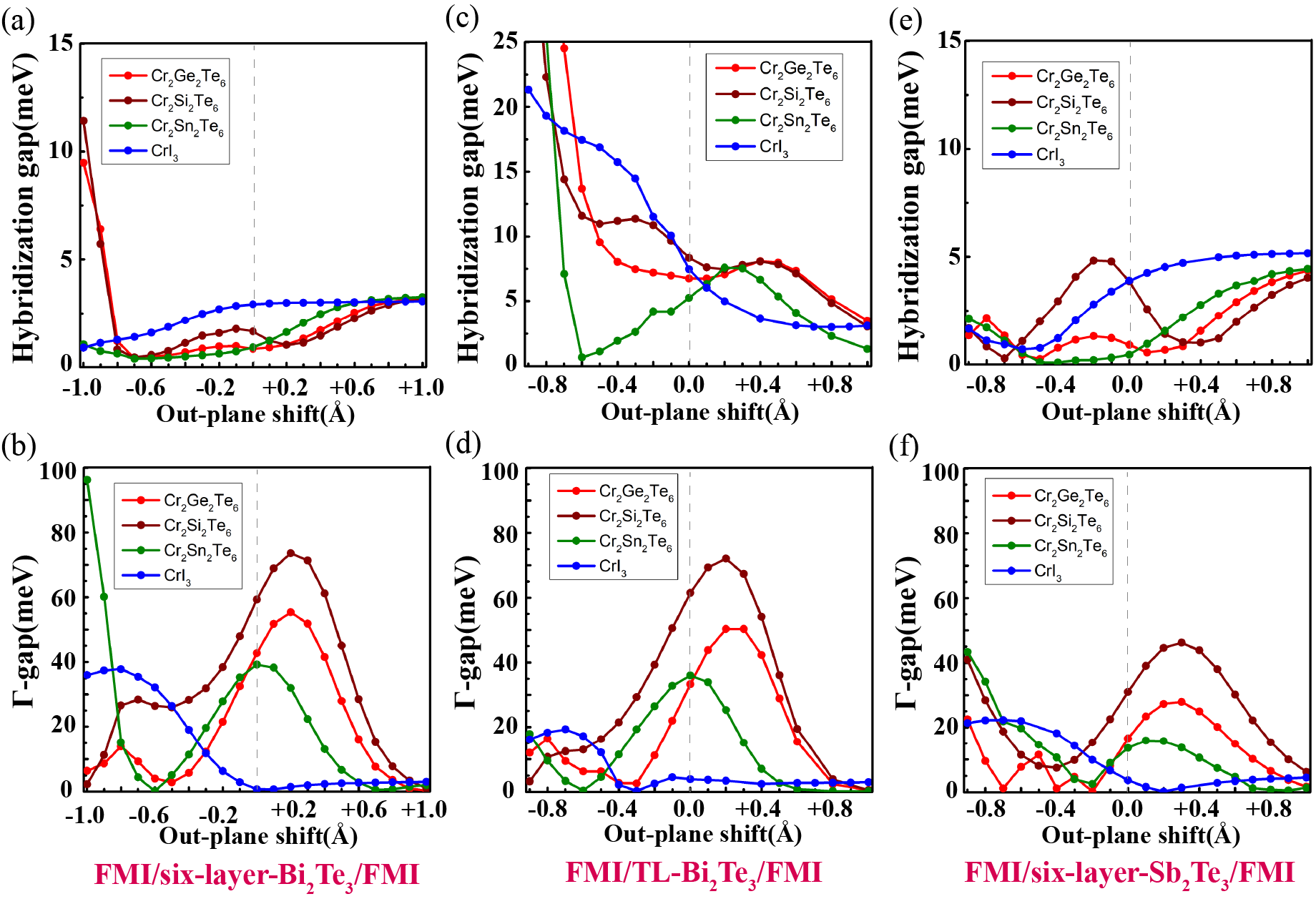}
		\caption{The evolution curves between the hybridization ($\Gamma$-point) gaps and the vdW-spacings when TI = BT or ST. (a) The hybridization gaps and (b) the $\Gamma$-point of FMI/six-layer-BT/FMI; (c) the hybridization gaps and (d) the $\Gamma$-point of FMI/TL-BT/FMI; (e) the hybridization gaps and (f) the $\Gamma$-point of FMI/six-layer-ST/FMI. Red, wine, olive and blue curves stand for the FMI layer as CGT, Cr$_2$Si$_2$Te$_6$, Cr$_2$Sn$_2$Te$_6$ and CrI$_3$ respectively. Light-gray dashed vertical line in each figure denotes the balanced distance of vdW-spacing (0.0Å).}
		\label{fig16:H-BT}
	\end{figure*}
\end{center}

\begin{center}
	\begin{figure*}
		\centering
		\includegraphics[width=0.56\linewidth]{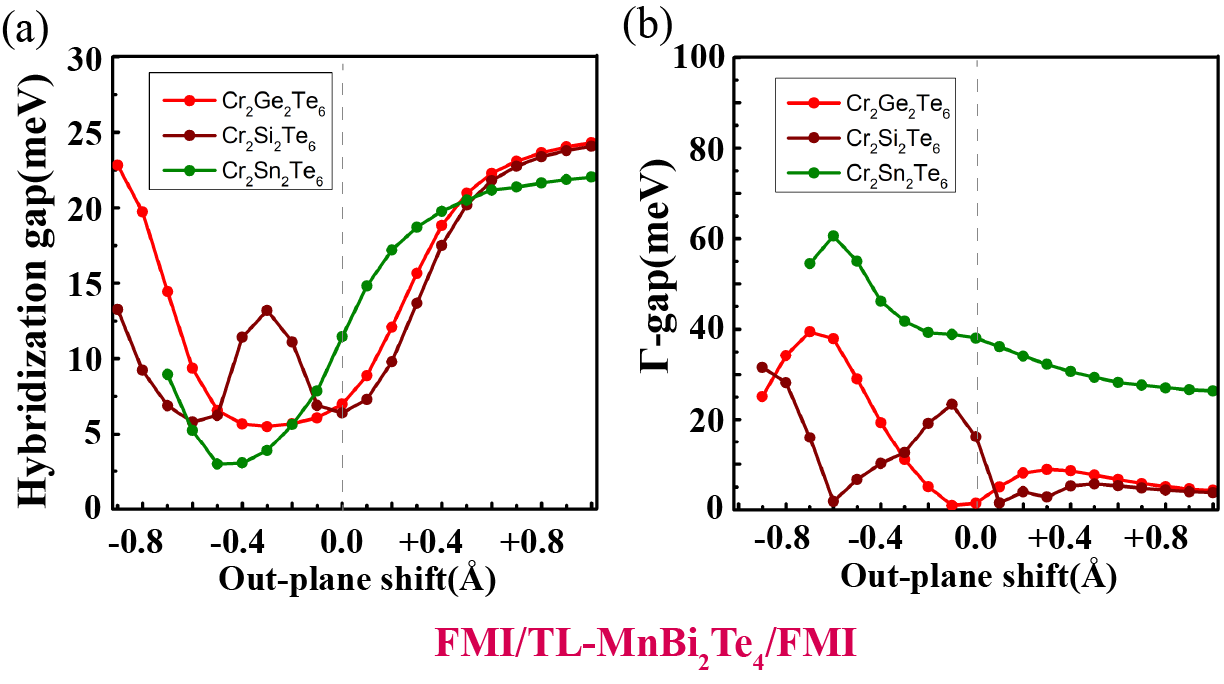}
		\caption{The evolution curves between the hybridization ($\Gamma$-point) gaps and the vdW-spacings when TI = MBT. (a) The hybridization gaps and (b) the $\Gamma$-point of FMI/TL-MBT/FMI. Red, wine, olive and blue curves stand for the FMI layer as CGT, Cr$_2$Si$_2$Te$_6$, Cr$_2$Sn$_2$Te$_6$ and CrI$_3$ respectively. Light-gray dashed vertical line in each figure denotes the balanced distance of vdW-spacing (0.0Å).}
		\label{fig17:H-MBT}
	\end{figure*}
\end{center}

\begin{center}
	\begin{figure*}
		\centering
		\includegraphics[width=1\linewidth]{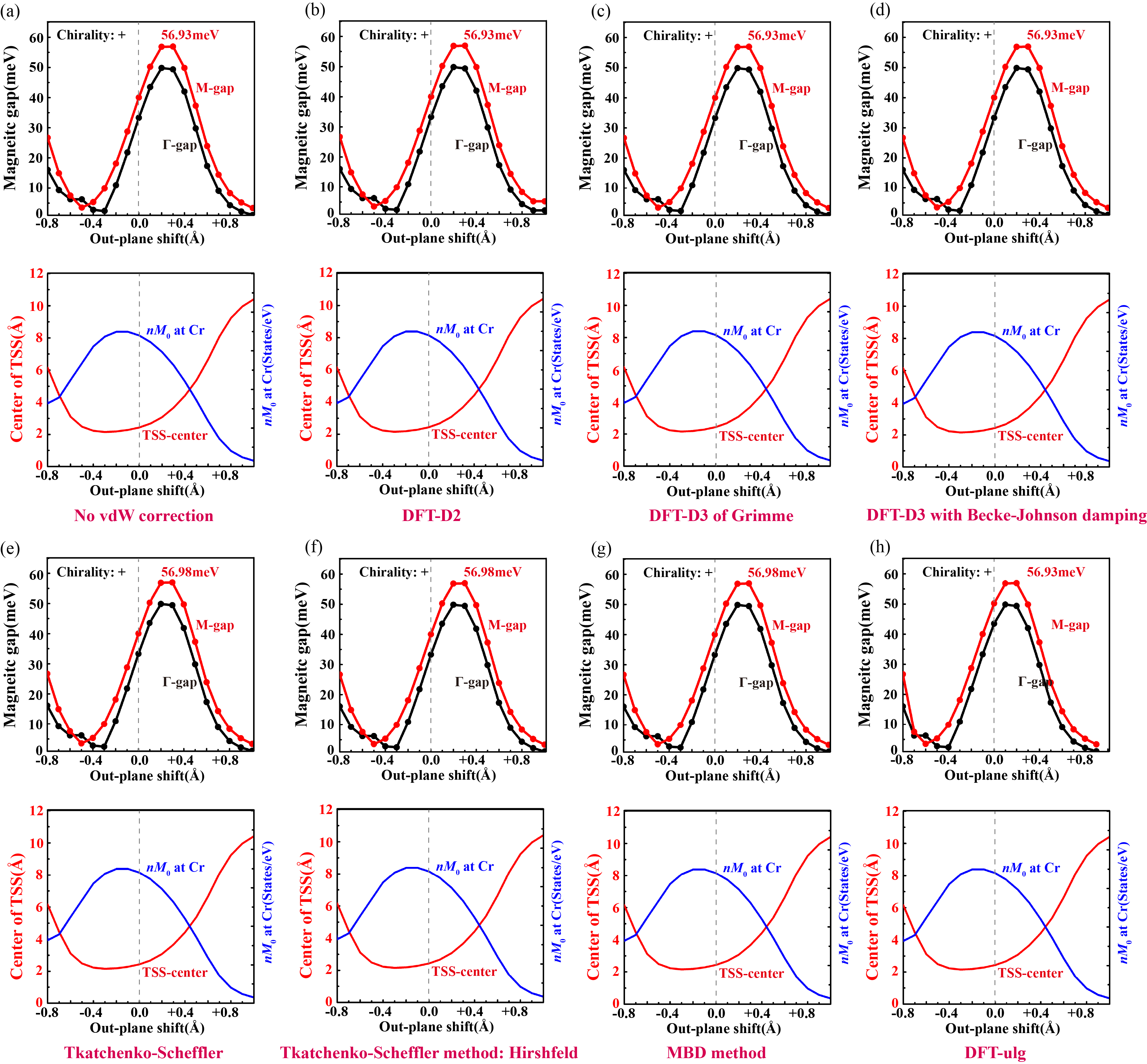}
		\caption{The evolution behaviors of the magnetic gaps, the $\Gamma$-point gaps, the center positions of TSS and the value of $\textbf{\textit{n}}\textbf{\textit{M}}_\textbf{0}$ at the Cr-atom position of CGT/TL-BT/CGT via adopting various vdW corrections by continuously varying the inter-vdW distances. From (a) to (h), computations are implemented under vdW corrections of DFT-D2, DFT-D3 method of Grimme with zero-damping function, DFT-D3 method with Becke-Johnson damping function, Tkatchenko-Scheffler method, Tkatchenko-Scheffler method with iterative Hirshfeld partitioning,  MBD energy method and DFT-ulg method respectively. Each figure includes the evolution behaviors of the magnetic gap (red curve) and the $\Gamma$-point gap (black curve) in the upper panel, and the evolution behaviors in the center position of TSS (red curve) and the values of $\textbf{\textit{n}}\textbf{\textit{M}}_\textbf{0}$ at the Cr atom (blue curve) in the lower panel.}
		\label{fig18:vdW_correction}
	\end{figure*}
\end{center}

\begin{center}
	\begin{figure*}
		\centering
		\includegraphics[width=1\linewidth]{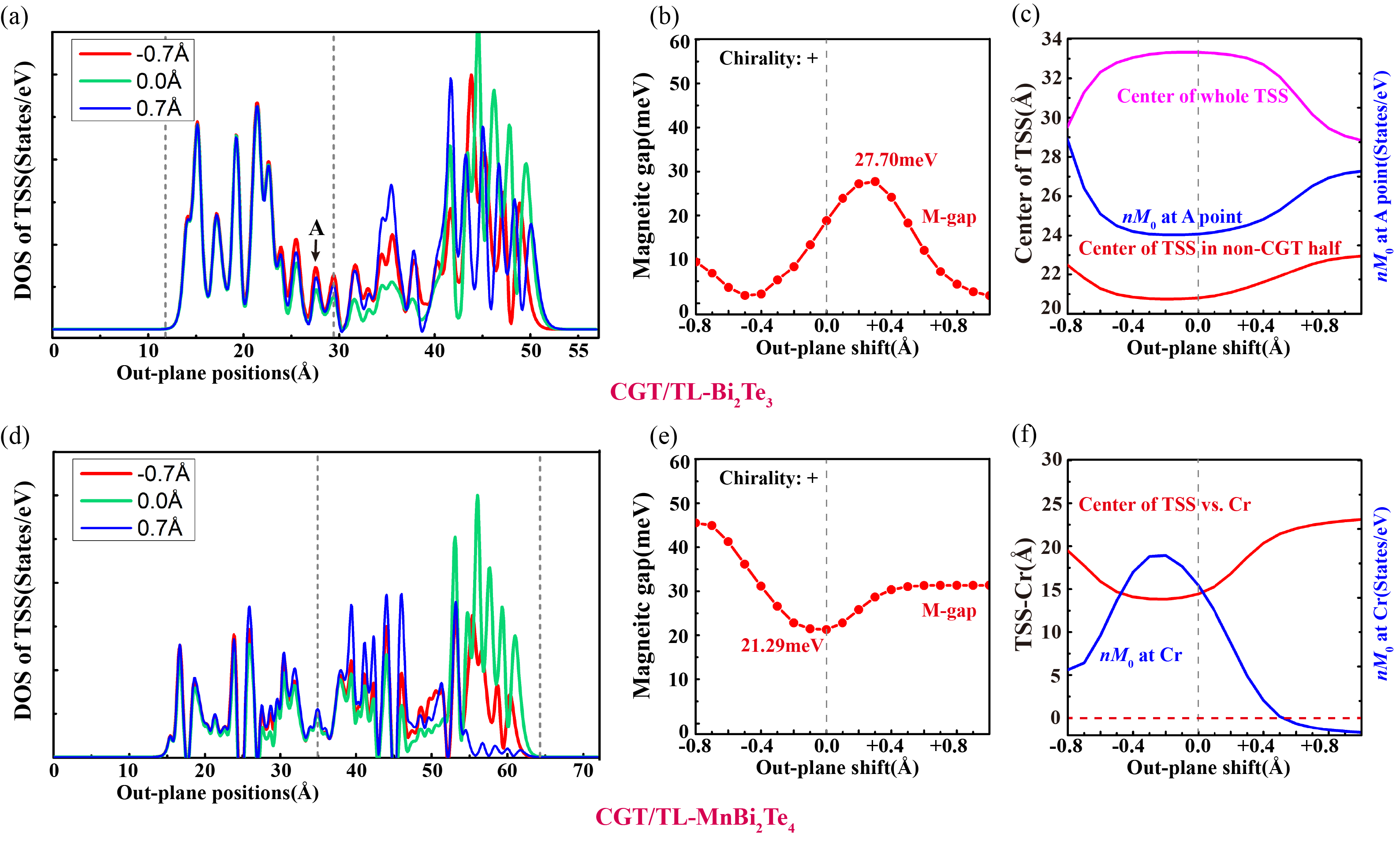}
		\caption{Inter-vdW distance varied magnetic gaps and the real-space distributions of TSSs within bi-heterostructure CGT/TL-BT and CGT/TL-MBT. (a) The real-space distributions of TSS under inter-vdW distances of -0.7Å (red curve), 0.0Å (green curve) and +0.7Å (blue curve) along $c$ axis. The CGT layer is arranged near to the right edge of TL BT layer, positioned at around 42Å to 52Å. Gray dashed line includes the half part that contributes the final magnetic gap in this system. (b) The evolution of the magnetic gaps of CGT/TL-BT by varying the inter-vdW distances. (c) The evolution of the center of the whole TSS (pink curve), the center of the half part TSS that contributes the final magnetic gap (red curve) and the value of $\textbf{\textit{n}}\textbf{\textit{M}}_\textbf{0}$ at the position “A” noted in (a) (blue curve). (d)-(f) are analogous to (a)-(c), but under the basis of CGT/TL-MBT. In (f), red curve shows the distances between the center of half TSS that contributes the final magnetic gap and the Cr positions; blue curve is the values of $\textbf{\textit{n}}\textbf{\textit{M}}_\textbf{0}$ at the Cr atoms.}
		\label{fig19:bi-hetero}
	\end{figure*}
\end{center}

As mentioned in Sec. \ref{iii:vdW-spacing}, the evolution behaviors of the magnetic gaps mainly rely on the out-of-vdW-plane distributions of TSS. Practically, the global gap of the whole Brillouin Zone determines the final potential of gap-corresponded factor of high temperature QAH state. For most FMI/TI building blocks, the global gaps are exactly the direct gap at the $\Gamma$ point, except for the case TI = BT within which the “M”-shaped valence band destroy the $\Gamma$-based direct gap. Considering that the hybridization gaps cancel out a portion of the magnetic gaps under non-zero Chern insulating regime (see Eq.\ref{eq3}), restraining the hybridization gap becomes also a crucial topic in pursuing larger $\Gamma$-based gaps and global gaps. Intuitively, we can select thicker layers of two dimensional building blocks to suppress the hybridization gaps, for instance, by arranging six layers of Bi(Sb)$_2$Te$_3$ as TI layer instead of three layers. Distinctly, the comparison carried out between FMI/six-layer-BT/FMI and FMI/TL-BT/FMI authenticates the thicker-layer-based stronger inhibition effect towards the hybridization gaps, seen from Figs. \ref{fig16:H-BT}(a) and \ref{fig16:H-BT}(b). The case when TI = six-layer-ST also suppresses the hybridization gaps in most vdW-spacing distances, shown in Fig. \ref{fig16:H-BT}(c), but with a little bit larger hybridization-gap values compared to that when TI = six-layer-BT, stemming from the smaller electronegativity difference between Sb and Te. 

The evolution performance of the hybridization gaps coincides with the distributing behaviors of TSS by varying the vdW-spacing distance, choosing thick TI layers. Apparently, for FMI as CGT and Cr$_2$Sn$_2$Te$_6$, the hybridization gap decreases firstly, then re-increase when diminishing the vdW-spacing distance, corresponding to the extraction-backflow process of the TSS (see Figs. \ref{fig3:nM0}-\ref{fig5:CGT-MBT}). More complicated behavior emerges in the case FMI as Cr$_2$Si$_2$Te$_6$, in which an obvious hump exists around the balanced vdW-distance, which is also related to its abnormal TSS-distributing behaviors [Fig. \ref{fig3:nM0}(a)]. These behaviors are valid when TI = TL MBT, see Fig. \ref{fig17:H-MBT}
(a). CrI$_3$ casts much weaker interaction strength with TIs, therefore only the decreasing process of the hybridization gap appears in the discussed vdW-spacing range (from -1.0Å to +1.0Å). Furthermore, thinner layer of TI causes more intricate interactions between the TSSs of the two edges, bringing about much elusive behaviors of hybridization gaps, shown in Fig. \ref{fig16:H-BT}(b) of which TI = TL BT.

Owing to the relatively small value of the hybridization gaps acting as the subtraction part, the $\Gamma$-gaps also maintain the rough evolution behaviors of the magnetic gaps, shown in Figs. \ref{fig16:H-BT}(b), \ref{fig16:H-BT}(d), \ref{fig16:H-BT}(f) and \ref{fig17:H-MBT}(b). When the $\Gamma$-gap decreases to zero, topological phase transition occurs, dividing distinct Chern numbers. Hereinafter all the locations of the phase transition point that relating to the values of vdW-spacing distance are listed separately in the next paragraph. Thicker layer of TIs maintains the nonzero Chern insulating characters immune to the much weaker FMI-TI hybridization strength in most conditions, meanwhile thinner layer of TIs is beneficial for constructing Chern-number tunable devices. 

In detail, for the condition that TI = six-layer BT, nonzero Chern-insulating phase dominates the whole range of vdW-spacing distance (-1.0Å$\sim$+1.0Å) under the arrangement that FMI = CGT and Cr$_2$Si$_2$Te$_6$, howbeit the phase transition from trivial to nonzero Chern insulating phase occurs at +0.7Å and 0.0Å for that of Cr$_2$Sn$_2$Te$_6$ and CrI$_3$ respectively. Similar manifestations appear via selecting six-layer ST, with the phase transitioning at +0.9Å (for Cr$_2$Sn$_2$Te$_6$) and +0.2Å (for CrI$_3$) respectively. Replacing the TI layer by TL BT fails to induce trivial-nontrivial phase transition even in the case of Cr$_2$Sn$_2$Te$_6$ on account of larger magnetic gaps induced by the other edge of FMI layer, meanwhile generates a smaller vdW-spacing-distance-related phase transition point in that of CrI$_3$ (-0.3Å). Markedly, we’re only capable of verifying the chirality (sign) of the Chern number, or it’s zero, given with the criterion illustrated in Figs. \ref{fig1:m_gap}(b) and \ref{fig1:m_gap}(c). The certain value of the Chern number needs to be ascertained by adopting local-density-of-state investigations based on TBHs, which is thrown into analysis in Appendix \ref{E:stacking-order}, majorly discussed in Fig. \ref{fig26:LDOS}.

\section{Robustness of non-monotonic evolutions in the magnetic gaps \label{C:Robustness}}

The non-monotonic evolving behavior of the magnetic gaps are totally immune to the selection of certain vdW correction method. Depicted in Fig. \ref{fig18:vdW_correction}, opting CGT/TL-BT/CGT as the representative system, we choose seven kinds of vdW corrections accompanied with no-vdW-correction to verify the robustness of the non-monotonic evolutions. The seven vdW corrections are serially mentioned in Sec. \ref{ii:methods}. Obviously, these vdW correction methods almost have no influence on the non-monotonic evolving behaviors of the magnetic gaps, the extraction-backflow nature of TSS along with the increasing of the inter-vdW hybridization strength, and even the largest magnetic gap values, totally falling in 56.93meV and 56.98meV. Hence, the non-monotonic evolving behaviors of the magnetic gaps in FMI/TI heterostructures is totally robust against various vdW correction methods.

Considering the expedient fabrication of FMI/TI bi-heterostructure building blocks, we also test the validity of the non-monotonic evolving behavior of the magnetic gaps in this situation. In these inversion symmetry broken systems, the final gap is determined by the smaller one comparing those at the two edges. Similarly, selecting two representative building blocks as CGT/TL-BT and CGT/TL-MBT, mainly discussed in the results exhibited in Fig. \ref{fig19:bi-hetero}, we succeed to confirm the robustness of the non-monotonic behavior even under these building bases with the inversion symmetry broken.

For the case of CGT/TL-BT, its final magnetic gap is determined by the edge without CGT layer, due to its smaller magnetic gap motivated by the CGT layer far in the other edge. We detailly test the evolution behaviors of TSS in this half part, circumscribed in Fig. \ref{fig19:bi-hetero}(a) with gray dashed lines. Intuitively, varying the inter-vdW distances implements little influence on the TSS distributions within this part. However, moderate inter-vdW distance between CGT and TL BT absorbs a little bit part of TSS near the center of TL BT, denoted as position “A” in Fig. \ref{fig19:bi-hetero}(a). This weak change is obviously revealed in the $\textbf{\textit{n}}\textbf{\textit{M}}_\textbf{0}$ values at position “A”, depicted as the blue curve in Fig. \ref{fig19:bi-hetero}(c). The reduction of the $\textbf{\textit{n}}\textbf{\textit{M}}_\textbf{0}$ values around position “A” also induces the little left shift of the center of TSS in the non-CGT half part, shown as the red curve in Fig. \ref{fig19:bi-hetero}(c). This little variation reduces the final magnetic gaps.

Hence, we’re capable of supplying a description of non-monotonic evolving behaviors in this building block. Due to the far distance between Cr atoms and the TSS distributing among the non-CGT half part, only the kinetic exchange mechanism takes effect. As the CGT layer pushed near to the TL BT layer firstly, the TSS distribution of the non-CGT half part remains unchanged, yielding the increasement of the magnetic gap. When the inter-vdW distance between CGT and TL BT decreases to the moderate extent (-0.5Å to +0.2Å), the absorption of TSS around the position “A” hampers the kinetic exchange strength, causing the magnetic gap descent. At the small inter-vdW extent (\textless -0.5Å), the backflow, or that is to say, the diving of TSS back to the TL BT layer re-increase the $\textbf{\textit{n}}\textbf{\textit{M}}_\textbf{0}$ values around position “A” and then, the value of the magnetic gap. Consequently, the non-monotonic evolving behavior of the magnetic gaps is still valid in this case [Fig. \ref{fig19:bi-hetero}(b)] but with much small values, with the maximum value as 27.7meV at +0.2Å.

The case of CGT/TL-MBT gives a much simpler description of the non-monotonic evolving mechanism [Figs. \ref{fig19:bi-hetero}(d)-\ref{fig19:bi-hetero}(f)]. Stemming from the opposite chirality of the magnetic gaps contributed by Cr and Mn, the neighboring of CGT layer cripples the magnetic gap in the CGT-neared edge, but leaving the magnetic gap almost unchanged in the non-CGT-neared edge. Hence, the final magnetic gap is determined by the CGT-neared edge, sharing the very analogous evolving behaviors with that of CGT/TL-MBT/CGT: decreasing-increasing characters. Finally, we can provide a summary that the non-monotonic evolution character is still strongly valid under the bases of FMI/TI and FMI/magnetic-TI bi-heterostructures.

\section{Magnetic-gap evolutions in other conditions \label{D:M-Gap_others}}

\begin{figure*}
	\centering
	\includegraphics[width=0.6\linewidth]{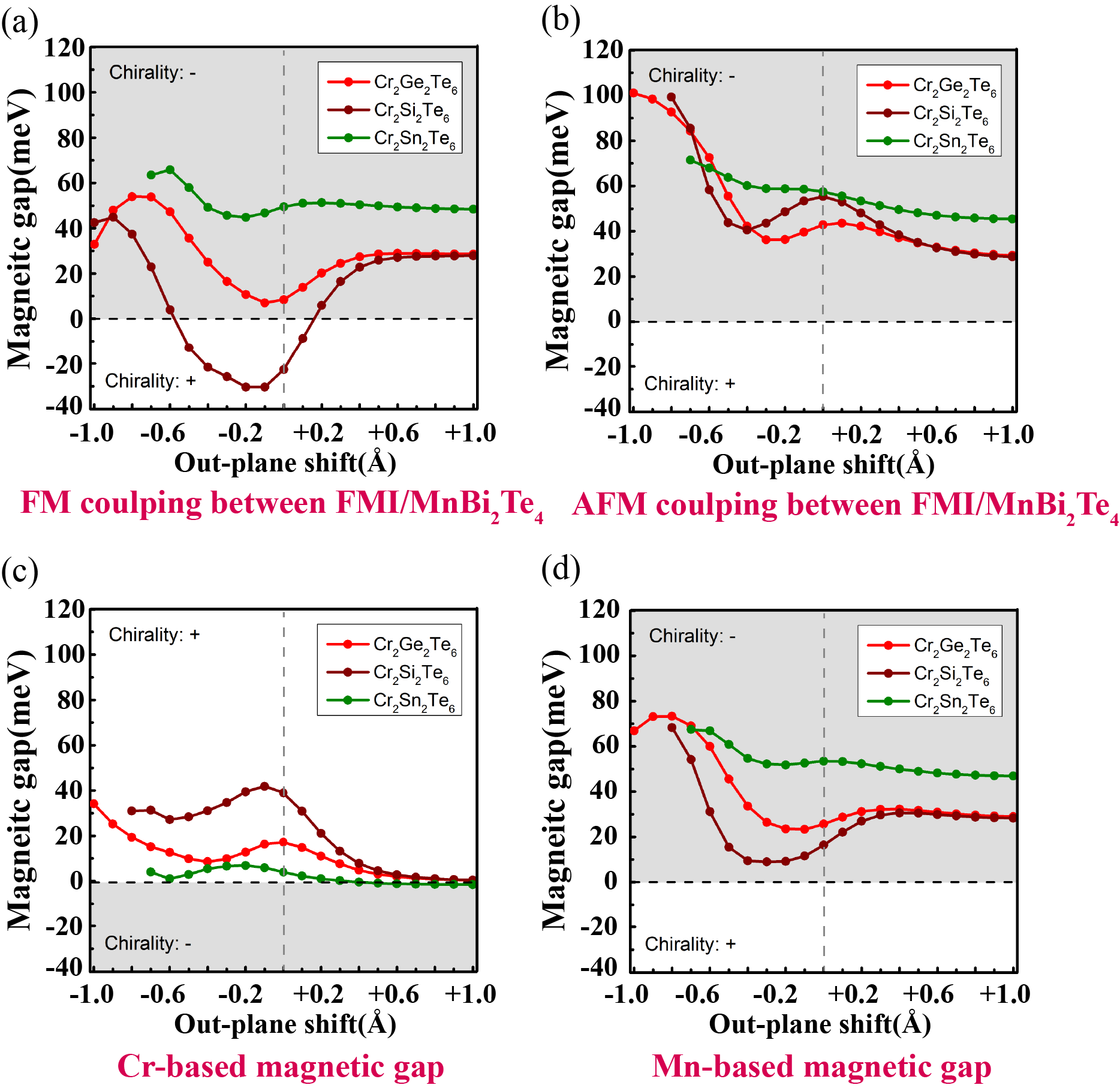}
	\caption{The evolution curves between various magnetic gaps and the vdW-spacings under the case of FMI/TL-MBT/FMI with both FM and AFM couplings between FMI and TL MBT. The magnetic-gap developments under (a) the FM coupling and (b) the AFM coupling between FMI and MBT layers. The extractions of (c) the Cr-based magnetic gaps and (d) the Mn-based magnetic gaps that evolve with vdW-spacing distances respectively. In all the figures, red, wine and olive curves are corresponding to FMIs as CGT, Cr$_2$Si$_2$Te$_6$ and Cr$_2$Sn$_2$Te$_6$, meanwhile white and light-gray zones respectively stand for the magnetic gaps with the positive and negative chirality. Light-gray dashed vertical line in each figure is positioned at the balanced distance of vdW-spacing (0.0Å).}
	\label{fig20:FMI-MBT-magneticgaps}
\end{figure*}

\begin{table}[h]
	\caption{\label{tab4:FMI-FMIs}Interlayer magnetic coupling strength between the two FMI layers in FMI/six-layer-BT/FMI, FMI/TL-BT/FMI and FMI/six-layer-ST/FMI. Positive and negative values are corresponding to FM and AFM couplings respectively, all derived from the equation: $E_{\rm ex}$ = $E_{\rm AFM}$ – $E_{\rm FM}$ per unit cell (UC) and in unit of meV.}
	\begin{ruledtabular}
		\begin{tabular}{ccc}
			TI-layer & FMI-layer & Exchange energy (meV/UC)
			\\ \hline
			\multirow{4}{*}{six-layer BT} & CGT & +0.009 \\
			& Cr$_2$Si$_2$Te$_6$ & +0.009 \\
			& Cr$_2$Sn$_2$Te$_6$ & +0.013 \\
			& CrI$_3$ & +0.12 \\ \hline
			\multirow{4}{*}{TL BT} & CGT & -0.155 \\
			& Cr$_2$Si$_2$Te$_6$ & -0.158 \\
			& Cr$_2$Sn$_2$Te$_6$ & -0.083 \\
			& CrI$_3$ & -0.044 \\ \hline
			\multirow{4}{*}{six-layer ST} & CGT & +0.007 \\
			& Cr$_2$Si$_2$Te$_6$ & -0.022 \\
			& Cr$_2$Sn$_2$Te$_6$ & -0.013 \\
			& CrI$_3$ & +0.006 \\
			
		\end{tabular}
	\end{ruledtabular}
\end{table}

\begin{table}[h]
	\caption{\label{tab5:FMI-MBTs}Interlayer magnetic coupling strength between FMI layers and Mn(Ni)BT layers in FMI/TL-MBT/FMI with AFM couplings between MBT layers, FMI/MBT/BL-BT/MBT/FMI and FMI/NiBT/BL-BT/MBT/FMI. Positive and negative values are corresponding to FM and AFM couplings respectively, all derived from the equation: $E_{\rm ex}$ = $E_{\rm AFM}$ – $E_{\rm FM}$ per UC and in unit of meV.}
	\begin{ruledtabular}
		\begin{tabular}{ccc}
			TI-layer & FMI-layer & Exchange energy (meV/UC)
			\\ \hline
			\multirow{4}{*}{TL MBT} & CGT & +48.071 \\
			& Cr$_2$Si$_2$Te$_6$ & +53.040 \\
			& Cr$_2$Sn$_2$Te$_6$ & +35.842 \\
			& CrI$_3$ & +34.357 \\ \hline
			\multirow{4}{*}{MBT/BL-BT/MBT} & CGT & +51.789 \\
			& Cr$_2$Si$_2$Te$_6$ & +57.262 \\
			& Cr$_2$Sn$_2$Te$_6$ & +38.499 \\
			& CrI$_3$ & +35.360 \\ \hline
			\multirow{3}{*}{NiBT/BL-BT/NiBT} & CGT & +79.088 \\
			& Cr$_2$Si$_2$Te$_6$ & +84.685 \\
			& Cr$_2$Sn$_2$Te$_6$ & +60.870 \\			
		\end{tabular}
	\end{ruledtabular}
\end{table}

\begin{table}[h]
	\caption{\label{tab6:MBT-MBTs}Interlayer magnetic coupling strength between the nearest Mn(Ni)BT layers in FMI/TL-MBT/FMI, FMI/MBT/BL-BT/MBT/FMI and FMI/NiBT/BL-BT/MBT/FMI, within all of which FM couplings are set between FMI and Mn(Ni)BT. Positive and negative values are corresponding to FM and AFM couplings respectively, all derived from the equation: $E_{\rm ex}$ = $E_{\rm AFM}$ – $E_{\rm FM}$ per UC and in unit of meV.}
	\begin{ruledtabular}
		\begin{tabular}{ccc}
			TI-layer & FMI-layer & Exchange energy (meV/UC)
			\\ \hline
			\multirow{4}{*}{TL MBT} & CGT & +1.622 \\
			& Cr$_2$Si$_2$Te$_6$ & +1.606 \\
			& Cr$_2$Sn$_2$Te$_6$ & +1.826 \\
			& CrI$_3$ & +1.431 \\ \hline
			\multirow{4}{*}{MBT/BL-BT/MBT} & CGT & +0.025 \\
			& Cr$_2$Si$_2$Te$_6$ & +0.014 \\
			& Cr$_2$Sn$_2$Te$_6$ & +0.035 \\
			& CrI$_3$ & +0.025 \\ \hline
			\multirow{3}{*}{NiBT/BL-BT/NiBT} & CGT & -0.115 \\
			& Cr$_2$Si$_2$Te$_6$ & -0.145 \\
			& Cr$_2$Sn$_2$Te$_6$ & -0.165 \\			
		\end{tabular}
	\end{ruledtabular}
\end{table}

\begin{figure*}
	\centering
	\includegraphics[width=0.6\linewidth]{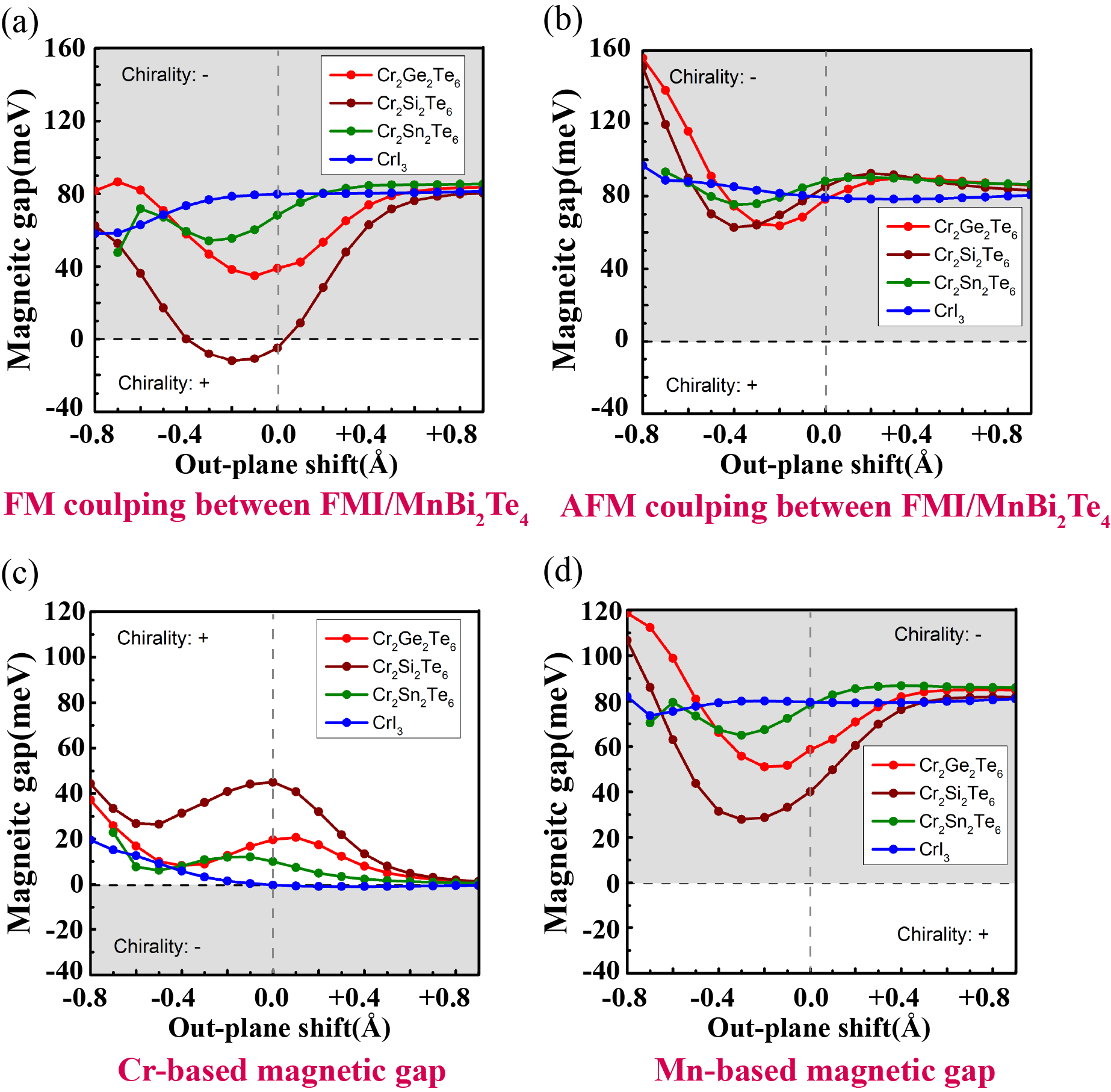}
	\caption{The evolution curves between various magnetic gaps and the vdW-spacings under the case of FMI/MBT/BL-BT/MBT/FMI with both FM and AFM couplings between FMI and the nearest MBT-layer. The magnetic-gap developments under (a) the FM coupling and (b) the AFM coupling between FMI and MBT layers. The extractions of (c) the Cr-based magnetic gaps and (d) the Mn-based magnetic gaps that evolve with vdW-spacing distances respectively. In all the figures, red, wine, olive and blue curves are corresponding to FMIs as CGT, Cr$_2$Si$_2$Te$_6$, Cr$_2$Sn$_2$Te$_6$ and CrI$_3$, meanwhile white and light-gray zones respectively stand for the magnetic gaps with the positive and negative chirality. Light-gray dashed vertical line in each figure is positioned at the balanced distance of vdW-spacing (0.0Å).}
	\label{fig21:FMI-MBT-2BT}
\end{figure*}

\begin{figure*}
	\centering
	\includegraphics[width=0.6\linewidth]{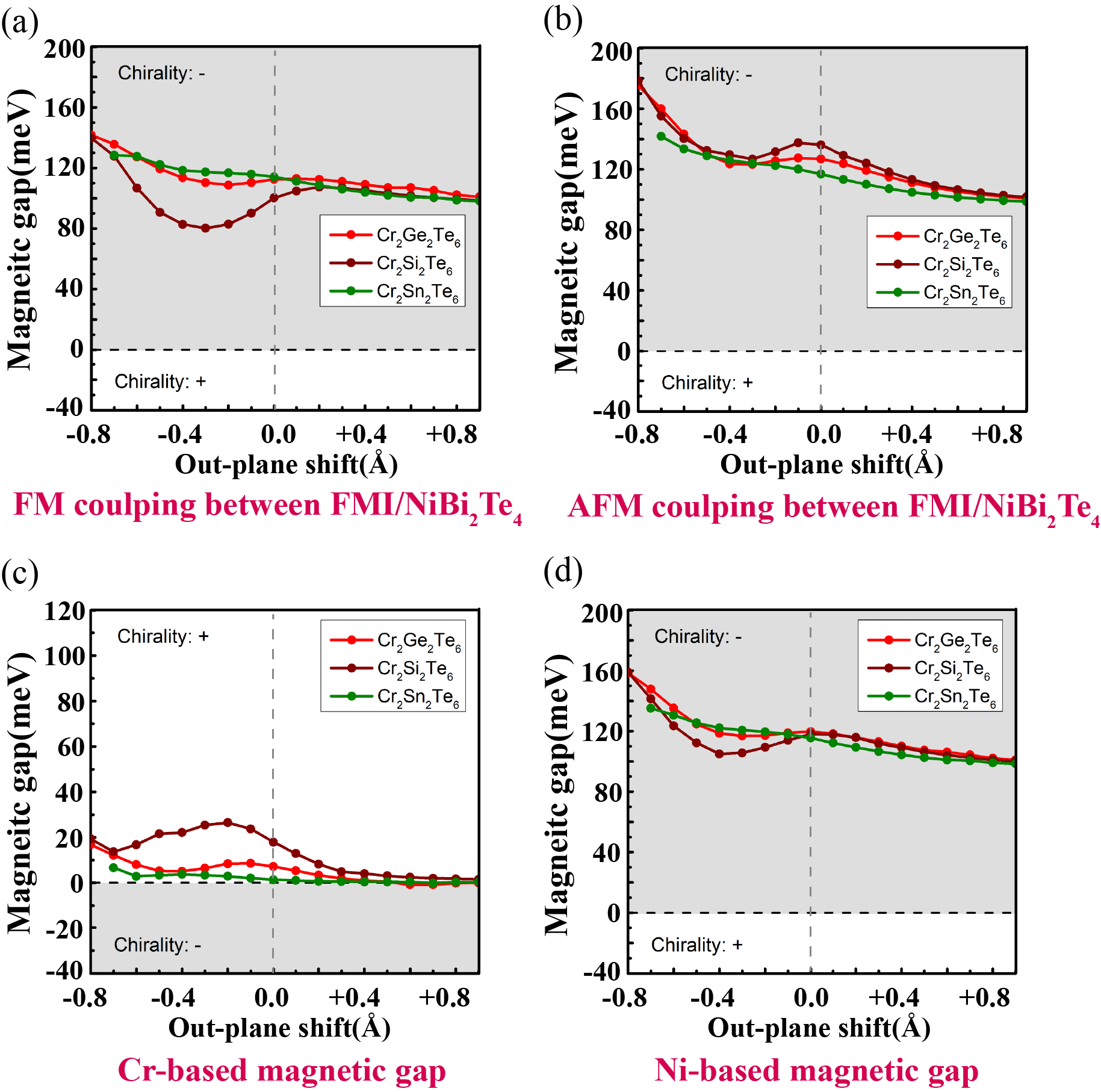}
	\caption{The evolution curves between various magnetic gaps and the vdW-spacings under the case of FMI/NiBT/BL-BT/NiBT/FMI with both FM and AFM couplings between FMI and the nearest NiBT-layer. The magnetic-gap developments under (a) the FM coupling and (b) the AFM coupling between FMI and NiBT layers. The extractions of (c) the Cr-based magnetic gaps and (d) the Ni-based magnetic gaps that evolve with vdW-spacing distances respectively. In all the figures, red, wine and olive curves are corresponding to FMIs as CGT, Cr$_2$Si$_2$Te$_6$ and Cr$_2$Sn$_2$Te$_6$, meanwhile white and light-gray zones respectively stand for the magnetic gaps with the positive and negative chirality. Light-gray dashed vertical line in each figure is positioned at the balanced distance of vdW-spacing (0.0Å).}
	\label{fig22:FMI-NiBT-2BT}
\end{figure*}

\begin{figure*}
	\centering
	\includegraphics[width=0.8\linewidth]{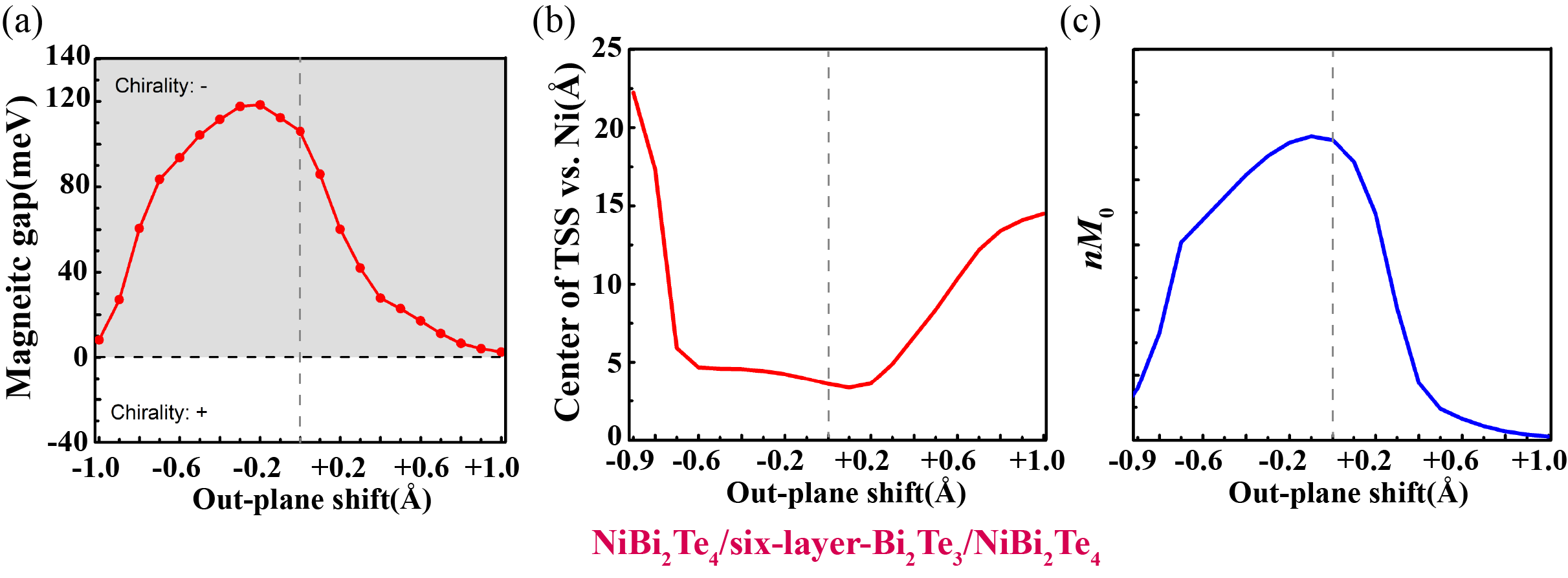}
	\caption{The evolution curves between various magnetic gaps and the vdW-spacings under the case of NiBT/six-layer-BT/NiBT. (a) The magnetic gaps, (b) the distance between the mean weighted center of TSS vs. Ni-atomic layer, (c) the values of $\textbf{\textit{n}}\textbf{\textit{M}}_\textbf{0}$ that evolves with the vdW-spacing distances. In (a), white and light-gray zones respectively stand for the magnetic gaps with the positive and negative chirality. In (b), the position of Ni-atomic layer is set as 0.0Å. Light-gray dashed vertical line in each figure is positioned at the balanced distance of vdW-spacing (0.0Å). }
	\label{fig23:NiBT-6BT}
\end{figure*}

Cr and Mn (Ni) contributes the opposite chirality of magnetic gaps selecting TI layer as magnetic topological insulators. Depicted in Fig. \ref{fig20:FMI-MBT-magneticgaps}, markedly the AFM-coupling based systems possesses large magnetic gaps than those of FM-coupling conditions around the whole range of vdW-spacing distances, indicating that Cr implements the opposite sign of magnetic gap compensating that generated by Mn. Via simple appropriate additions and subtracts, the separated magnetic gaps supplied by Cr and Mn can be extracted, both exhibited in Figs. \ref{fig20:FMI-MBT-magneticgaps}(c) and \ref{fig20:FMI-MBT-magneticgaps}(d) respectively with vdW-spacing modulated developments.

For the case of CrI$_3$, the highly confused bands mixed around the Fermi level hampers the extraction of the TSS-based four bands at $\Gamma$-point, therefore we omit this condition. The Cr$_2$Sn$_2$Te$_6$-based system suffers the same problem when the vdW-spacing distance is less than -0.7Å. Cr-based magnetic gap finds itself as the same behaviors as the increasing, decreasing and re-increasing process by decreasing vdW-spacing distance [Fig. \ref{fig20:FMI-MBT-magneticgaps}(c)], in the meantime Mn-based magnetic gap experiences conversely [Fig. \ref{fig20:FMI-MBT-magneticgaps}(d)] owing to the extraction-backflow character of TSS.

Noticeably, CGT and Cr$_2$Sn$_2$Te$_6$ fails to drive the negative sign of residue magnetic gaps to positive, while Cr$_2$Si$_2$Te$_6$ succeeds [Fig. \ref{fig20:FMI-MBT-magneticgaps}(a)], inspiring us to design Chern-number-chirality tunable devices in virtue of the competitive nature rooting in the opposite sign of magnetic gaps based on different magnetic elements.

In consideration of eliminating the contributions from the medial MBT layer, we artificially construct a more complicated heterostructure: FMI/MBT/BL-BT/MBT/FMI, with FM-coupling between FMI and MBT, to purely search the evolution behaviors of multiple-based magnetic gaps, results revealed in Fig. \ref{fig21:FMI-MBT-2BT}. Obviously, the negative-chirality of magnetic gaps increases entirely compared to that of FMI/TL-MBT/FMI, with the similar performance. Replacing Mn with Ni evokes much larger negative signed magnetic gaps, shown in Fig. \ref{fig22:FMI-NiBT-2BT}. Even opting Cr$_2$Si$_2$Te$_6$ fails to cancel the Ni-based magnetic gap totally by varying the vdW-spacing distance [Fig. \ref{fig22:FMI-NiBT-2BT}(a)]. In the NiBT-based case, choosing FMI by CrI$_3$ suffers severely from the convergence disaster, therefore we also abandon this condition.

Noteworthily, we acquiescently set FM-coupling as the magnetic ground state between FMI and Mn(Ni)BT when TI contains magnetism, and between two FMI layers when TI contains no magnetism. First of all, thick TI layer substantially weakens the interlayer magnetic couplings between the two edge layers of FMI to even ignorable value, all listed in Table \ref{tab4:FMI-FMIs}. Observably, the exchange values fall below 0.12meV in all conditions, indicating that we can regard these systems as no magnetic coupling. Similarly, BL BT intercalated nature also attenuates interlayer couplings between MBT-layers that below 0.04meV and between NiBT-layers that below 0.17meV respectively, shown in Table \ref{tab6:MBT-MBTs}. Nevertheless, FMI and Mn(Ni)BT form sizable strength of FM couplings stemming from $d$-electron occupation below 5 in Cr and equal or above 5 in Mn or Ni \cite{li2020tunable,tang2023intrinsic,fu2020exchange}, seen from Table \ref{tab5:FMI-MBTs}. Hence, we can utilize FM couplings both between two FMI layers, and between FMI and Mn(Ni)BT layers regarding as the ground states.

Additionally, we also check the building block: opting NiBT as FMI, and six-layer BT as TI, to investigate its magnetic-gap evolving performance. Monolayer NiBT loses the topological character suffering from the quantum confinement, acting as topologically trivial FMI accordingly \cite{liu2010oscillatory,bernevig2006quantum,otrokov2017highly}. Obviously depicted in Fig. \ref{fig23:NiBT-6BT}, in all the vdW-spacing range the chirality of the magnetic gaps provides the negative sign, dominated by the same chirality of kinetic exchange and Coulomb exchange mechanisms. The existence of non-monotonic evolving properties of the magnetic gaps indicates the universally available and applicable feature in the guidance put forward in this work. Figs. \ref{fig23:NiBT-6BT}(b) and \ref{fig23:NiBT-6BT}(c) exhibit the mean weighted positions (center) of the TSS, and the values of $\textbf{\textit{n}}\textbf{\textit{M}}_\textbf{0}$ that evolve with the vdW-spacing distances. Markedly, the maximum of the magnetic gap (-0.2Å) appears later than the most extracted state of TSS (+0.1Å) and even later than the peak of $\textbf{\textit{n}}\textbf{\textit{M}}_\textbf{0}$ (-0.1Å) as the vdW-spacing descends. This phenomenon intimates that if the sign of Coulomb exchange and kinetic exchange interactions meets the same, no chirality inversion appears, in which the extremal value of the magnetic gap comes later when the peak value of $\textbf{\textit{n}}\textbf{\textit{M}}_\textbf{0}$ also comes no prior to the happening of maximal TSS extraction.

\section{Other related outcomes under stacking-order-shifts   \label{E:stacking-order}}

\begin{figure*}
	\centering
	\includegraphics[width=0.9\linewidth]{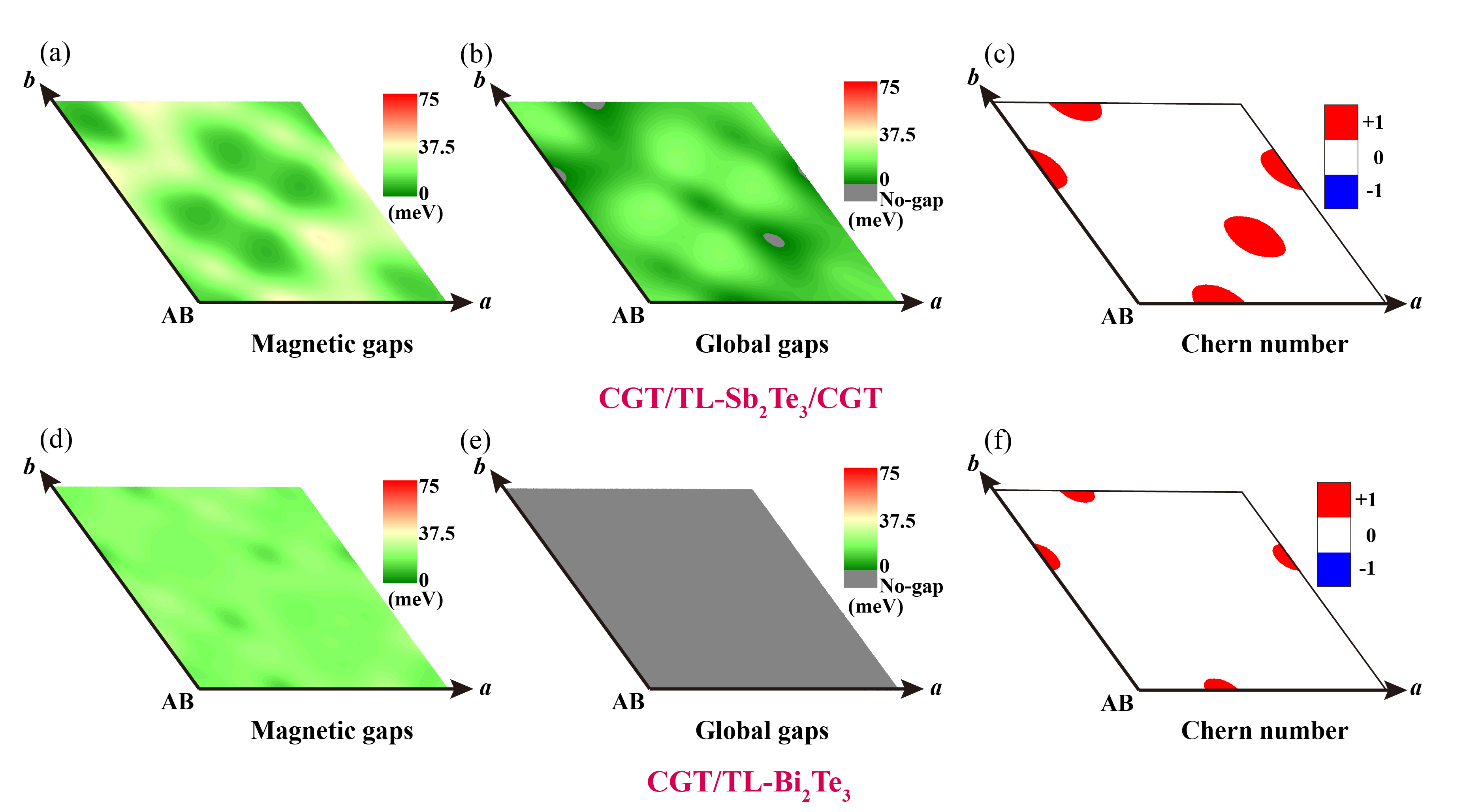}
	\caption{Magnetic-gap, global-gap and Chern-number manipulations in CGT/TL-ST/CGT and CGT/TL-BT induced by stacking-order-shifts. Distributions of (a) the magnetic gaps, (b) the global gaps and (c) the Chern numbers via employing stacking-order-shifts of FMI layers under the building-block of CGT/TL-ST/CGT. In gap-distributions, green-yellow-red color bar is adopted, with the gray color standing for “no-gap” zones. In Chern-number distributions, blue, white and red colors stand for the Chern number as -1, 0, +1 respectively. (d)-(f) are similar to (a)-(c), but under the building-block of CGT/TL-BT. }
	\label{fig24:CGT-3ST-stacking}
\end{figure*}

\begin{figure*}
	\centering
	\includegraphics[width=0.7\linewidth]{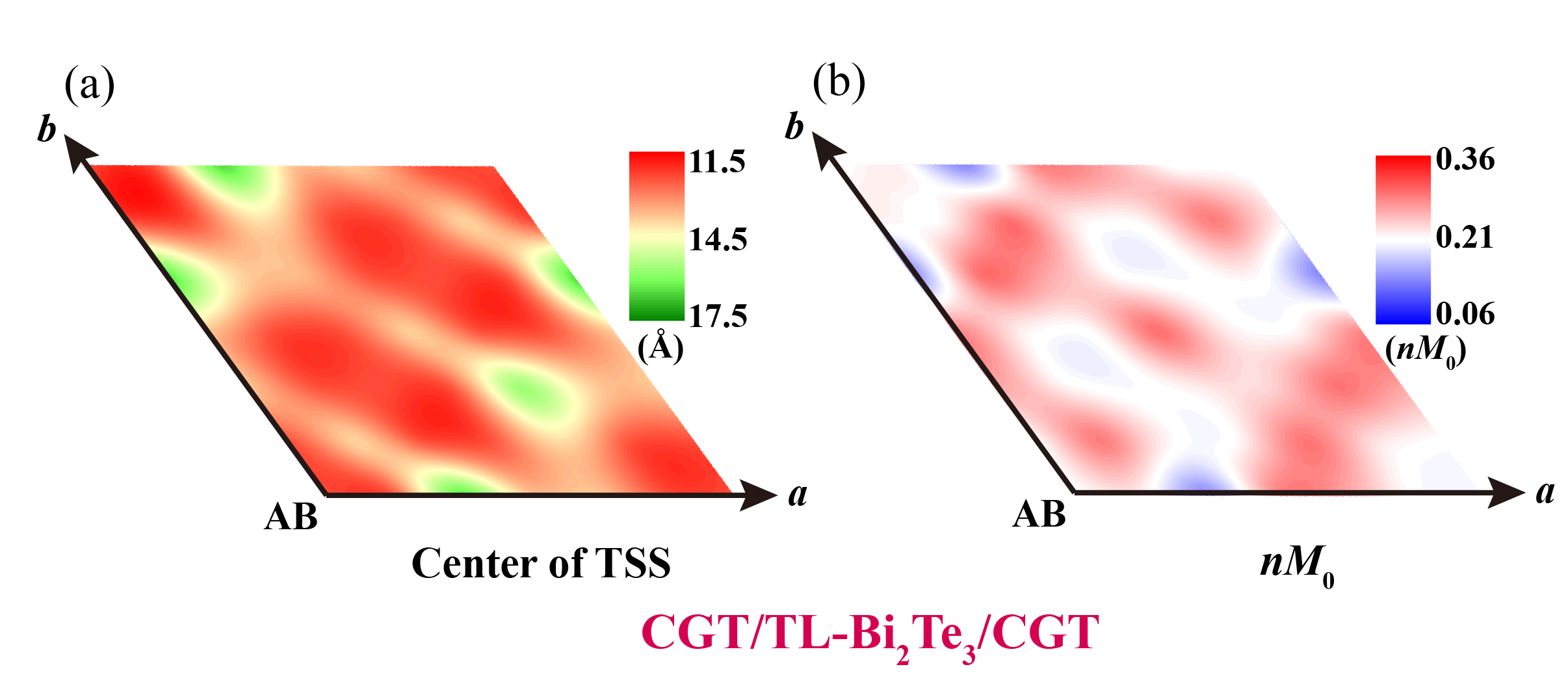}
	\caption{(a) The weighted mean position evolutions of one half of TSS and (b) the value of $\textbf{\textit{n}}\textbf{\textit{M}}_\textbf{0}$ in CGT/TL-BT/CGT distributing along stacking-order-shifts. In TSS-center-distributions, green-yellow-red color bar is adopted corresponding to the “far to near” distances between the center of TSS and the layer of CGT. Positive value means the center of TSS located at the right side of Cr. In $\textbf{\textit{n}}\textbf{\textit{M}}_\textbf{0}$ distributions, blue-white-red color bar describes its value from low to high.}
	\label{fig25:TSS-stacking}
\end{figure*}

\begin{figure*}
	\centering
	\includegraphics[width=0.9\linewidth]{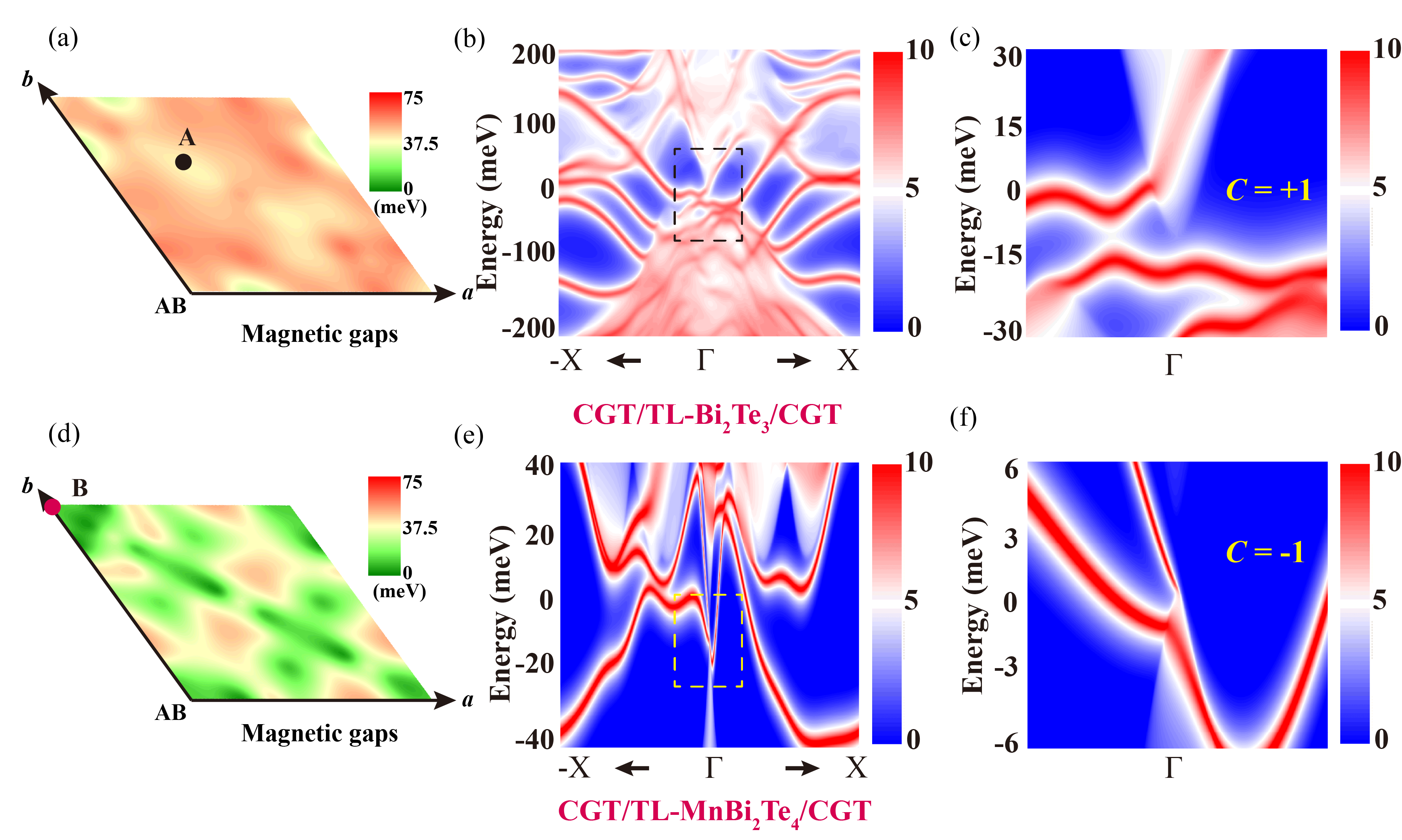}
	\caption{Verifying the Chern numbers of CGT/TL-BT/CGT and CGT/TL-MBT/CGT. (a) The magnetic-gap distributions of CGT/TL-BT/CGT. “Point A” means the stacking-order-point that adopted to compute the topological edge states exhibited in (b) large-region pattern and (c) zoom-in pattern respectively. (d)-(f) are similar with (a)-(c), but under the case of CGT/TL-MBT/CGT in which “Point B” is utilized. }
	\label{fig26:LDOS}
\end{figure*}

\begin{figure*}
	\centering
	\includegraphics[width=0.9\linewidth]{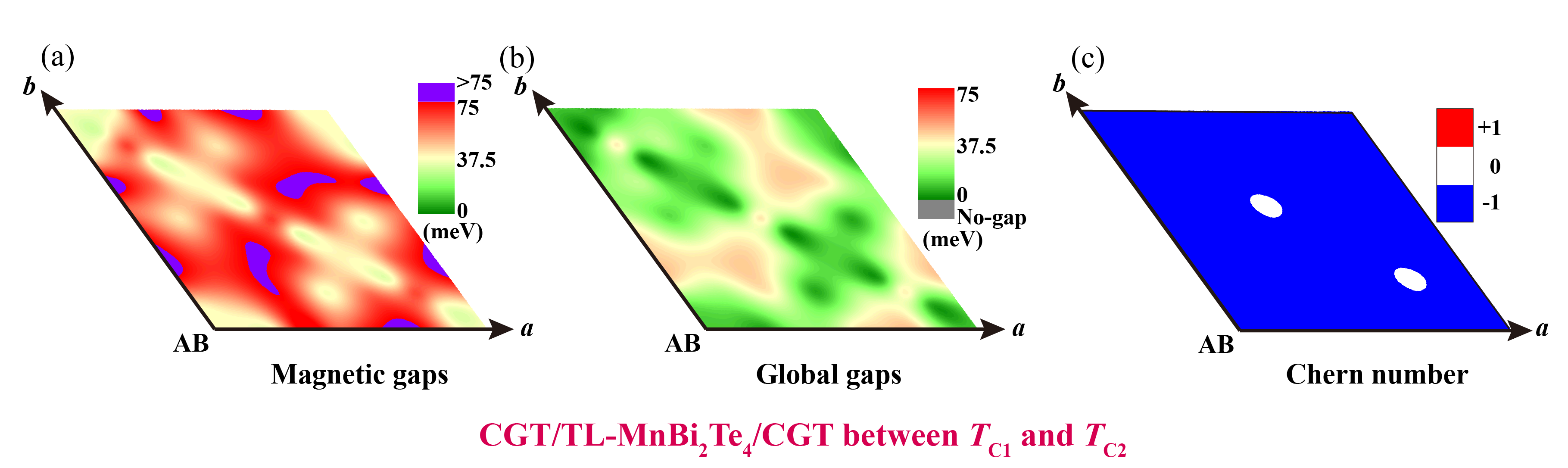}
	\caption{(a) The magnetic gaps, (b) the global gaps and (c) the Chern number distributions of CGT/TL-MBT/CGT between $T_{\rm C1}$ and $T_{\rm C2}$. In (a), purple regions are associated with the magnetic gaps higher than 75meV. }
	\label{fig27:High-T CGT-3MBT}
\end{figure*}

\begin{figure*}
	\centering
	\includegraphics[width=0.43\linewidth]{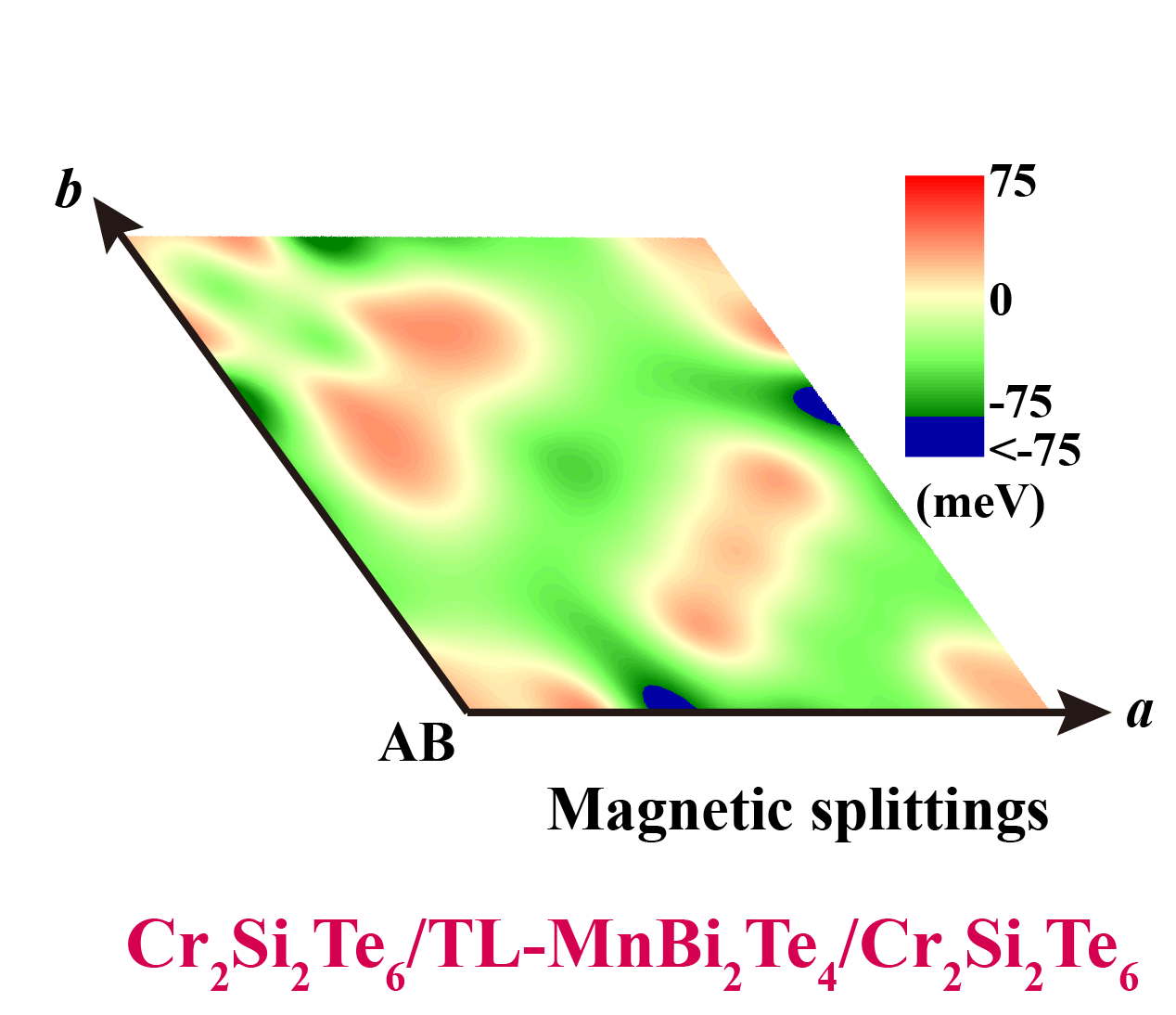}
	\caption{Magnetic Zeeman splitting values of Cr$_2$Si$_2$Te$_6$/TL-MBT/Cr$_2$Si$_2$Te$_6$ that considers the chirality. In this mapping pattern, no-splitting denotes with yellow-colored region, meanwhile the red (green) color is related to positive (negative) chirality of the magnetic splitting. Dark-blue region illustrates the negative-chirality of the magnetic gap higher than 75meV. }
	\label{fig28:Zeeman-splitting}
\end{figure*}

In this section, we list the residue results of mapping distributions related to those detailly discussed within Figs. \ref{fig6:CGT-BT}-\ref{fig9:CrSiTe3-MBT-stacking} Considering TL ST as TI layer firstly in Figs. \ref{fig24:CGT-3ST-stacking}(a)-(c) with FMI as CGT, conquering the “M”-shaped valence band that emerged in BT releases the global gap in most regions [Fig. \ref{fig24:CGT-3ST-stacking}(b)], but its larger hybridization gap almost totally destroys its Chern-insulating features, only leaving few non-trivial islands that corresponding to the no-global gap regions. By constructing bilayer-like CGT/TL-BT heterostructures instead of sandwich structures, the breaking of inversion symmetry totally ruins the global gap in all the stacking-order region [Fig. \ref{fig24:CGT-3ST-stacking}(e)], with the even increasing of the hybridization gaps consumes Chern-insulating regions except for only the “$AA$”-stacking point along [100] direction of stacking-order-shift. Hence, the above two systems fail to make themselves excellent candidates for tunable magnetic gaps of Chern insulators.

We also cast the distributions including the weighted mean position, or that’s to say, the center of TSS (one half) and value of “$\textbf{\textit{n}}\textbf{\textit{M}}_\textbf{0}$” in Fig. \ref{fig25:TSS-stacking}, in correspondence with those of CGT/TL-BT/CGT shown in Fig. \ref{fig6:CGT-BT}. Similar with vdW-spacing variations, the red-color zone doesn’t coincide with each other between the center of TSS [Fig. \ref{fig25:TSS-stacking}(a)] and the value of “$\textbf{\textit{n}}\textbf{\textit{M}}_\textbf{0}$” [Fig. \ref{fig25:TSS-stacking}(b)]. Besides, the large magnetic-gap region [Fig. \ref{fig6:CGT-BT}(d)] approximately fits with the medial values of TSS-center region [Fig. \ref{fig25:TSS-stacking}(a)], directly indicating that stacking-order-shifts perform as an analogous method with the vdW-spacing to reform the real-space TSS-distributions, and then, the value of the magnetic gaps.

In the interest of verifying the concrete value of Chern number, we select one stacking-order point of CGT/TL-BT/CGT [shown in Fig. \ref{fig26:LDOS}(a) noted as “Point A”] positioning at the 1/3 point along [11$\overline{0}$] direction, and the normal stacking-order point of CGT/TL-MBT/CGT [depicted in Fig. \ref{fig26:LDOS}(d) noted as “Point B”]. Topological edge-state behaves complicatedly in both cases, but we can even distinguish the total Chern number after neutralizing left- and right-handed chiral edge states, and provides the conclusion that $C$ = +1 for the former and $C$ = -1 for the latter cases respectively, seen from Figs. \ref{fig26:LDOS}(b) and \ref{fig26:LDOS}(e).

The absence of the magnetism in the medial MBT-layer of CGT/TL-MBT/CGT significantly enhances both the magnetic gaps and the global gaps in the whole stacking-order zone [Figs. \ref{fig27:High-T CGT-3MBT}(a) and Fig. \ref{fig27:High-T CGT-3MBT}(b)], broadening the nonzero-Chern insulating zones accordingly drawn in Fig. \ref{fig27:High-T CGT-3MBT}(c). Among the temperature between $T_{\rm C1}$ [11K, see Fig. \ref{fig11:FMI-3BT-FMT-Cv}(c)] and $T_{\rm C2}$ [36K, see Fig. \ref{fig11:FMI-3BT-FMT-Cv}(c)], the confinements from the interlayer magnetic coupling between MBT and the intralayer couplings within the medial MBT-layer break, leading to the above condition. Hence, if the TI layer contains weak AFM-coupling nature, moderate external temperature is capable of amplifying magnetic gaps independence of the stacking orders.

In the Fig.  \ref{fig9:CrSiTe3-MBT-stacking} of main text, choosing Cr$_2$Si$_2$Te$_6$ as the FMI layer, the antagonism between the Mn-based and Cr-based magnetic gaps brings impediment to observing the extent of magnetic splitting. Considering the chirality, we draw the magnetic-splitting mapping along the stacking order in Fig. \ref{fig28:Zeeman-splitting}. Red (green) regions are in correspondence with positive (negative) sign of the magnetic gaps, or that is to say, large (small) gaps opened by Cr. Noticeably, these red regions fit well with the positive Chern number regions revealed in Fig. \ref{fig9:CrSiTe3-MBT-stacking}(c), howbeit the green regions are exactly the negative or zero Chern number zones in Fig. \ref{fig9:CrSiTe3-MBT-stacking}(c).

\section{Other related outcomes of critical temperatures  \label{F:Tc}}

\begin{figure*}
	\centering
	\includegraphics[width=0.35\linewidth]{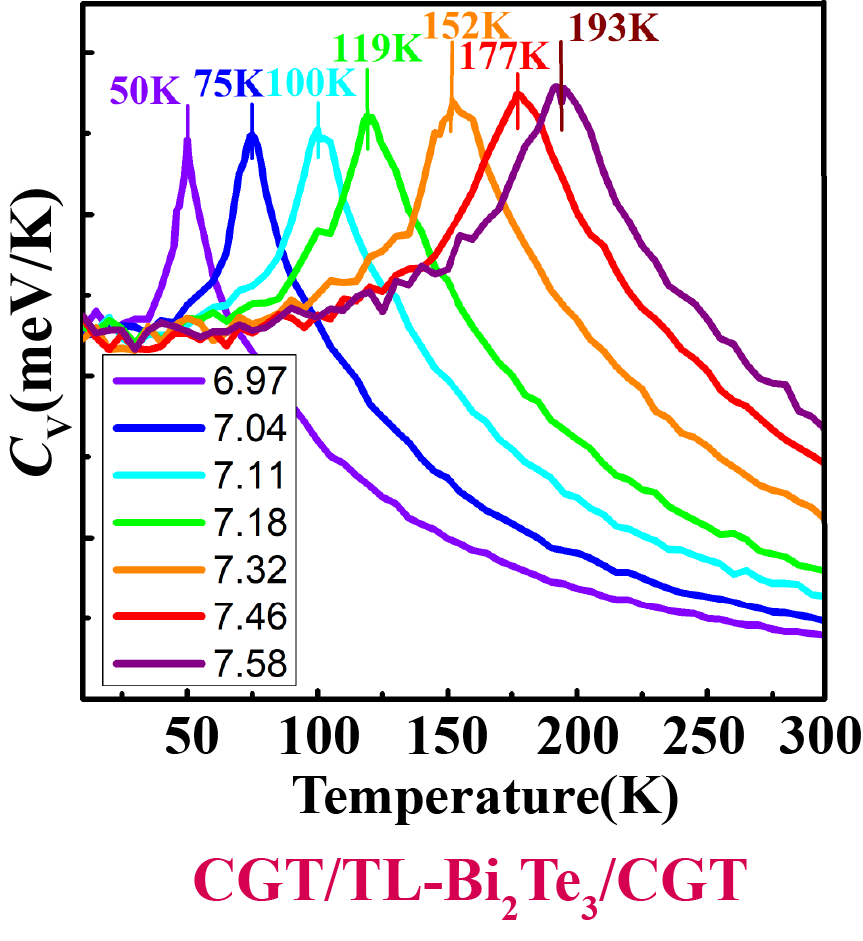}
	\caption{$C_{\rm V}$-$T$ curves of monolayer CGT under various in-plane lattice constants. The violet, blue, cyan, green, orange, red and purple curves are linked to in-plane lattice constants as 6.97Å, 7.04Å, 7.11Å, 7.18Å, 7.32Å, 7.46Å and 7.58Å respectively.}
	\label{fig29:CvT-CGT}
\end{figure*}

\begin{figure*}
	\centering
	\includegraphics[width=1\linewidth]{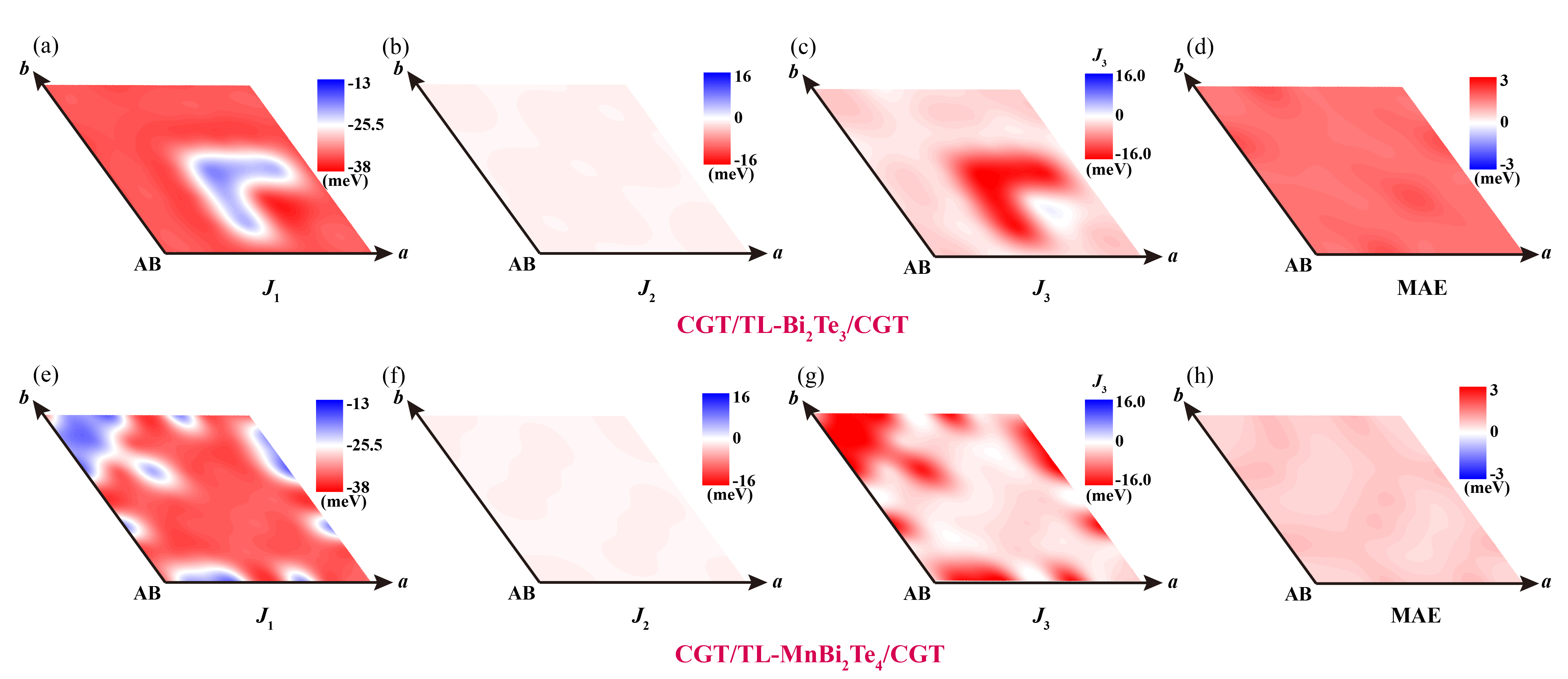}
	\caption{$J_1$, $J_2$, $J_3$ (for intralayer couplings of CGT) and MAE distributions of CGT/TL-BT/CGT and CGT/TL-MBT/CGT along stacking-order-shift. Distributions of (a) $J_1$, (b) $J_2$, (c) $J_3$ and (d) MAE of CGT/TL-BT/CGT. In first-order intralayer couplings ($J_1$) within CGT layer, blue-white-red colors stand for the FM-coupling strength from weak to strong. In $J_2$ and $J_3$, blue-white-red colors stand for the AFM-coupling, no-coupling to FM-couplings. Negative (positive) value means the FM (AFM)-coupling. In MAE, the red color spread in the whole stacking-order region indicates the out-of-plane magnetism as the ground state. (e)-(h) are same to (a)-(d) but under the condition of CGT/TL-MBT/CGT.}
	\label{fig30:J-mapping}
\end{figure*}

\begin{figure*}
	\centering
	\includegraphics[width=0.6\linewidth]{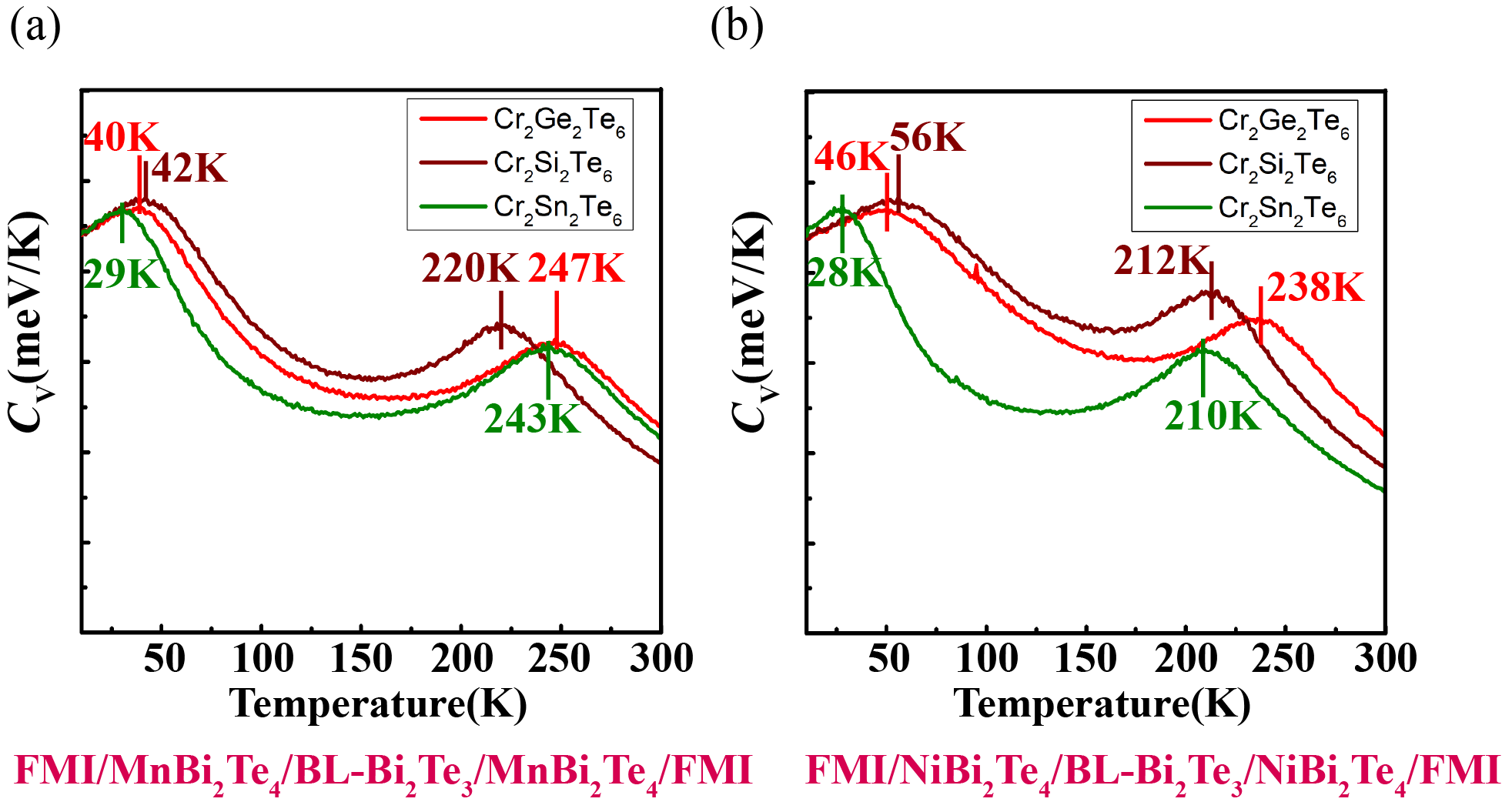}
	\caption{$C_{\rm V}$-$T$ curves of (a) FMI/MBT/BL-BT/MBT/FMI and (b) FMI/NiBT/BL-BT/NiBT/FMI. Red, wine and olive curves are corresponding to FMIs as CGT, Cr$_2$Si$_2$Te$_6$ and Cr$_2$Sn$_2$Te$_6$ respectively.}
	\label{fig31:CvT-CGT-MBT-2BT}
\end{figure*}

For monolayer CGT, the Curie temperature increases significantly relying on the increasing of in-plane lattice constant, depicted in Fig. \ref{fig29:CvT-CGT}. Obviously, at the same in-plane lattice constant, for instance, \textbf{\textit{a}} = 7.58Å, monolayer-CGT possess the Curie temperature as 193K, lower than that grown upon BT with 244K. This discrepancy mainly originates from the enhancement of $J_3$ when the substrate of BT layer exists. Analogously, via stacking-order-shift, the $T_{\rm C}$ of CGT itself mainly rests on the value of $J_3$ exactly, comparing Fig. \ref{fig30:J-mapping}(c) with Fig. \ref{fig10:CGT-BT-Tc}(c), and Fig. \ref{fig30:J-mapping}(g) with Fig. \ref{fig12:FMI-MBT-Tc2}(c). Looking into Fig. \ref{fig30:J-mapping}, $J_2$ and MAE values almost have no variations along the stacking-order-shifts, meanwhile $J_1$ behaves oppositely with $J_3$. This phenomenon unmasks the critical role of the third-nearest-interactions within CGT itself that determines its final $T_{\rm C}$, of equal importance with that of $J_1$.

Finally, we estimate the $C_{\rm V}$-$T$ curves of FMI/M(Ni)BT/BL-BT/M(Ni)BT/FMI in Fig. \ref{fig31:CvT-CGT-MBT-2BT} in the absence of medial magnetic TI layer that containing the opposite directed magnetism. The two peaks in each curve stand for the $T_{\rm C}$ of interlayer magnetic coupling between CGT and Mn(Ni)BT, and CGT itself. No obvious increasement appears compared to that of CGT/TL-MBT/CGT [Fig. \ref{fig11:FMI-3BT-FMT-Cv}(c)]. Even choosing NiBT that induces much stronger interlayer couplings cause no marked improvement of the two $T_{\rm C}$ points. Consequently, the critical temperatures, especially the highest $T_{\rm C}$ points of these system mainly depend on the CGT layer, in which both the in-plane lattice constant and the emerge of “substrate” like BT, MBT, perform the dominant roles.

\newpage
\nocite{*}
\bibliography{Main_text}

\end{document}